\documentclass[12pt]{article}
\usepackage[utf8]{inputenc}
\usepackage[export]{adjustbox}
\usepackage{amssymb}
\usepackage{authblk}
\usepackage{bbm}
\usepackage{pgfpages}
\usepackage{graphicx}
\usepackage{multicol}
\usepackage{csquotes}
\usepackage{multirow}
\usepackage{graphicx} 
\usepackage{booktabs} 
\usepackage{authblk}
\usepackage{caption}
\usepackage{subcaption}
\usepackage{stackengine}
\usepackage{amsmath}
\usepackage{mathabx}
\PassOptionsToPackage{hyphens}{url}\usepackage{hyperref}
\usepackage{comment}
\usepackage[capposition=top]{floatrow}
\usepackage{hyperref}
\usepackage{colortbl}
\hypersetup{
    colorlinks=true,
    linkcolor=blue,
    filecolor=blue,      
    urlcolor=blue,
    citecolor=blue,
}
\usepackage[capposition=top]{floatrow}

\usepackage[spanish,english]{babel}

\usepackage{setspace}

\usepackage{stackrel}
\usepackage{natbib} 

\usepackage[a4paper, margin=0.75in]{geometry}
\def\footnoterule{\vfill 
   \kern-3pt\hrule width 2truein \kern 2.6pt}
   
\title{Non-Linearities in International Spillovers of the ECB's Monetary Policy\\
\begin{large}
    The Case of Non-ERM II Countries and Anti-Fragmentation Policy.
\end{large}}

\author[1]{Iones Kelanemer Holban\footnote{ 855 Sherbrooke St W, Montreal, Quebec H3A 2T7, Canada.}}

\affil[1]{McGill University }

\date{\today}

\begin{document}
\onehalfspacing
\date{}
\maketitle

\begin{abstract}
We investigate the presence of sign and size non-linearities in the impact of the European Central Bank's Anti-Fragmentation Policy on non-ERM II, EU countries. After identifying three orthogonal monetary policy shock using the method of \cite{fanelli2022sovereign}, we then select an optimal specification and estimate both linear and non linear impulse response functions using local projections (\cite{dufour_reneault_1998}, \cite{Goncalves2021}). The choice of non-linear transformations to separate sign and size effects is based on \cite{CaravelloMartinezWP}. Lastly we compare the linear model to the non-linear ones using a battery of Wald tests and find significant evidence of sign non-linearities in the international spillovers of ECB policy. 

\medskip

\end{abstract}

\newpage
\section{Introduction} \label{sec:introduction}
In a world where financial markets are connected internationally, monetary policy spills over to neighbouring countries. In the context of American monetary policy the question, has been studied originally by \cite{fleming1962domestic}, \cite{mundell1963capital} and \cite{dornbusch1976expectations}. More recent papers such as \cite{JAROCINSKI2022} and \cite{Camara2023US} use a high-frequency identification strategy to look at the spillovers to the European Union and Emerging Markets respectively. The European central bank has also been shown to produce international spillovers. For example, \cite{JAROCINSKI2022} finds that ECB conventional monetary policy produce small spillovers to the US and \cite{Miranda-Agrippo-Nenova2022} finds spillovers from unconventional monetary policy. \\[\baselineskip]
In this paper we study the impact of identified monetary policy shocks from the European Central Bank on EU countries that are not part of the European Exchange Rate Mechanism from 2002 to 2019. That is, countries in the EU but outside the Euro-zone, that have a free floating currency. This is related to \cite{hajekHorvath2018}, which studies the spillovers to non euro-zone eu countries. However, instead of proxying conventional and unconventional monetary policy with the shadow rate, we use the identification strategy of \cite{fanelli2022sovereign} to recover three orthogonal monetary policy shocks. \textbf{1)} A pure monetary policy shock, which represents the actual effect of monetary policy (borrowing costs). \textbf{2)} An information shock, which represents the information about the current and future state of the economy that can be inferred from monetary policy (overheating, stagnation). \textbf{3)} A spread shock, which captures anti-fragmentation policies of the ECB. We are interested in the presence of non-linearities in the effects of these monetary policy shocks motivated by the fact the ECB's main refinancing rate was close to the zero lower bound for nearly a decade (2012-2022). \\[\baselineskip]
In section 1.1, we go over the European Central Bank's anti-fragmentation policy and the implication of the zero-lower bound on the non-linear behavior of monetary policy. In section 2, we give a detailed description of our data. Section 3, goes over our econometric methodology, including the postulated underlying data generating process, the linear and non-linear models and the model selection. Section 4, provides the results of the impulse response functions estimation and lastly, section 5, provides inference on the non-linear specifications, both about the presence of non-linearities and the significance of the IRF itself .

\subsection{Anti-Fragmentation and the Zero Lower Bound}
Fragmentation in the sovereign debt markets in the EU refers to the phenomenon where bond spreads between countries with higher levels of debt to GDP (Greece, Portugal, Spain, Italy) and Germany increase in periods of monetary contractions (or transition from expansion to a more neutral stance). For example, the Portuguese-German spread in the 10-year bond rate reached nearly $14 \%$ and more famously, the Greek-German spread reached $32\%$\footnote{\url{https://www.europarl.europa.eu/RegData/etudes/STUD/2022/733984/IPOL_STU(2022)733984_EN.pdf}}. Sovereign spreads indicate different market expectations of default risks between countries. Therefore, their presence suggests that the European financial markets are not fully integrated. Contractionary monetary policy from the ECB will affect riskier countries more adversely. Meaning, the transmission of monetary policy is harmed (\cite{Wyplosz2022}). To combat this, the ECB began intervening in sovereign debt markets, starting in 2010, with the securities market program (SMP), as detailed in \cite{Dabrowski2022}. With this program, Euro-zone countries' central banks could purchase government assets to lower the cost of borrowing in their respective sovereign debt markets. In Greece, the program led to a decrease in the 5 year rate of up to $0.21\%$ for every billion euros in purchases without however, increasing the quantitaty of money in circulation as: ``Any additional monetary base created [...] was to be sterilised by the ECB in its open-market operations" (\cite{Dabrowski2022}). Other programs eventually replaced the SMP such as the Outright Monetary Transaction (OMT) program and the Emergency Liquidity assistance (ELA), all with the same goal of reducing fragmentation risk. In July 2022, the ECB unveiled the Transmission Protection Instrument (TPI), a program allowing the purchase of governmental securities with maturities of one to ten years\footnote{\url{https://www.ecb.europa.eu/press/pr/date/2022/html/ecb.pr220721~973e6e7273.en.html}}.\\[\baselineskip]
Estimating the spillovers of anti-fragmentation and conventional monetary policy is complicated by the fact that since 2012, the ECB's refinancing rate is near the zero-lower bound (ZLB). Recent research suggests that dynamic stochastic general equilibrium models should be estimated with non-linear methods to replicate outcomes near the ZLB \cite{atkinsonRichterThrockmorton2020}, \cite{fernandezGordonGuerronRubio2015}, \cite{wuXia2016}. Other papers, such as \cite{caggianocastelnuovopellegrino2017},  use non-linear transformation of variables in vector auto regressive models to study the behavior of the economy near the ZLB. Our paper is therefore related in the sense that we use the ZLB to justify investigating the presence of non-linearities in the effects of monetary policy.

\section{Data} \label{sec:data_methodology}

\subsection{Country Specific Variables}
We build an unbalanced panel of European Union countries at the monthly frequency from the period of January 2002 to February 2023 (Appendix D, presents the results for the pre-pandemic sub-sample January 2002 to December 2019). The panel includes countries that satisfies the following conditions: \textbf{1)} the country is in the European Union, \textbf{2)} the country has a free floating currency that is not the Euro, that is, countries that are not part of the European Exchange Rate Mechanism (ERM 2). There are 5 countries that satisfy these conditions: Romania, Poland, Hungary, Czechia and Sweden. 
Table 1, shows the composition of the unbalanced panel. \\
\begin{table}[h!]
    \centering
    \small
    \begin{tabular}{c|c|c}
\textbf{Country} &\textbf{Entry Date}			&	\textbf{Exit Date} \\ \hline \hline
Czechia	&	2002M01	&	2023M02	\\
Hungary	&	2002M01	&	2023M02	\\
Poland	&	2002M01	&	2023M02	\\
Romania	&	2005M04	&	2017M09	\\
Sweden	&	2002M01	&	2017M05\\ \hline \hline
    \end{tabular}
    \caption{Composition of the Panel}
    \label{tab:country_list_Adv}
\end{table}\\
For each country we first collect 4 endogenous variables from the IMF dataset: a seasonally adjusted industrial production index, the consumer price index, the unemployment rate and a measure of the long term interest rate. We remove the seasonality of unemployment by projecting it on the orthogonal complement of the column space of a matrix whose columns are monthly dummies.\\[\baselineskip]
We then add a trade-valued real exchange rate to every country in the panel, which is sourced from Bruegel\footnote{\url{https://www.bruegel.org/publications/datasets/real-effective-exchange-rates-for-178-countries-a-new-database}} . This measure is a weighted average based on a basket of 120 trading partners of each country. 
\subsection{Euro Area Control Variables}
We collect 4 control variables for the euro area. An aggregate of the consumer price index is taken from the IMF dataset. Seasonally adjusted unemployement rate and industrial production is taken from the OECD dataset. Lastly, we also include the 120 countries trade valued real exchange rate for the Euro from Bruegel. However, the trading partner weights are computed based on the first 12 countries of the Eurozone. Our panel's countries are not part of this list. 

\subsection{Identified Shocks}
To extract our identified shocks we follow the three-shocks methodology of \cite{fanelli2022sovereign}. They extract three exogenous shocks from high-frequency movements of prices on a portfolio of European financial assets. These shocks are 1) a pure monetary policy shock, 2) a central bank information shock and 3) a spread shock attributed to the ECB's anti-fragmentation policy. The high-frequency surprises are calculated as the difference between the median quote price in a 10 minute window 20 minute before the start of an ECB press conference and the median quote in a 10 minute window 55 minutes after the start of the conference. This difference will be interpreted as a ``surprise" unanticipated by the market. They postulate that the three exogenous shocks are related to these high-frequency surprises by the static factor model:\\
\begin{small}
\begin{align*}
\Delta_t= &\Lambda f_t+e_t\\
\begin{bmatrix}
\Delta OIS_{1y,t} \\
\Delta OIS_{2y,t} \\
\Delta OIS_{5y,t} \\
\Delta OIS_{10y,t} \\
\Delta \left(IT_{2y,t}-OIS_{2y,t}\right) \\
\Delta \left(IT_{5y,t}-OIS_{5y,t}\right) \\
\Delta \left(IT_{10y,t}-OIS_{10y,t}\right) \\
\Delta STOXX50_t
\end{bmatrix}=&
\begin{bmatrix}
 + & + & + \\
+ & + & + \\
+ & + & + \\
+ & \bullet & + \\
+ & - & -- \\
+ & - & -- \\
+ & - & -- \\
- & + & +
\end{bmatrix} 
\begin{bmatrix}
    f_{mp} \\
    f_{inf} \\
    f_{spread} 
\end{bmatrix}
+e_t
\end{align*}
\end{small}
$\Delta OIS_{iy,t}$, is the surprise in the price of overnight indexed swaps on the ``i" years interest rate, $\Delta IT_{iy,t}$ is the surprise on the rate of Italian bonds of ``i" year maturity and $\Delta STOXX50$ is the surprise in the STOXX50 a Eurozone stock exchange index. These variables come from the EA-MPD dataset established by \cite{Altavilla2019}. \\[\baselineskip]
The matrix of factor loadings $\Lambda$ and the factors $f_t$ are jointly estimated by maximum likelyhood. However, if $\Lambda^*$ and $f_t^*$ are the maximum likelyhood estimates, post multiplying $\Lambda$ by an orthonormal matrix $Q$ and premultiplying $f_t$ by the transpose of Q also satisfies the same equation. Therefore, the authors introduce sign restrictions on the factor loadings which allow set identification of theoretically sound factors. $+$ and $-$ indicates positive and negative co movement respectively. $\bullet$ represents the absence of restrictions and $--$ indicates that the magnitude of the movement must be greater than all other factors.  To sum up, the pure monetary policy shock ($f_{mp}$) coincides with higher rates, decreases in the stock index and an increase in the spreads (due to reasons outlined in section 1.1). A positive information shock ($f_{inf}$) announces to the markets that the economy is overheating. Therefore, it is associated with an increase in the stock indexes and a decrease in the spreads. Lastly, a positive spread shock ($f_{spread}$) should lower the spreads and increase the stock indexes.\\[\baselineskip]
We estimate the factor model using data from january 2002 up to November 2023, the latest date available in the EA-MPD dataset at the time. As in \cite{fanelli2022sovereign}, we randomly generate orthonormal matrices Q. If $\Lambda Q$ satisfies the sign restrictions, it is kept. Denote $\Lambda^*Q^*$ to be the median rotated factor loading to satisfy the sign restrictions. $(Q^*)^\prime f_t$ are our factors.\\[\baselineskip]
These three factors are the identified exogenous shock whose impact we will study. We transform them to monthly time series by assigning to each month the value of the shock from a conference that happened during that month. Months without any ECB conferences have values of 0. Exceptionally august 2006 had two ECB conferences  while September 2006 had none. Since, the second conference was on the 31st of august we assign it to the month of september 2006.\\
In the next section we will outline our econometric methodology.
\section{Econometric methodology}
In this section we outline the main components of our econometric methodology. We start with a description of the hypothesized underlying structural model. We then describe the estimation method, the model selection and lastly, our strategy to recover possible sign and size effects of the identified shocks.
\subsection{The Structural Model}
We postulate that the data generating process, for each country j, is a structural vector auto-regressive model of the following form (similar to the one studied in \cite{Goncalves2021} and the VARX of \cite{PesaranGVAR2007}):
\begin{align}
   A_0 w_{j,t} =a+A(L)w_{j,t-1}+C(L)f(x_t)+\eta_{j,t}
\end{align}
Where $w_{j,t}=\langle x_t,y_{j,t},z_t\rangle$, $x_t$ is a vector of identified shocks, $y_{j,t}$ is a vector of country specific endogenous variables and $z_t$ are the euro-area endogenous variables. 
\begin{align}
    A_0 = \begin{bmatrix} \mathbf{I} & \mathbf{0} & \mathbf{0} \\ - A_{0,21} & \mathbf{I}   & - A_{0,23}  \\ - A_{0,31} & - A_{0,32}   & \mathbf{I} 
    \end{bmatrix}
\end{align} 
The elements of $y_{j,t}$ and $z_t$ do not impact $x_t$ contemporaneously but $x_t$ does affects $y_{j,t}$ and $z_t$ contemporaneously. We do not restrict the contemporaneous relationship between $y_{j,t}$ and $z_t$ as we only require a block-recursive DGP to identify the impulse response to a shock in $x_t$.
\begin{align}
    A(L)=\begin{bmatrix}
       \mathbf{0} & \mathbf{0}& \mathbf{0} \\
        A_{21}(L) & A_{22}(L) & A_{23}(L)  \\
         A_{31}(L) & \mathbf{0} & A_{33}(L)  \\
    \end{bmatrix}
\end{align}
Where $A_{21}(L)$, $A_{22}(L) $ and $A_{23}(L) $ are polynomials of the lag operator of order $p$, $q$ and $r$ respectively. The first row of sub-matrices of $A(L)$ are zero matrices as our shocks are exogenous and observed.\\
As in \cite{Goncalves2021}, we assume the following structure for $C(L)$:
\begin{align}
    C(L)=\begin{bmatrix}
        \mathbf{0} \\
        C_2(L)\\
        C_3(L) 
    \end{bmatrix}
\end{align}
By their construction, the identified shocks are uncorrelated between themselves and have unit variances. 
\begin{align}
    \eta_{j,t}=\langle u_t, \epsilon_{j,t}, v_t \rangle \sim i.i.d \left( \begin{bmatrix}
        0 \\ 0 \\ 0
    \end{bmatrix},  \begin{bmatrix}
        \mathbf{I} & \mathbf{0} & \mathbf{0} \\
       \mathbf{0} & \Sigma_{2,2} & \Sigma_{2,3}\\
       \mathbf{0} & \Sigma_{3,2} & \Sigma_{3,3} \\
    \end{bmatrix} \right)
\end{align}
The object of interest in our analysis is the impulse response to a shock in $u_t$ to $y_t$. Since, the shocks are observed, estimating the impulse response conditional on the sample is akin to estimating the unconditional impulse response function (\cite{Goncalves2021}).
\subsection{The Linear Local Projection and Model Selection}
We estimate the impulse responses of interest using the method of local projections (\cite{dufour_reneault_1998} and  \cite{jorda2005estimation}). In the first baseline specification we treat $C_2(L)$ as a zero matrix, completely ignoring non-linearities. We refer to this as the linear specification which yields the following local projection.
\begin{align}
     y_{j,k,t+h}= & \alpha_{j,k,h}+ \sum_{i=0}^p \psi_{j,h,i} x_{t-i}+\sum_{i=1}^q \phi_{j,h,i} y_{k,t-i} +\sum_{i=1}^r\beta_{j,h,i} z_{t-i}+I_1\beta_{j,h}t+I_1I_2\gamma_{j,h}t^2+\epsilon_{j,k,t}, \\ 
      h\in\{0,...,H\} & \quad  k \in \{1,...,5\} \quad j \in \{1,...,5\} \quad \langle p,q,r\rangle \in \mathbf{N}^3 \quad \langle I_1,I_2 \rangle  \in \{0,1\}^2 \quad t \in \{0,...,T\} \notag
\end{align}
The unconditional impulse response function  for a shock $\delta $ will then be:
\begin{align*}
    IRF_{j,h,\delta}=\psi_{j,h,0} \delta
\end{align*}
\textbf{Notation}, in equation (6), j is the subscript for the country specific endogenous variable, k the country, t the time, h the horizon and i the lag. That is,$y_{j,k,t+h}$ corresponds to the endogenous variable j for country k at time t+k, while $y_{k,t-i}$ is the 5 by 1 vector of endogenous variable for country k at time t-i. $\alpha_{j,k,h}$ is a equation-country-horizon specific intercept. $x_{t-i}$ is the 3 by 1 vector of shocks and $z_{t-i}$ is the 4 by 1 vector of Euro-Area endogenous variables. $x_{t-i}$ and $z_{t-i}$ are common to every country and every equation j. $t$ and $t^2$ are the linear and quadratic time trends and $I_1$ and $I_2$ are either 0 or 1 and will be used in model selection.\\[\baselineskip]
\textbf{Model Selection} We select the optimal number of lags p, q and r as well as the presence of a linear and quadratic time trend at every horizon $h$ using the information criteria of \cite{Akaike1973}.\\[\baselineskip]
Denote by $\hat{u}^h(p,q,r,I_1,I_2)$ the residuals from estimating equation (6) with OLS. Define the sets: $A=\{2,3,4,5,6\}$ and $B=\{0,1\}$. We select $p^*(h)$, $q^*(h)$, $r^*(h)$, $I_1^*(h)$ and $I_2^*(h)$ to solve the following problem at each $h\in\{0,..,H\}$:
\begin{align}
    \underset{\{p,q,r\} \in A^3 ,\{I_1,I_2\}\in B^2}{\text{max}} n*ln((\hat{u}^h(p,q,r,I_1,I_2)^\prime \hat{u}^h(p,q,r,I_1,I_2))/n)-2(p+q+r+I_1+I_1I_2+5) 
\end{align}
Appendix  A, provides the detailed results of this model selection, while appendix B, provides the results if we instead used the BIC criterion of \cite{Schwarz1978}  .\\[\baselineskip]
We then estimate both the standard linear local projection and the local projection augmented with non-linear transformation of the shocks using the optimal selection  $p^*(h)$, $q^*(h)$, $r^*(h)$, $I_1^*(h)$ and $I_2^*(h)$.

\subsection{Uncovering Non-Linearities}
To investigate the presence of non-linearities in the impulse response functions we augment our local projections with non-linear transformation of the shocks. That is, we specify a $f(x)$ and we let $C_2(L)$, be non-zero. We are interested in two types of non-linearities. Sign based non-linearity and Size based non-linearity. Our choice of function $f(x)$ to disentangle the two effects will be taken from \cite{CaravelloMartinezWP}. We then compute the impulse responses using the plug in non linear local projection method of \cite{Goncalves2021}. \\[\baselineskip]
\cite{CaravelloMartinezWP} define \textbf{Sign Non-Linearities} as when the magnitude of the effect of a shock depending on the sign of the shock. To estimate the sign effect they propose augmenting the local projection with the absolute value of the shock: $f(x_t)=|x_t|$. They demonstrate that this function will annihilate size non-linearities and recover only the sign effect. For example, a positive shock will have an immediate impact of: $\Gamma_{j,0,0}+\psi_{j,0,0}$, while for a negative shock it will be: $\Gamma_{j,0,0}-\psi_{j,0,0}$. This result depends on the symmetry of the distribution of shocks around 0, which is discussed in appendix C. We call the following the sign specification local projection:
\begin{align}
    y_{j,k,t+h}= & \alpha_{j,k,h}+ \sum_{i=0}^{p^*(h)} \psi_{j,h,i} x_{t-i}+\sum_{i=0}^{p^*(h)} \Gamma_{j,h,i}|x_{t-i}|+\sum_{i=1}^{q^*(h)} \phi_{j,h,i} y_{k,t-i} \notag \\ &+\sum_{i=1}^{r^*(h)}\beta_{j,h,i} z_{t-i} 
    +I_1^*\beta_{j,h}t+I_1^*(h)I_2^*(h)\gamma_{j,h}t^2+\epsilon_{j,k,t}  \\
      h\in\{0,...,H\} &   \quad  k \in \{1,...,5\} \quad j \in \{1,...,5\} \quad  t \in \{0,...,T\} \notag
\end{align}
Where $p^*(h), q^*(h), r^*(h), I_1^*(h), I_2^*(h)$ are the optimal parameters selected in for the linear specification.\\[\baselineskip]
\clearpage
\noindent
\textbf{Size Non-Linearities } occur when the marginal effect of the shock depends on the magnitude of the shock. For example, if increasing the interest rate twice by $1\%$ does not have the same average effect as increasing it once by $2\%$. Size non-linearities are trickier to report as the scaled impulse response depends on the magnitude of the shock. The authors recommend picking a threshold $\bar{b}$ on the shock's magnitude to partition them into small and large. The non-linear transformation of the shocks will then be: $f(x_t)=\mathbf{I}\{x_{t} \varleq -\bar{b}\}(x_{t}+\bar{b})+\mathbf{I}\{x_{t} \vargeq \bar{b}\}(x_{t}-\bar{b})$. This non-linear transformation will capture the average marginal effect for the part of the shock above the magnitude threshold. 
We denote this specification the size local projection:
\begin{align}
    y_{j,k,t+h}= & \alpha_{j,k,h}+ \sum_{i=0}^{p^*(h)} \psi_{j,h,i} x_{t-i}+\sum_{i=0}^{p^*(h)} \Gamma_{j,h,i}\left[\mathbf{I}\{x_{t-i} \varleq -\bar{b}\}(x_{t-i}+\bar{b})+\mathbf{I}\{x_{t-i} \vargeq \bar{b}\}(x_{t-i}-\bar{b})\right] \notag \\ & +\sum_{i=1}^{q^*(h)} \phi_{j,h,i} y_{k,t-i} +\sum_{i=1}^{r^*(h)}\beta_{j,h,i} z_{t-i} 
    +I_1^*\beta_{j,h}t+I_1^*(h)I_2^*(h)\gamma_{j,h}t^2+\epsilon_{j,k,t}  \\
      h\in\{0,...,H\} &   \quad  k \in \{1,...,5\} \quad j \in \{1,...,5\} \quad  t \in \{0,...,T\} \notag
\end{align}
Although we chose $f(x_t)=|x_t|$ and $f(x_t)=\mathbf{I}\{x_{t} \varleq -\bar{b}\}(x_{t}+\bar{b})+\mathbf{I}\{x_{t} \vargeq \bar{b}\}(x_{t}-\bar{b})$, any even function could have worked in place of $f(x_t)=|x_t|$ and any odd function could have worked in place of $f(x_t)=\mathbf{I}\{x_{t} \varleq -\bar{b}\}(x_{t}+\bar{b})+\mathbf{I}\{x_{t} \vargeq \bar{b}\}(x_{t}-\bar{b})$. We sum up \cite{CaravelloMartinezWP}, analysis on the choice of function (and borrowing their notation). Define $G(\epsilon_t,h)=H(\epsilon_t,h)+\beta(h)$ to be the marginal effect of a shock $\epsilon_t$ on an outcome variable  $y_{t+h}$, where $H(\epsilon_t,h)$ is the non-linear part of the marginal effect and $\beta(h)$ is the linear part. From \cite{Yitzhaki1996}, we know the coefficient on the non-linear transformation can be written as: $\Gamma_{j,h,0}=\int_{-\infty}^\infty w_f(a)E(H(a,h))da$, which is a weighted average of the expected non-linear marginal effects over the distribution of the shocks.  We can decompose $E(H(a,h))$ into the sum of an odd component, which causes a sign effect and an even component, which causes a size effect. The weights $w_f$, depend on the choice of function included in the linear projection. An odd function will induce even weights and an even function will induce odd weights. To tie everything together, the integral over an interval symmetric around 0 (see appendix C) of the product of an odd and an even function is 0 (they are orthogonal). Therefore, an odd function will recover the size effect and annihilate the sign effect and an even function will recover the sign effect and annihilate the size effect. We will use this result to draw conclusion on the presence of each type of non-linearities.\\[\baselineskip]
As \textbf{the impulse response function} is non-linear, its shape depends on the magnitude of the shock, $\delta$, and the initial value of $x_t$. For example, with a size non-linearity if $x_t=0.5$ and $\delta=1$ the response will be different than if $x_t=2$ and $\delta=1$. \cite{Goncalves2021} provides us with a method to report the unconditional impulse response using a plug in local projection estimator. We outline their procedure here:\\[\baselineskip]
1) We estimate the local projections in the standard way which yields for each equation j and each horizon h a pair $\hat{\psi}_{j,h,0}$ and $\hat{\Gamma}_{j,h,0}$. \\[\baselineskip]
2) We estimate the quantity: $A_\delta= E(f(x_{t}+\delta)-f(x_t))$ with $\hat{A}_\delta= \frac{1}{T}\sum_{t=1}^T \left[f(x_t+\delta)-f(x_t)\right] $ for each of the three shocks.\\[\baselineskip]
3) The impulse response function will be: $IRF(h,j)=\hat{\psi}_{j,h,0} \delta+\hat{\Gamma}_{j,h,0} \hat{A}_\delta$.\\[\baselineskip]
This method relies on the assumption that the shocks are not serially correlated. Estimating AR(1) models of each shocks yields non-significant coefficients that are less than 0.1. \\[\baselineskip]
Lastly we also report conditional impulse responses for the sign non-linearity specification. These are the impulse responses conditional on $x_t$ being non-negative (non-positive) and being disturbed by $\delta$ that is positive (negative). We compute them as: $IRF(h,j|x_t\geq 0, \delta>0)=\hat{\Gamma}_{j,h,0}+\hat{\psi}_{j,h,0}$ and $IRF(h|x_t\leq 0, \delta<0)=\hat{\Gamma}_{j,h,0}-\hat{\psi}_{j,h,0}$ respectively.\\[\baselineskip]
The \textbf{inference} on the non-linear models will be two-pronged. We first look at the significance of the coefficient on the non-linear transformation of the shock, to establish the presence of non-linearities. We then test the significance of both the unconditional and conditional IRFs, to determine if they are non-zero. The inference on the non-linear models will be discussed in great details in section 5.

\newpage
\section{IRF Estimation Results} \label{sec:main_results} 
In this section we provide the estimated impulse response functions. We start with the linear specification and move to both the sign and size non-linear specifications.
\subsection{The Linear Model}
\begin{figure}[b!]
    \centering
    \includegraphics[scale=0.27]{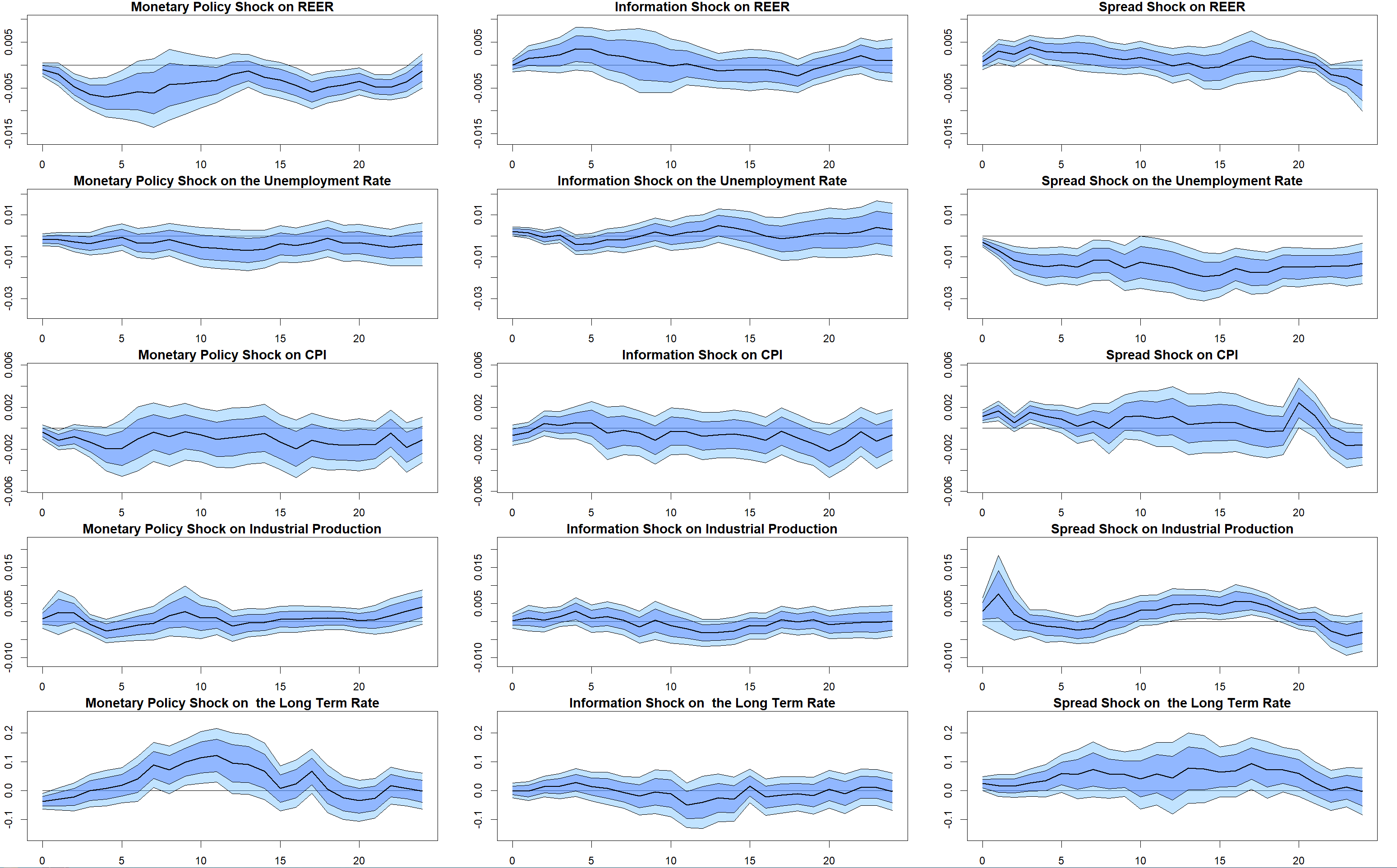}
    \caption{Results from the linear specification (6)}
\floatfoot{The horizontal axis represents the months since the shock (h). The rows are in order: the real exchange rate, the unemployment rate, the consumer price index, the index of industrial production and the long term interest rate. The columns are in order: the pure monetary policy shock, the information shock and the spread shock. The real exchange rate, consumer price index and industrial production index are in log points, while the unemployment rate and long term interest rate are in percentages. The dark blue lines represent the 1 standard deviation confidence interval while the light blue are $90\%$. The standard errors are computed using the method of \cite{Arellano1987} with the ``Jackknife" $HC_3$, HCCME.}.
\end{figure}
\noindent
\textbf{Figure 1}, provides the estimated impulse response functions of the linear specification. The pure monetary policy shock has significant, but temporary, effects on the real exchange rate and the CPI.  The information shock has an quick but temporary impact on the real exchange rate and a delayed (and less significant) impact on CPI. Lastly, the spread shock has a very significant negative impact on the unemployment rate and positive impact on industrial production. For nominal variables, impact on CPI, the REER and the long term rate is positive and significant.
\subsection{The Non-Linear Model, Sign Non-Linearities}

\begin{figure}[!b]
    \centering
    \includegraphics[scale=0.27]{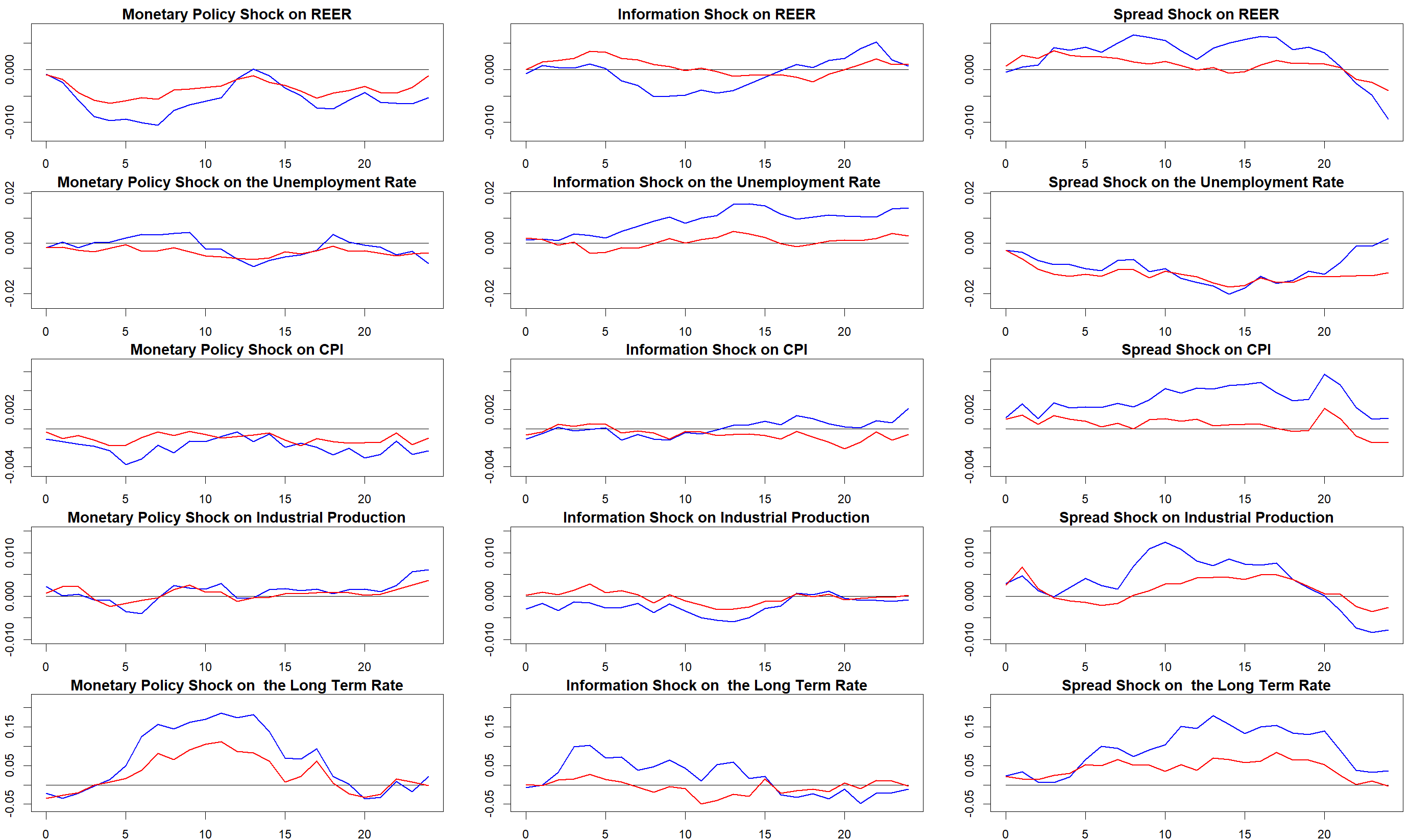}
    \caption{Impulse responses of the sign specification à la \cite{Goncalves2021}}
    \floatfoot{The red line is the original linear impulse response function, the blue line is the \cite{Goncalves2021} non-linear IRF. The horizontal axis represents the months since the shock (h). The rows are in order: the real exchange rate, the unemployment rate, the consumer price index, the index of industrial production and the long term interest rate. The columns are in order: the pure monetary policy shock, the information shock and the spread shock. The real exchange rate, consumer price index and industrial production index are in log points, while the unemployment rate and long term interest rate are in percentages.}
\end{figure}
\noindent
In \textbf{figure 2}, we plot in blue our estimate of the unconditional impulse response function to a 1 standard deviation disturbance in the sign-effect non-linear specification. In red is the original linear impulse response function to a 1 standard deviation shock. When comparing both we see that for most graphs the two IRFs nearly coincide, implying for them that on average both specification will estimate very similar IRFs. However, the monetary shock on the long term rate, is a lot stronger (temporarily) starting from horizon 5, when estimated with the sign-specification. For the information shock, the linear model estimates an effect on unemployment and industrial production that is close to zero, while the non-linear model estimates a much larger and positive IRF for unemployment and a negative IRF for industrial production.  This implies that the effect of the information shock on real variables contains a sign effect. Lastly, the non-linear model estimates a larger, but temporary effect of the spread shock on industrial production, the real exchange rate,the consumer price index and the long term rate.

\begin{figure}[!b]
    \centering
    \includegraphics[scale=0.27]{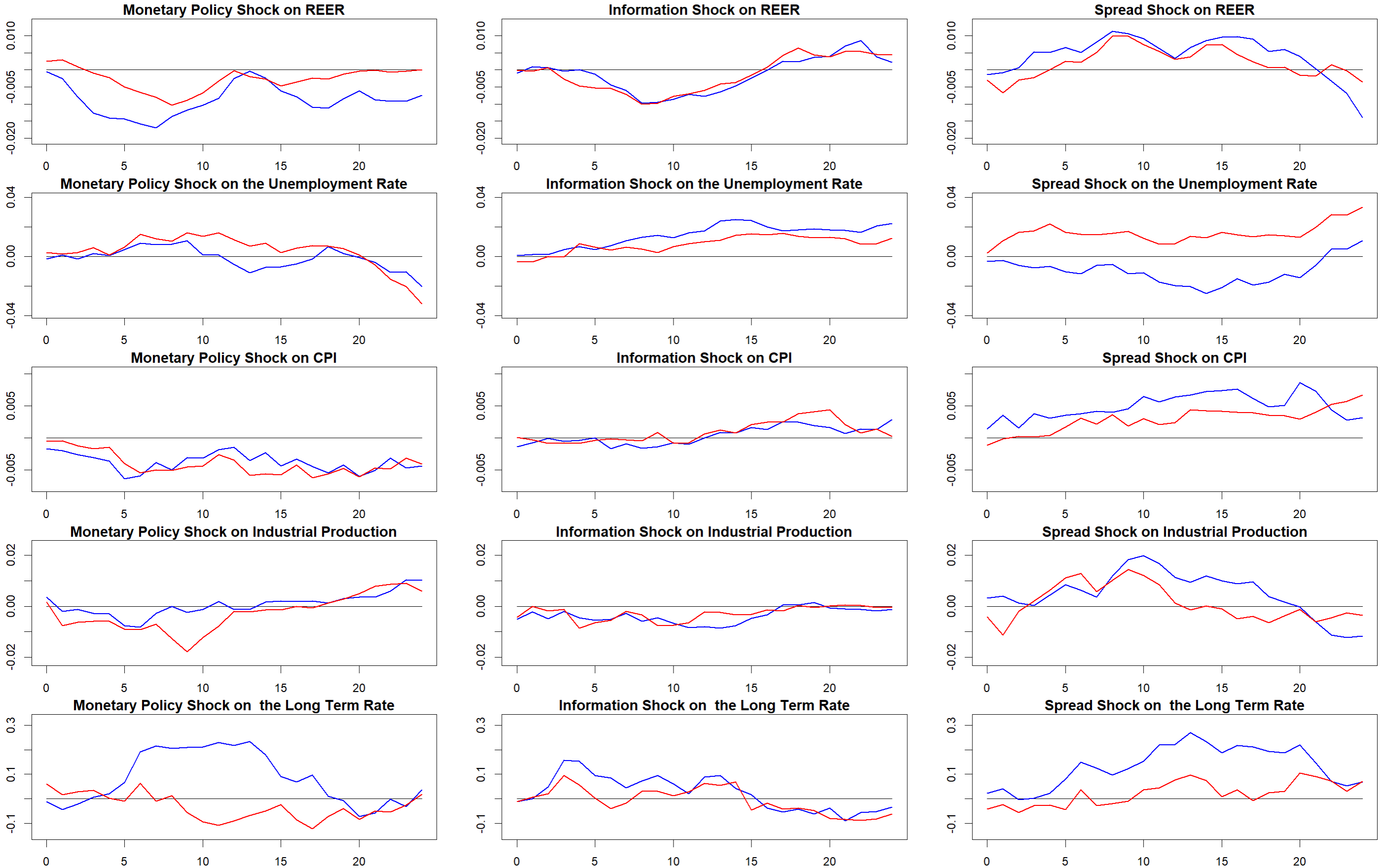}
    \caption{Conditional IRFs of the sign-effect specification }
    \floatfoot{The rows are in order: the real exchange rate, the unemployment rate, the consumer price index, the index of industrial production and the long term interest rate. The columns are in order: the pure monetary policy shock, the information shock and the spread shock. The real exchange rate, consumer price index and industrial production index are in log points, while the unemployment rate and long term interest rate are in percentages.}
\end{figure}
\noindent
\textbf{Figure 3}, compares two conditional impulse response functions estimated with the sign-effect specification. These are:
\begin{align*}
    \text{In blue:  IRF}(h,j|x_t \vargeq 0, \delta>0)= & \hat{\Gamma}_{j,h,0}+\hat{\psi}_{j,h,0} \\
       \text{In red:  (IRF}(h,j|x_t \varleq 0, \delta<0))*(-1)= & (\hat{\Gamma}_{j,h,0}-\hat{\psi}_{j,h,0} )*(-1)
\end{align*}
The blue IRFs are very similar to the unconditional \cite{Goncalves2021} IRFs as the standard deviation and the estimated $\hat{A}$ are both close to 1. We can see that for monetary policy shocks, a (conditional) positive shock has a large negative effect on REER and a large positive effect on the long term rate, while a negative shock has a more restrained effect on both. For industrial production, it is the opposite, negative monetary policy shocks have more negative impacts. Positive information shocks have stronger permanent effects on unemployment. For the spread shock the main difference is in its effect on the unemployment rate. Positive and negative shocks have opposite marginal effects. A positive spread shocks also has a much larger effect on the long term rate. All other conditional Impulse response functions are similar when comparing negative and positive shocks.

\subsection{The Non-Linear Model, Size Non-Linearities}

\begin{figure}[!b]
    \centering
    \includegraphics[scale=0.27]{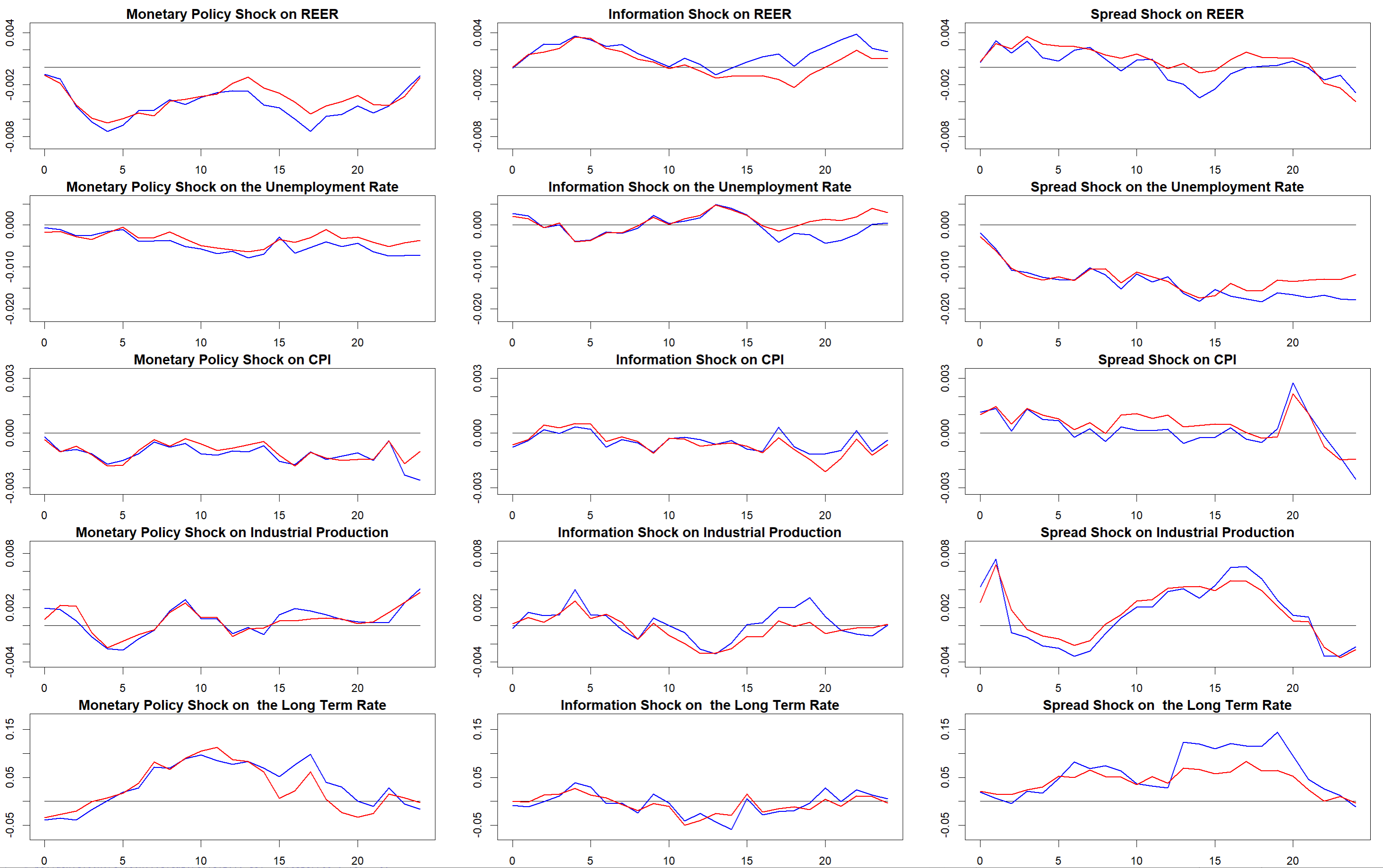}
    \caption{Impulse responses of the size specification à la \cite{Goncalves2021}}
    \floatfoot{The rows are in order: the real exchange rate, the unemployment rate, the consumer price index, the index of industrial production and the long term interest rate. The columns are in order: the pure monetary policy shock, the information shock and the spread shock. The real exchange rate, consumer price index and industrial production index are in log points, while the unemployment rate and long term interest rate are in percentages.}
\end{figure}
\noindent
\textbf{Figure 4} plots the \cite{Goncalves2021} unconditional impulse response functions for the size-effect specification in blue. We set the threshold $\bar{b}$ for each shock to be such that $P(|x_t|\varleq\bar{b})\approx 0.6$. The shock used in computing of $\hat{A}$ is still 1 standard deviation. In red is the original linear impulse response function. The monetary policy shock has a slightly larger temporary negative effect on the REER for horizon 13 to 22. The information shock has a smaller effect on the unemployment rate, CPI and Industrial production when it is estimated with the non-linear specification. The spread shock has a smaller effect on CPI and a larger effect on the long term rate at longer horizons.For all other impulse response functions, the linear and non-linear  specification coincide. This suggests that there are no important size effects for a 1 standard deviation shock.

\begin{figure}[!t]
    \centering
    \includegraphics[scale=0.27]{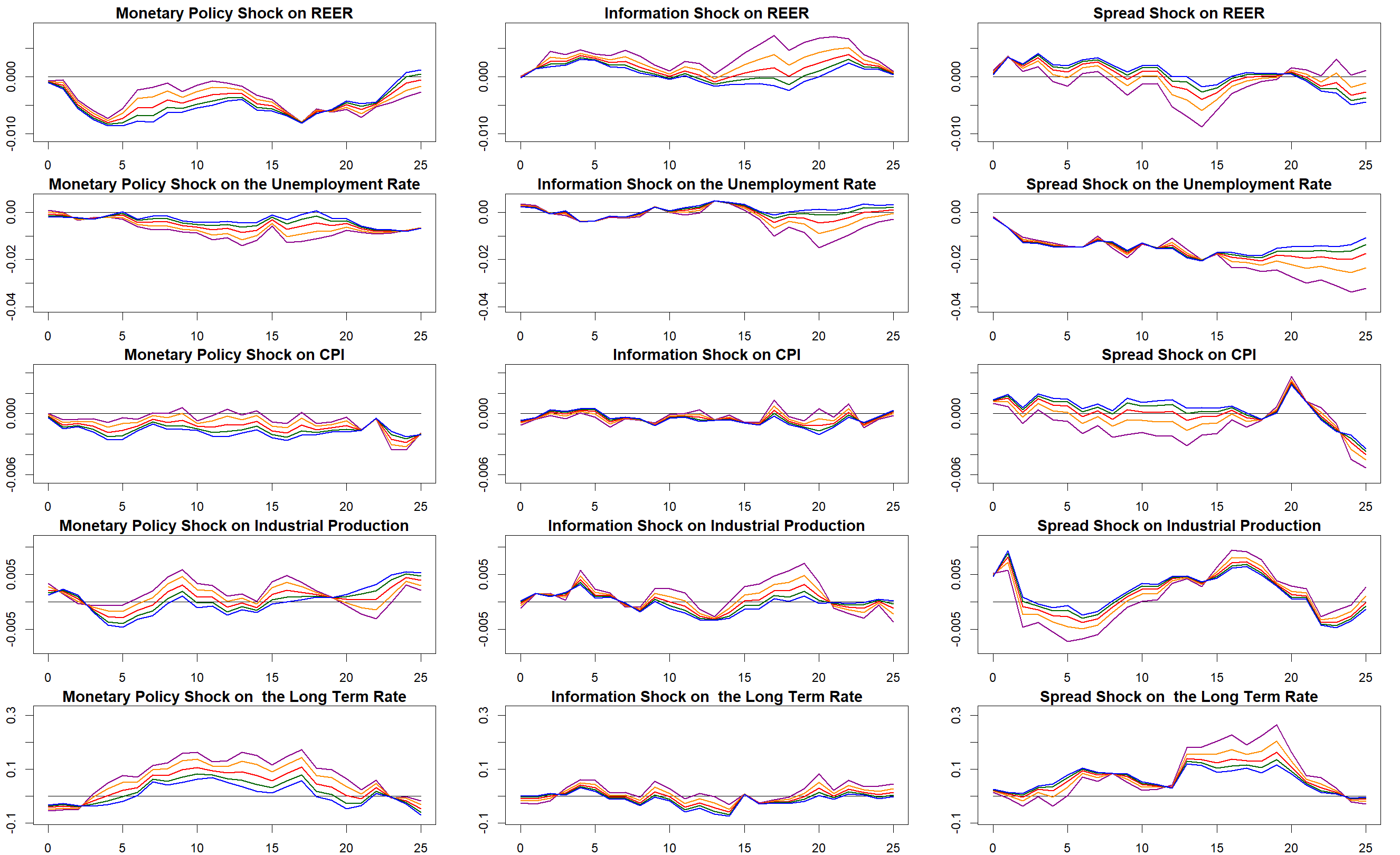}
    \caption{Impulse responses of the size specification à la \cite{Goncalves2021}}
    \floatfoot{ The magenta, orange, red, green and blue line are for shocks that are 0.5, 0.75, 1, 1.25 and 1.5 times the standard deviation respectively.
    The rows are in order: the real exchange rate, the unemployment rate, the consumer price index, the index of industrial production and the long term interest rate. The columns are in order: the pure monetary policy shock, the information shock and the spread shock. The real exchange rate, consumer price index and industrial production index are in log points, while the unemployment rate and long term interest rate are in percentages.}
\end{figure}
\noindent
\textbf{Figure 5} presents an alternative way of illustrating the size effect using the method of \cite{Goncalves2021}. Each line is an impulse response of the following form:
\begin{align*}
    IRF(h,a)=\frac{1}{a}\left(\hat{\psi}_{j,h,0}*a*\sigma+\sum_{t=1}^T \frac{1}{T}\left[f(x_t+a*\sigma)-f(x_t) \right]*a*\hat{\Gamma}_{j,h,0}  \right)
\end{align*}
Where $a\in \{0.5,0.75,1,1.25,1.5\}$.
We can see that for many impulse responses the shape depends greatly on the size of the shock. This suggest that even if we did not see any differences in the impulse response between the linear and non-linear specification for the average shock (1 standard deviation), it could appear for different size shocks. For example, monetary policy shocks on the REER have bigger marginal effects for larger shocks for the first year ($h<12$), but smaller effects for the second year.  The information shock's marginal effect on REER and Unemployment is larger for smaller shocks. Lastly, the spread shock's marginal effect on the REER, the CPI and Industrial Production is larger for smaller magnitude shocks. Therefore, if there are size effects, they are concave in the magnitude of the shock. It is important to remember that the significance level of the coefficient of the non-linear transformation (and therefore the presence of non-linearities) does not depend on the size of the shock. 
\clearpage

\section{Inference on the Non-Linear Models}
\begin{table}[h!]
\begin{footnotesize}
\centering
\scalebox{0.8}{ \begin{tabular}{|r|r|rrrrrrrrrrrrr|}
  \hline
&h= & 0 & 1 & 2 & 3 & 4 & 5 & 6 & 7 & 8 & 9 & 10 & 11 & 12 \\ 
  \hline
\multirow{ 3}{*}{REER}&Monetary &  &  & \cellcolor{green!50} & \cellcolor{green!50} & \cellcolor{green!50} & \cellcolor{green!50} & \cellcolor{green!50} & \cellcolor{green!50} & \cellcolor{green!50} & \cellcolor{green!50} & \cellcolor{green!50} & \cellcolor{green!50} &  \\ 
 & Information &  &  &  &  &  &  & \cellcolor{green!50} & \cellcolor{green!50} & \cellcolor{green!50} & \cellcolor{green!50} & \cellcolor{green!50} & \cellcolor{green!50} & \cellcolor{green!50} \\ 
 & Spread & \cellcolor{green!50} & \cellcolor{green!50} &  &  &  & \cellcolor{green!50} & \cellcolor{yellow!50} & \cellcolor{green!50} & \cellcolor{green!50} & \cellcolor{green!50} & \cellcolor{green!50} & \cellcolor{green!50} &  \\ 
   \hline\multirow{ 3}{*}{Unemp.}&  Monetary &  &  &  &  &  & \cellcolor{yellow!50} & \cellcolor{green!50} & \cellcolor{green!50} & \cellcolor{green!50} & \cellcolor{green!50} &  & \cellcolor{yellow!50} &  \\ 
 & Information &  &  &  &  & \cellcolor{green!50} & \cellcolor{green!50} & \cellcolor{green!50} & \cellcolor{green!50} & \cellcolor{green!50} & \cellcolor{green!50} & \cellcolor{green!50} & \cellcolor{green!50} & \cellcolor{green!50} \\ 
&  Spread &  & \cellcolor{yellow!50} &  &  &  &  &  &  &  &  &  &  &  \\ 
   \hline \multirow{ 3}{*}{CPI} & Monetary & \cellcolor{green!50} & \cellcolor{green!50} & \cellcolor{green!50} & \cellcolor{green!50} & \cellcolor{green!50} & \cellcolor{green!50} & \cellcolor{green!50} & \cellcolor{green!50} & \cellcolor{green!50} & \cellcolor{green!50} & \cellcolor{green!50} &  &  \\ 
&  Information &  &  &  &  &  &  &  &  &  &  &  &  &  \\ 
 & Spread &  & \cellcolor{green!50} &  & \cellcolor{green!50} & \cellcolor{green!50} & \cellcolor{green!50} & \cellcolor{green!50} & \cellcolor{green!50} & \cellcolor{green!50} & \cellcolor{green!50} & \cellcolor{green!50} & \cellcolor{green!50} & \cellcolor{green!50} \\ 
   \hline\multirow{ 3}{*}{Industry} &  Monetary & \cellcolor{yellow!50} & \cellcolor{yellow!50} & \cellcolor{green!50} & \cellcolor{green!50} & \cellcolor{green!50} & \cellcolor{green!50} & \cellcolor{green!50} & \cellcolor{green!50} & \cellcolor{yellow!50} & \cellcolor{green!50} & \cellcolor{green!50} &  &  \\ 
 & Information & \cellcolor{green!50} &  & \cellcolor{green!50} &  & \cellcolor{green!50} & \cellcolor{green!50} & \cellcolor{green!50} &  & \cellcolor{yellow!50} & \cellcolor{yellow!50} & \cellcolor{green!50} & \cellcolor{green!50} & \cellcolor{green!50} \\ 
 & Spread &  &  &  &  & \cellcolor{green!50} & \cellcolor{green!50} & \cellcolor{green!50} & \cellcolor{yellow!50} & \cellcolor{green!50} & \cellcolor{green!50} & \cellcolor{green!50} & \cellcolor{green!50} & \cellcolor{green!50} \\ 
   \hline \multirow{ 3}{*}{LT Rate} & Monetary &  &  &  &  &  &  & \cellcolor{green!50} & \cellcolor{yellow!50} & \cellcolor{yellow!50} &  &  &  &  \\ 
 & Information &  &  &  & \cellcolor{green!50} & \cellcolor{green!50} &  &  &  &  &  &  &  &  \\ 
 & Spread &  &  &  &  &  &  & \cellcolor{green!50} &  &  &  &  & \cellcolor{green!50} & \cellcolor{green!50} \\ 
   \hline
  
&h= & 13 & 14 & 15 & 16 & 17 & 18 & 19 & 20 & 21 & 22 & 23 & 24 & 25  \\ 
  \hline
\multirow{ 3}{*}{REER} &Monetary &  &  & \cellcolor{green!50} & \cellcolor{green!50} & \cellcolor{green!50} & \cellcolor{green!50} & \cellcolor{green!50} &  & \cellcolor{green!50} & \cellcolor{green!50} & \cellcolor{green!50} & \cellcolor{yellow!50} & \cellcolor{yellow!50} \\ 
&  Information & \cellcolor{green!50} & \cellcolor{yellow!50} &  &  &  & \cellcolor{green!50} & \cellcolor{yellow!50} & \cellcolor{yellow!50} & \cellcolor{green!50} & \cellcolor{green!50} & \cellcolor{yellow!50} &  &  \\ 
 & Spread & \cellcolor{green!50} & \cellcolor{green!50} & \cellcolor{green!50} & \cellcolor{green!50} & \cellcolor{green!50} &  & \cellcolor{yellow!50} &  &  &  &  & \cellcolor{green!50} & \cellcolor{green!50} \\ 
   \hline\multirow{ 3}{*}{Unemp.} &  Monetary &  &  &  &  &  &  &  &  &  & \cellcolor{yellow!50} & \cellcolor{green!50} & \cellcolor{green!50} & \cellcolor{green!50} \\ 
 & Information & \cellcolor{green!50} & \cellcolor{green!50} & \cellcolor{green!50} & \cellcolor{green!50} & \cellcolor{green!50} & \cellcolor{green!50} & \cellcolor{green!50} & \cellcolor{green!50} & \cellcolor{green!50} &  & \cellcolor{yellow!50} & \cellcolor{green!50} & \cellcolor{yellow!50} \\ 
 & Spread &  &  &  &  &  &  &  &  &  & \cellcolor{green!50} & \cellcolor{green!50} & \cellcolor{green!50} & \cellcolor{green!50} \\ 
   \hline\multirow{ 3}{*}{CPI} &  Monetary & \cellcolor{green!50} & \cellcolor{green!50} & \cellcolor{green!50} & \cellcolor{green!50} & \cellcolor{green!50} & \cellcolor{green!50} & \cellcolor{green!50} & \cellcolor{green!50} & \cellcolor{green!50} & \cellcolor{green!50} & \cellcolor{green!50} & \cellcolor{green!50} & \cellcolor{green!50} \\ 
&  Information &  &  &  &  & \cellcolor{green!50} & \cellcolor{green!50} & \cellcolor{green!50} & \cellcolor{green!50} &  &  &  &  &  \\ 
 & Spread & \cellcolor{green!50} & \cellcolor{green!50} & \cellcolor{green!50} & \cellcolor{green!50} & \cellcolor{green!50} & \cellcolor{green!50} & \cellcolor{green!50} & \cellcolor{green!50} & \cellcolor{green!50} & \cellcolor{green!50} & \cellcolor{green!50} & \cellcolor{green!50} & \cellcolor{green!50} \\ 
   \hline\multirow{ 3}{*}{Industry} &  Monetary &  &  &  &  &  &  &  & \cellcolor{yellow!50} & \cellcolor{green!50} & \cellcolor{green!50} & \cellcolor{green!50} & \cellcolor{green!50} &  \\ 
&  Information & \cellcolor{green!50} & \cellcolor{green!50} & \cellcolor{yellow!50} &  &  &  &  &  &  &  &  &  &  \\ 
&  Spread &  & \cellcolor{green!50} & \cellcolor{green!50} &  &  &  &  &  & \cellcolor{green!50} & \cellcolor{green!50} & \cellcolor{green!50} & \cellcolor{green!50} & \cellcolor{green!50} \\ 
   \hline\multirow{ 3}{*}{LT Rate} & Monetary &  &  &  &  &  &  &  & \cellcolor{yellow!50} &  &  &  &  &  \\ 
 & Information &  &  &  &  &  &  &  &  & \cellcolor{green!50} & \cellcolor{yellow!50} & \cellcolor{yellow!50} &  &  \\ 
 & Spread & \cellcolor{green!50} & \cellcolor{green!50} & \cellcolor{yellow!50} & \cellcolor{green!50} & \cellcolor{yellow!50} & \cellcolor{yellow!50} & \cellcolor{green!50} & \cellcolor{green!50} & \cellcolor{green!50} &  &  &  & \cellcolor{yellow!50} \\ 
   \hline
   \end{tabular}}
\end{footnotesize}
    \caption{Significance of the Non-Linear Component of the Sign Effect Specification}
\floatfoot{We report the significance level of the estimated coefficient $\hat{\Gamma}_{j,h,0}$. The non-linear transformation is the absolute value of the shocks. White cells refer to p-values greater than 0.1, yellow cells are p-values between 0.1 and 0.05 and green cells are p-values smaller than 0.05.}
\end{table}
\noindent
\textbf{Table 2} reports the results of Wald tests (\cite{Wald1943}) of the significance of the coefficient on the non-linear transformation of the shocks. As in \cite{CaravelloMartinezWP}, the results of these tests allow us to establish the presence of non-linearities at horizon h. We find multi-horizon significant sign-effects of every type of shock on every outcome variable with the exception of the monetary shock on the long term rate, the information shock on CPI and the long term rate and the spread shock on unemployement. We describe the structure of the test bellow as we will present results on different version of it. We test the hypotheses, $$H_{0}^{1,h}:\hat{\Gamma}^{monetary}_{j,h,0}=0  \quad H_0^{2,h}: \hat{\Gamma}^{information}_{j,h,0}=0 \quad H_0^{3,h}:\hat{\Gamma}^{spread}_{j,h,0} =0$$
To do so we conduct the Wald test:
\begin{align*}
\hat{\beta}_{j,h}= & \langle\hat{\psi}^{monetary}_{j,h,0},\hat{\psi}^{information}_{j,h,0},\hat{\psi}^{spread}_{j,h,0},\hat{\Gamma}^{monetary}_{j,h,0},\hat{\Gamma}^{information}_{j,h,0},\hat{\Gamma}^{spread}_{j,h,0}, ... \rangle\\
R_1 = & \langle 0,0,0,1,0,0,...\rangle, \quad R_2 =  \langle 0,0,0,0,1,0,...\rangle, \quad R_3 =  \langle 0,0,0,0,0,1,...\rangle\\
W_{i,j,h}= & (R_i\hat{\beta_{j,h}}^\prime)^\prime (R_i \hat{\Omega} R_i^\prime)^{-1}(R\hat{\beta_{j,h}}^\prime)
\end{align*}
Then for the Wald statistic $W_{i,j,h}\overset{a}{\sim}\chi(1)$, for hypotheses $i=\{1,2,3\}$ and every horizon $h$. The matrix $\hat{\Omega}$ is the covariance matrix of $\beta$ estimated using the method of \cite{Arellano1987} with $HC_3$.

\begin{table}[t!]
\begin{footnotesize}
\centering
\scalebox{0.8}{ \begin{tabular}{|r|r|rrrrrrrrrrrrr|}
  \hline
&h= & 0 & 1 & 2 & 3 & 4 & 5 & 6 & 7 & 8 & 9 & 10 & 11 & 12 \\ 
  \hline
\multirow{ 3}{*}{REER} &Monetary & & \cellcolor{yellow!50} & & & & \cellcolor{yellow!50} & \cellcolor{green!50} & \cellcolor{green!50} & \cellcolor{green!50} & \cellcolor{yellow!50} & \cellcolor{green!50} & \cellcolor{green!50} & \cellcolor{yellow!50} \\ 
&  Information & & & \cellcolor{green!50} & & & & & & & & & & \cellcolor{yellow!50} \\ 
&  Spread & & & & & & & & & & \cellcolor{yellow!50} & & & \cellcolor{green!50} \\ 
\hline\multirow{ 3}{*}{Unemp.} &  Monetary & \cellcolor{green!50} & & & & & & & \cellcolor{green!50} & \cellcolor{yellow!50} & & & \cellcolor{green!50} & \cellcolor{yellow!50} \\ 
&  Information & & & & & & & & & & & & & \\ 
&  Spread & & & & & & & & & & & & & \\ 
  \hline\multirow{ 3}{*}{CPI} &  Monetary & & \cellcolor{green!50} & & \cellcolor{green!50} & \cellcolor{green!50} & \cellcolor{green!50} & & & \cellcolor{yellow!50} & \cellcolor{green!50} & & \cellcolor{green!50} & \cellcolor{green!50} \\ 
&  Information & & & & & & & & & & & & & \\ 
&  Spread & & \cellcolor{green!50} & \cellcolor{green!50} & \cellcolor{green!50} & \cellcolor{green!50} & \cellcolor{green!50} & \cellcolor{green!50} & \cellcolor{green!50} & \cellcolor{green!50} & \cellcolor{green!50} & \cellcolor{green!50} & \cellcolor{green!50} & \cellcolor{green!50} \\ 
\hline\multirow{ 3}{*}{Industry} &  Monetary & \cellcolor{yellow!50} & & & & \cellcolor{yellow!50} & \cellcolor{green!50} & \cellcolor{green!50} & \cellcolor{green!50} & \cellcolor{green!50} & \cellcolor{green!50} & \cellcolor{green!50} & \cellcolor{green!50} & \cellcolor{yellow!50} \\ 
&  Information & & & & & & & & & & & \cellcolor{yellow!50} & \cellcolor{yellow!50} & \\ 
&  Spread & & \cellcolor{green!50} & \cellcolor{green!50} & & \cellcolor{yellow!50} & \cellcolor{green!50} & \cellcolor{green!50} & \cellcolor{green!50} & & & & & \\ 
\hline\multirow{ 3}{*}{LT Rate} &  Monetary & \cellcolor{green!50} & & & \cellcolor{yellow!50} & \cellcolor{green!50} & \cellcolor{green!50} & \cellcolor{green!50} & \cellcolor{yellow!50} & \cellcolor{green!50} & \cellcolor{green!50} & \cellcolor{green!50} & & \cellcolor{yellow!50} \\ 
&  Information & \cellcolor{green!50} & & & & & & & & & & & & \\ 
&  Spread & & & \cellcolor{yellow!50} & & \cellcolor{green!50} & \cellcolor{yellow!50} & & & & & & & \\ 
   \hline

 &h=& 13 & 14 & 15 & 16 & 17 & 18 & 19 & 20 & 21 & 22 & 23 & 24 & 25 \\ 
  \hline
\multirow{ 3}{*}{REER} &Monetary & & & & & & & & & & & \cellcolor{yellow!50} & \cellcolor{green!50} & \cellcolor{green!50} \\ 
 & Information & & \cellcolor{yellow!50} & \cellcolor{green!50} & \cellcolor{green!50} & \cellcolor{green!50} & \cellcolor{green!50} & \cellcolor{green!50} & \cellcolor{green!50} & \cellcolor{green!50} & \cellcolor{green!50} & & & \\ 
 & Spread & \cellcolor{green!50} & \cellcolor{green!50} & \cellcolor{yellow!50} & & & & & & & & \cellcolor{green!50} & \cellcolor{green!50} & \cellcolor{green!50} \\ 
 \hline\multirow{ 3}{*}{Unemp.} & Monetary & \cellcolor{green!50} & \cellcolor{yellow!50} & & \cellcolor{green!50} & \cellcolor{green!50} & \cellcolor{green!50} & \cellcolor{yellow!50} & & & & & & \\ 
 & Information & & & & & & & & \cellcolor{green!50} & \cellcolor{green!50} & \cellcolor{yellow!50} & & & \\ 
 & Spread & & & & & & & & \cellcolor{yellow!50} & \cellcolor{green!50} & \cellcolor{green!50} & \cellcolor{green!50} & \cellcolor{green!50} & \cellcolor{green!50} \\ 
  \hline\multirow{ 3}{*}{CPI}  & Monetary & \cellcolor{yellow!50} & \cellcolor{yellow!50} & & & \cellcolor{yellow!50} & & & & & & & & \\ 
 & Information & & & & & & & & \cellcolor{green!50} & & & & & \\ 
 & Spread & \cellcolor{green!50} & \cellcolor{yellow!50} & & & & & & & & & & \cellcolor{yellow!50} & \\ 
\hline\multirow{ 3}{*}{Industry}  & Monetary & & & & \cellcolor{yellow!50} & & & & & \cellcolor{green!50} & \cellcolor{green!50} & \cellcolor{green!50} & & \\ 
 & Information & & & & \cellcolor{yellow!50} & \cellcolor{yellow!50} & \cellcolor{green!50} & \cellcolor{green!50} & & & & & & \\ 
 & Spread & & & & & & & & & & & & & \\ 
\hline\multirow{ 3}{*}{LT Rate}  & Monetary & \cellcolor{green!50} & \cellcolor{green!50} & \cellcolor{green!50} & \cellcolor{green!50} & \cellcolor{green!50} & \cellcolor{green!50} & \cellcolor{green!50} & \cellcolor{green!50} & & & & & \\ 
 & Information & & & & & & & & \cellcolor{yellow!50} & & & & & \\ 
 & Spread & & & \cellcolor{green!50} & \cellcolor{green!50} & & \cellcolor{green!50} & \cellcolor{green!50} & & & & & & \\ 
   \hline
   \end{tabular}}
\end{footnotesize}
    \caption{Significance of the Non-Linear Component of the Size Effect Specification}
\floatfoot{We report the significance level of the estimated coefficient $\hat{\Gamma}_{j,h,0}$. The non-linear transformation is $f(x_t)=\mathbf{I}\{x_{t} \varleq -\bar{b}\}(x_{t}+\bar{b})+\mathbf{I}\{x_{t} \vargeq \bar{b}\}(x_{t}-\bar{b})$ with $\bar{b}$ chosen such that $P(|x_t|<\bar{b})=0.6$. White cells refer to p-values greater than 0.1, yellow cells are p-values between 0.1 and 0.05 and green cells are p-values smaller than 0.05.}
\end{table}
\noindent
\textbf{Table 3} reports the result of the same Wald test as in table 2, but on the size effect specification. That is: 
\begin{align*}
    R_1 = & \langle 0,0,0,1,0,0,...\rangle, \quad R_2 =  \langle 0,0,0,0,1,0,...\rangle, \quad R_3 =  \langle 0,0,0,0,0,1,...\rangle
\end{align*}
The results are more muted then with the sign effect. For example, there are very little size effects of any shock on the REER and the unemployement rate for the first 12 months. The information shock has no size effect on industrial production. The persistent and significant non-linearities are that of the monetary shock on the long term rate and the spread shock on the CPI.

\begin{table}[h!]
\begin{footnotesize}
\centering
\scalebox{0.8}{ \begin{tabular}{|r|r|rrrrrrrrrrrrr|}
  \hline

&h= & 0 & 1 & 2 & 3 & 4 & 5 & 6 & 7 & 8 & 9 & 10 & 11 & 12 \\ 
  \hline
\multirow{ 3}{*}{REER} &Monetary &  &  & \cellcolor{green!50} & \cellcolor{green!50} & \cellcolor{green!50} & \cellcolor{green!50} & \cellcolor{green!50} & \cellcolor{green!50} & \cellcolor{green!50} & \cellcolor{green!50} & \cellcolor{yellow!50} &  &  \\ 
&  Information &  &  &  &  &  &  &  &  &  &  &  &  &  \\ 
 & Spread &  &  &  & \cellcolor{green!50} & \cellcolor{yellow!50} & \cellcolor{yellow!50} &  & \cellcolor{green!50} & \cellcolor{green!50} & \cellcolor{yellow!50} & \cellcolor{yellow!50} &  &  \\ 
  \hline \multirow{ 3}{*}{Unemp.} & Monetary &  &  &  &  &  &  &  &  &  &  &  &  &  \\ 
  &Information &  &  &  &  &  &  &  & \cellcolor{yellow!50} & \cellcolor{green!50} & \cellcolor{green!50} &  & \cellcolor{yellow!50} & \cellcolor{green!50} \\ 
  & Spread & \cellcolor{yellow!50} &  & \cellcolor{green!50} & \cellcolor{green!50} & \cellcolor{green!50} & \cellcolor{green!50} & \cellcolor{green!50} &  &  & \cellcolor{green!50} & \cellcolor{yellow!50} & \cellcolor{green!50} & \cellcolor{green!50} \\ 
  \hline \multirow{ 3}{*}{CPI}&  Monetary &  & \cellcolor{yellow!50} & \cellcolor{yellow!50} & \cellcolor{yellow!50} & \cellcolor{yellow!50} & \cellcolor{green!50} & \cellcolor{yellow!50} &  &  &  &  &  &  \\ 
  &Information &  &  &  &  &  &  &  &  &  &  &  &  &  \\ 
 & Spread & \cellcolor{green!50} & \cellcolor{green!50} &  & \cellcolor{green!50} & \cellcolor{green!50} & \cellcolor{green!50} & \cellcolor{green!50} & \cellcolor{green!50} & \cellcolor{green!50} & \cellcolor{green!50} & \cellcolor{green!50} & \cellcolor{green!50} & \cellcolor{green!50} \\ 
  \hline \multirow{ 3}{*}{Industry}  &Monetary &  &  &  &  &  &  &  &  &  &  &  &  &  \\ 
  &Information &  &  &  &  &  &  &  &  &  &  &  &  & \cellcolor{yellow!50} \\ 
  &Spread & \cellcolor{green!50} & \cellcolor{yellow!50} &  &  &  & \cellcolor{yellow!50} &  &  & \cellcolor{yellow!50} & \cellcolor{green!50} & \cellcolor{green!50} & \cellcolor{green!50} & \cellcolor{green!50} \\ 
  \hline \multirow{ 3}{*}{LT Rate}  &Monetary &  &  &  &  &  &  & \cellcolor{green!50} & \cellcolor{green!50} & \cellcolor{green!50} & \cellcolor{green!50} & \cellcolor{green!50} & \cellcolor{green!50} & \cellcolor{green!50} \\ 
  &Information &  &  &  & \cellcolor{green!50} & \cellcolor{green!50} &  &  &  &  &  &  &  &  \\ 
  &Spread &  &  &  &  &  &  & \cellcolor{yellow!50} & \cellcolor{green!50} &  & \cellcolor{yellow!50} & \cellcolor{yellow!50} & \cellcolor{green!50} & \cellcolor{green!50} \\ 

  \hline
 &  h= & 13 & 14 & 15 & 16 & 17 & 18 & 19 & 20 & 21 & 22 & 23 & 24 & 25 \\ 
  \hline
\multirow{ 3}{*}{REER}&Monetary &  &  &  & \cellcolor{yellow!50} & \cellcolor{green!50} & \cellcolor{green!50} & \cellcolor{green!50} &  & \cellcolor{green!50} & \cellcolor{green!50} & \cellcolor{green!50} & \cellcolor{green!50} &  \\ 
&  Information &  &  &  &  &  &  &  &  &  &  &  &  &  \\ 
 & Spread &  &  &  &  &  &  &  &  &  &  &  & \cellcolor{green!50} & \cellcolor{yellow!50} \\ 
  \hline \multirow{ 3}{*}{Unemp.}  &Monetary &  &  &  &  &  &  &  &  &  &  &  &  &  \\ 
  &Information & \cellcolor{green!50} & \cellcolor{green!50} & \cellcolor{green!50} & \cellcolor{yellow!50} &  &  &  &  &  &  &  &  &  \\ 
  &Spread & \cellcolor{green!50} & \cellcolor{green!50} & \cellcolor{yellow!50} &  &  &  &  &  &  &  &  &  &  \\ 
  \hline \multirow{ 3}{*}{CPI}  &Monetary &  &  &  &  &  &  &  & \cellcolor{yellow!50} &  &  &  &  &  \\ 
  &Information &  &  &  &  &  &  &  &  &  &  &  &  &  \\ 
  &Spread & \cellcolor{green!50} & \cellcolor{green!50} & \cellcolor{green!50} & \cellcolor{green!50} & \cellcolor{green!50} & \cellcolor{green!50} & \cellcolor{green!50} & \cellcolor{green!50} & \cellcolor{green!50} &  &  &  &  \\ 
  \hline\multirow{ 3}{*}{Industry}  &Monetary &  &  &  &  &  &  &  &  &  &  &  & \cellcolor{yellow!50} &  \\ 
  &Information & \cellcolor{yellow!50} &  &  &  &  &  &  &  &  &  &  &  &  \\ 
  &Spread & \cellcolor{green!50} & \cellcolor{green!50} & \cellcolor{green!50} & \cellcolor{green!50} & \cellcolor{green!50} & \cellcolor{yellow!50} &  &  &  & \cellcolor{green!50} & \cellcolor{yellow!50} & \cellcolor{green!50} &  \\ 
  \hline \multirow{ 3}{*}{LT Rate}  &Monetary & \cellcolor{green!50} & \cellcolor{yellow!50} &  &  &  &  &  &  &  &  &  &  &  \\ 
  &Information &  &  &  &  &  &  &  &  &  &  &  &  &  \\ 
  &Spread & \cellcolor{green!50} & \cellcolor{green!50} & \cellcolor{green!50} & \cellcolor{green!50} & \cellcolor{green!50} & \cellcolor{green!50} & \cellcolor{green!50} & \cellcolor{green!50} & \cellcolor{yellow!50} &  &  &  &  \\ 
   \hline
   \end{tabular}}
\end{footnotesize}
    \caption{Sign-Effect Impulse Response Function \`a la \cite{Goncalves2021}}
    \floatfoot{We report the significance level of the \cite{Goncalves2021} IRF with the sign effect specification.  White cells refer to p-values greater than 0.1, yellow cells are p-values between 0.1 and 0.05 and green cells are p-values smaller than 0.05.}
\end{table}

\noindent
\textbf{Table 4}, reports our attempt at testing the significance of the \cite{Goncalves2021} with a Wald test. This corresponds to the the following three hypotheses at each horizon, $$H_0^{i,h}:\hat{\psi}_{j,h,0}^i\delta^{i}+\hat{A}^i\hat{\Gamma}_{j,h,0}^i=0, \quad i \in \{monetary,information,spread\}$$
To do so, we construct the following matrix of linear restrictions:
\begin{align*}
    R_1= & \langle \delta^{monetary},0,0,\hat{A}^{monetary}(\delta^{monetary}),0,0,...\rangle \\
    R_2= & \langle 0,\delta^{information},0,0,\hat{A}^{information}(\delta^{information}),0,...\rangle \\ 
     R_3= & \langle 0,0,\delta^{spread},0,0,\hat{A}^{spread}(\delta^{spread}),...\rangle  
     \end{align*}
Where: $\delta^i$ is the sample standard deviation of shock $i$ and $\hat{A}^i= \frac{1}{T}\sum_{t=1}^T \left[f(x_t^i+\delta^i)-f(x_t^i)\right] $, and recall that for the sign effect specification $f(x)=|x|$. The Wald Statistic is then constructed in the same way as before. This procedure is not exact as it does not take into account the estimation uncertainty in $\hat{A}^i$. Although we previously established the presence (or not) of non-linearities, the goal of this exercise is to see if these non-linearities do not ``cancel out" the linear effect such that the weighted sum of both is not significant. The \cite{Goncalves2021} IRF with the sign effect specification is significant for monetary shocks to the REER and the long term rate and the spread shock to CPI, unemployment and industrial production. The sign and magnitude of the impulse response is reported in \textbf{figure 2}.

\begin{table}[h!]
\begin{footnotesize}
\centering
\scalebox{0.8}{ \begin{tabular}{|r|r|rrrrrrrrrrrrr|}
 \hline
&h= & 0 & 1 & 2 & 3 & 4 & 5 & 6 & 7 & 8 & 9 & 10 & 11 & 12 \\ 
  \hline
\multirow{ 3}{*}{REER} &Monetary & & & \cellcolor{green!50} & \cellcolor{green!50} & \cellcolor{green!50} & \cellcolor{green!50} & & & & & & & \\ 
 & Information & & & & & & & & & & & & & \\ 
 & Spread & & \cellcolor{yellow!50} & & & & & & & & & & & \\ 
  \hline\multirow{ 3}{*}{Unemp.} & Monetary & & & & & & & & & & & & & \\ 
 & Information & \cellcolor{yellow!50} & & & & & & & & & & & & \\ 
 & Spread & & \cellcolor{green!50} & \cellcolor{green!50} & \cellcolor{green!50} & \cellcolor{green!50} & \cellcolor{green!50} & \cellcolor{green!50} & \cellcolor{yellow!50} & \cellcolor{green!50} & \cellcolor{green!50} & & \cellcolor{yellow!50} & \cellcolor{yellow!50} \\ 
\hline\multirow{ 3}{*}{CPI} & Monetary & & \cellcolor{yellow!50} & & & \cellcolor{yellow!50} & & & & & & & & \\ 
&  Information & & & & & & & & & & & & & \\ 
  &Spread & \cellcolor{green!50} & \cellcolor{green!50} & & \cellcolor{yellow!50} & & & & & & & & & \\ 
 \hline\multirow{ 3}{*}{Industry} &Monetary & & & & & & & & & & & & & \\ 
  &Information & & & & & & & & & & & & & \\ 
  &Spread & \cellcolor{yellow!50} & & & & & & & & & & & & \\ 
 \hline\multirow{ 3}{*}{LT Rate} &Monetary & \cellcolor{green!50} & & & & & & & & & \cellcolor{yellow!50} & \cellcolor{green!50} & \cellcolor{yellow!50} & \\ 
  &Information & & & & & & & & & & & & & \\ 
  &Spread & & & & & & & & & & & & & \\ 
   \hline
 &h=& 13 & 14 & 15 & 16 & 17 & 18 & 19 & 20 & 21 & 22 & 23 & 24 & 25 \\ 
  \hline
\multirow{ 3}{*}{REER} &Monetary & & \cellcolor{green!50} & \cellcolor{green!50} & \cellcolor{green!50} & \cellcolor{green!50} & \cellcolor{green!50} & \cellcolor{green!50} & \cellcolor{green!50} & \cellcolor{green!50} & \cellcolor{green!50} & & & \\ 
&  Information & & & & & & & & & & & & & \\ 
&  Spread & & & & & & & & & & & & & \\ 
  \hline\multirow{ 3}{*}{Unemp.} &  Monetary & & & & & & & & & & & & & \\ 
&  Information & & & & & & & & & & & & & \\ 
&  Spread & \cellcolor{green!50} & \cellcolor{green!50} & \cellcolor{green!50} & \cellcolor{green!50} & \cellcolor{green!50} & \cellcolor{green!50} & \cellcolor{green!50} & \cellcolor{green!50} & \cellcolor{green!50} & \cellcolor{green!50} & \cellcolor{green!50} & \cellcolor{green!50} & \cellcolor{green!50} \\ 
\hline\multirow{ 3}{*}{CPI} & Monetary & & & & & & & & & & & & \cellcolor{yellow!50} & \\ 
 & Information & & & & & & & & & & & & & \\ 
 & Spread & & & & & & & & & & & & \cellcolor{yellow!50} & \cellcolor{green!50} \\ 
\hline\multirow{ 3}{*}{Industry} & Monetary & & & & & & & & & & & & & \\ 
 & Information & & & & & & & & & & & & & \\ 
 & Spread & & & & \cellcolor{yellow!50} & \cellcolor{green!50} & \cellcolor{yellow!50} & & & & & & & \\ 
 \hline\multirow{ 3}{*}{LT Rate} & Monetary & & & & & \cellcolor{green!50} & & & & & & & & \\ 
 & Information & & & & & & & & & & & & & \\ 
  &Spread & \cellcolor{yellow!50} & \cellcolor{yellow!50} & \cellcolor{yellow!50} & \cellcolor{yellow!50} & \cellcolor{yellow!50} & \cellcolor{yellow!50} & \cellcolor{green!50} & & & & & & \\ 
   \hline
   \end{tabular}}
\end{footnotesize}
    \caption{Size-Effect Impulse Response Function \`a la \cite{Goncalves2021}}
\floatfoot{We report the significance level of the \cite{Goncalves2021} IRF with the size effect specification.  White cells refer to p-values greater than 0.1, yellow cells are p-values between 0.1 and 0.05 and green cells are p-values smaller than 0.05.}
\end{table}
\pagebreak
\noindent
\textbf{Table 5} reports results of the same test as in table 4, albeit on the size-effect specification, with $\bar{b}$ such that $P(|x_t|<\bar{b})=0.6$. That is we test the hypothesis:
$$H_0^{i,h}:\hat{\psi}_{j,h,0}^i\delta^{i}+\hat{A}^i\hat{\Gamma}_{j,h,0}^i=0, \quad i \in \{monetary,information,spread\}$$
The wald statistic is constructed in the same way as in table 4. As when we tested the significance of $\hat{\Gamma}_{j,h,0}^i$ alone, we find more muted results. The impulse response function is only significant (for multiple horizons) for monetary shocks to REER and for spread shocks to unemployment.

\begin{table}[h!]
\begin{footnotesize}
\centering
\scalebox{0.8}{ \begin{tabular}{|r|r|rrrrrrrrrrrrr|}
  \hline
&h= & 0 & 1 & 2 & 3 & 4 & 5 & 6 & 7 & 8 & 9 & 10 & 11 & 12 \\ 
  \hline
\multirow{ 3}{*}{REER} &Monetary &  &  & \cellcolor{green!50} & \cellcolor{green!50} & \cellcolor{green!50} & \cellcolor{green!50} & \cellcolor{green!50} & \cellcolor{green!50} & \cellcolor{green!50} & \cellcolor{green!50} & \cellcolor{yellow!50} &  &  \\ 
 & Information &  &  &  &  &  &  &  &  & \cellcolor{yellow!50} & \cellcolor{yellow!50} &  &  & \cellcolor{yellow!50} \\ 
  &Spread &  &  &  & \cellcolor{yellow!50} &  & \cellcolor{yellow!50} &  & \cellcolor{green!50} & \cellcolor{green!50} & \cellcolor{green!50} & \cellcolor{yellow!50} &  &  \\ 
\hline\multirow{ 3}{*}{Unemp.}  &Monetary &  &  &  &  &  &  &  &  &  &  &  &  &  \\ 
  &Information &  &  &  &  &  &  &  & \cellcolor{yellow!50} & \cellcolor{green!50} & \cellcolor{green!50} & \cellcolor{yellow!50} & \cellcolor{green!50} & \cellcolor{green!50} \\ 
  &Spread &  &  & \cellcolor{yellow!50} & \cellcolor{yellow!50} &  & \cellcolor{green!50} & \cellcolor{yellow!50} &  &  &  &  & \cellcolor{yellow!50} & \cellcolor{yellow!50} \\ 
  \hline\multirow{ 3}{*}{CPI}  &Monetary & \cellcolor{yellow!50} & \cellcolor{yellow!50} & \cellcolor{yellow!50} & \cellcolor{green!50} & \cellcolor{yellow!50} & \cellcolor{green!50} & \cellcolor{green!50} &  & \cellcolor{yellow!50} &  &  &  &  \\ 
  &Information &  &  &  &  &  &  &  &  &  &  &  &  &  \\ 
  &Spread & \cellcolor{green!50} & \cellcolor{green!50} &  & \cellcolor{green!50} & \cellcolor{green!50} & \cellcolor{green!50} & \cellcolor{green!50} & \cellcolor{green!50} & \cellcolor{green!50} & \cellcolor{green!50} & \cellcolor{green!50} & \cellcolor{green!50} & \cellcolor{green!50} \\ 
   \hline\multirow{ 3}{*}{Industry} &Monetary &  &  &  &  &  & \cellcolor{yellow!50} & \cellcolor{green!50} &  &  &  &  &  &  \\ 
  &Information & \cellcolor{yellow!50} &  &  &  &  & \cellcolor{yellow!50} &  &  &  &  &  & \cellcolor{yellow!50} & \cellcolor{yellow!50} \\ 
  &Spread &  &  &  &  &  & \cellcolor{green!50} &  &  & \cellcolor{green!50} & \cellcolor{green!50} & \cellcolor{green!50} & \cellcolor{green!50} & \cellcolor{green!50} \\ 
   \hline\multirow{ 3}{*}{LT Rate} &Monetary &  &  &  &  &  &  & \cellcolor{green!50} & \cellcolor{green!50} & \cellcolor{yellow!50} & \cellcolor{green!50} & \cellcolor{yellow!50} & \cellcolor{yellow!50} & \cellcolor{yellow!50} \\ 
  &Information &  &  &  & \cellcolor{green!50} & \cellcolor{green!50} &  &  &  &  &  &  &  &  \\ 
  &Spread &  &  &  &  &  &  & \cellcolor{yellow!50} & \cellcolor{yellow!50} &  &  & \cellcolor{yellow!50} & \cellcolor{green!50} & \cellcolor{green!50} \\ 
   \hline

&h= & 13 & 14 & 15 & 16 & 17 & 18 & 19 & 20 & 21 & 22 & 23 & 24 & 25 \\ 
  \hline
\multirow{ 3}{*}{REER} &Monetary &  &  &  & \cellcolor{yellow!50} & \cellcolor{green!50} & \cellcolor{green!50} & \cellcolor{yellow!50} &  & \cellcolor{yellow!50} & \cellcolor{yellow!50} & \cellcolor{yellow!50} &  &  \\ 
 & Information &  &  &  &  &  &  &  &  &  & \cellcolor{yellow!50} &  &  &  \\ 
 & Spread &  &  &  &  &  &  &  &  &  &  &  & \cellcolor{green!50} & \cellcolor{yellow!50} \\ 
\hline\multirow{ 3}{*}{Unemp.} & Monetary &  &  &  &  &  &  &  &  &  &  &  &  &  \\ 
 & Information & \cellcolor{green!50} & \cellcolor{green!50} & \cellcolor{green!50} & \cellcolor{green!50} &  &  &  &  &  &  &  &  &  \\ 
 & Spread & \cellcolor{yellow!50} & \cellcolor{green!50} &  &  &  &  &  &  &  &  &  &  &  \\ 
  \hline\multirow{ 3}{*}{CPI} & Monetary &  &  &  &  &  & \cellcolor{yellow!50} &  & \cellcolor{green!50} & \cellcolor{yellow!50} &  &  &  & \cellcolor{yellow!50} \\ 
 & Information &  &  &  &  &  &  &  &  &  &  &  &  &  \\ 
  &Spread & \cellcolor{green!50} & \cellcolor{green!50} & \cellcolor{green!50} & \cellcolor{green!50} & \cellcolor{green!50} & \cellcolor{green!50} & \cellcolor{green!50} & \cellcolor{green!50} & \cellcolor{green!50} &  &  &  &  \\ 
  \hline\multirow{ 3}{*}{Industry} & Monetary &  &  &  &  &  &  &  &  &  &  & \cellcolor{yellow!50} & \cellcolor{yellow!50} &  \\ 
  &Information & \cellcolor{yellow!50} & \cellcolor{yellow!50} &  &  &  &  &  &  &  &  &  &  &  \\ 
  &Spread & \cellcolor{green!50} & \cellcolor{green!50} & \cellcolor{green!50} & \cellcolor{green!50} & \cellcolor{green!50} &  &  &  & \cellcolor{yellow!50} & \cellcolor{green!50} & \cellcolor{yellow!50} & \cellcolor{green!50} & \cellcolor{yellow!50} \\ 
    \hline\multirow{ 3}{*}{LT Rate} &Monetary & \cellcolor{yellow!50} &  &  &  &  &  &  &  &  &  &  &  &  \\ 
  &Information &  &  &  &  &  &  &  &  &  &  &  &  &  \\ 
  & Spread & \cellcolor{green!50} & \cellcolor{green!50} & \cellcolor{green!50} & \cellcolor{green!50} & \cellcolor{green!50} & \cellcolor{green!50} & \cellcolor{green!50} & \cellcolor{green!50} & \cellcolor{yellow!50} &  &  &  &  \\ 
   \hline
   \end{tabular}}
\end{footnotesize}
    \caption{Conditional IRFs Positive Sign-Effect}
    \floatfoot{We report the significance level of the conditional IRFs for a positive shock with the sign effect specification.  White cells refer to p-values greater than 0.1, yellow cells are p-values between 0.1 and 0.05 and green cells are p-values smaller than 0.05.}
\end{table}
\noindent
\textbf{Table 6}, provides the results for significance tests on the sign-effect, positive-shock, conditional impulse response functions. That is we test the hypotheses:
$$H_0^{i,h}:\hat{\psi}_{j,h,0}^i+\hat{\Gamma}_{j,h,0}^i=0, \quad i \in \{monetary,information,spread\}$$
To do so we construct the restriction matrix:
$$R_1=\langle 1,0,0,1,0,0,...\rangle \quad R_2=\langle 0,1,0,0,1,0,...\rangle \quad R_3=\langle 0,0,1,0,0,1,...\rangle $$
We then construct the Wald statistic in the same way. We find that conditional on $x_t\vargeq0$ and $\delta>0$, there are significant effects of the monetary policy shock to REER, CPI and the Long term Rate, significant effects of the information shock to unemployment at horizon 11 to 18 and significant effects of the spread shocks to CPI, Industry and Unemployment.

\begin{table}[t]
\begin{footnotesize}
\centering
\scalebox{0.8}{ \begin{tabular}{|r|r|rrrrrrrrrrrrr|}
    \hline
&h= & 0 & 1 & 2 & 3 & 4 & 5 & 6 & 7 & 8 & 9 & 10 & 11 & 12 \\ 
  \hline
\multirow{ 3}{*}{REER}&Monetary &  &  &  &  &  &  &  &  &  &  &  &  &  \\ 
 & Information &  &  &  &  &  &  &  & \cellcolor{yellow!50} & \cellcolor{green!50} & \cellcolor{green!50} &  & \cellcolor{yellow!50} &  \\ 
 & Spread &  & \cellcolor{green!50} &  &  &  &  &  &  & \cellcolor{yellow!50} & \cellcolor{green!50} &  &  &  \\ 
  \hline\multirow{ 3}{*}{Unemp.} & Monetary &  &  &  &  &  &  & \cellcolor{green!50} & \cellcolor{yellow!50} &  & \cellcolor{green!50} &  &  &  \\ 
 & Information &  &  &  &  & \cellcolor{green!50} &  &  &  &  &  &  &  &  \\ 
 & Spread &  & \cellcolor{green!50} & \cellcolor{green!50} & \cellcolor{yellow!50} & \cellcolor{green!50} & \cellcolor{yellow!50} &  &  &  &  &  &  &  \\ 
  \hline\multirow{ 3}{*}{CPI} & Monetary &  &  &  &  &  &  & \cellcolor{yellow!50} &  & \cellcolor{yellow!50} &  &  &  &  \\ 
 & Information &  &  &  &  &  &  &  &  &  &  &  &  &  \\ 
 & Spread &  &  &  &  &  &  &  &  &  &  &  &  &  \\ 
\hline\multirow{ 3}{*}{Industry} & Monetary &  &  &  &  &  & \cellcolor{green!50} & \cellcolor{green!50} &  & \cellcolor{green!50} &  &  &  &  \\ 
 & Information & \cellcolor{yellow!50} &  &  &  & \cellcolor{yellow!50} & \cellcolor{yellow!50} &  &  &  &  &  &  &  \\ 
 & Spread &  &  &  &  &  & \cellcolor{green!50} & \cellcolor{green!50} & \cellcolor{yellow!50} & \cellcolor{green!50} & \cellcolor{yellow!50} & \cellcolor{green!50} & \cellcolor{yellow!50} &  \\ 
  \hline\multirow{ 3}{*}{LT Rate} & Monetary & \cellcolor{yellow!50} &  &  &  &  &  &  &  &  &  &  &  &  \\ 
 & Information &  &  &  & \cellcolor{yellow!50} &  &  &  &  &  &  &  &  &  \\ 
 & Spread &  &  &  &  &  &  &  &  &  &  &  &  &  \\ 
   \hline

 &h= & 13 & 14 & 15 & 16 & 17 & 18 & 19 & 20 & 21 & 22 & 23 & 24 & 25 \\ 
  \hline
\multirow{ 3}{*}{REER} &Monetary &  &  &  &  &  &  &  &  &  &  &  &  &  \\ 
&  Information &  &  &  &  &  &  &  &  & \cellcolor{yellow!50} &  &  &  &  \\ 
&  Spread &  & \cellcolor{yellow!50} & \cellcolor{green!50} &  &  &  &  &  &  &  &  &  &  \\ 
  \hline\multirow{ 3}{*}{Unemp.} &  Monetary &  &  &  &  &  &  &  &  &  &  &  & \cellcolor{green!50} & \cellcolor{green!50} \\ 
& Information &  &  &  &  &  &  &  &  &  &  &  &  &  \\ 
&  Spread &  &  &  &  &  &  &  &  & \cellcolor{yellow!50} & \cellcolor{green!50} & \cellcolor{green!50} & \cellcolor{green!50} & \cellcolor{green!50} \\ 
  \hline\multirow{ 3}{*}{CPI} &  Monetary & \cellcolor{green!50} & \cellcolor{yellow!50} & \cellcolor{yellow!50} &  & \cellcolor{yellow!50} & \cellcolor{yellow!50} &  & \cellcolor{yellow!50} &  &  &  &  &  \\ 
&  Information &  &  &  &  &  & \cellcolor{green!50} & \cellcolor{green!50} & \cellcolor{yellow!50} &  &  &  &  &  \\ 
&  Spread &  &  &  &  &  &  &  &  &  & \cellcolor{green!50} & \cellcolor{green!50} & \cellcolor{green!50} & \cellcolor{green!50} \\ 
\hline\multirow{ 3}{*}{Industry} &  Monetary &  &  &  &  &  &  &  &  &  & \cellcolor{yellow!50} & \cellcolor{yellow!50} &  &  \\ 
&  Information &  &  &  &  &  &  &  &  &  &  &  &  &  \\ 
&  Spread &  &  &  &  &  &  &  &  &  &  &  &  &  \\ 
  \hline\multirow{ 3}{*}{LT Rate} &  Monetary &  &  &  &  &  &  &  &  &  &  &  &  &  \\ 
&  Information &  &  &  &  &  &  &  &  &  &  &  &  &  \\ 
&  Spread &  &  &  &  &  &  &  &  &  &  &  &  &  \\ 
   \hline
   \end{tabular}}
\end{footnotesize}
    \caption{Conditional IRFs Negative Sign-Effect}
    \floatfoot{We report the significance level of the conditional IRFs for a negative shock with the sign effect specification. Yellow cells are p-values between 0.1 and 0.05 and green cells are p-values smaller than 0.05.}
\end{table}
\noindent
\textbf{Table 7}, provides the results for significance tests on the sign-effect, negative-shock, conditional impulse response functions. That is we test the hypotheses:
$$H_0^{i,h}:\hat{\Gamma}_{j,h,0}^i-\hat{\psi}_{j,h,0}^i=0, \quad i \in \{monetary,information,spread\}$$
Which imply the restriction matrices:
$$R_1=\langle -1,0,0,1,0,0,...\rangle \quad R_2=\langle 0,-1,0,0,1,0,...\rangle \quad R_3=\langle 0,0,-1,0,0,1,...\rangle $$
From the wald test we find that if $\delta<0$ and $x_t \varleq 0$, then there are significant effects of a monetary policy shock to the CPI and Industry, significant effects of an information shock to the REER, unemployement rate and the long term rate and significant effects of the spread shock to the REER, the unemployement rate and industrial production.

\section{Conclusion} \label{sec:conclusion}
We investigated the presence of non-linearities in the impulse responses of EU countries that are not part of the EU exchange rate mechanism to exogenous monetary policy shocks. Using the strategy of \cite{fanelli2022sovereign}, we identify three shocks, a pure monetary shock, a central bank information shock and a spread shock. Focusing on their impact on the real exchange rate, the unemployment rate, the consumer price index, an index of industrial production and the long term government bond rate, we estimate benchmark linear impulse response functions and conditional and unconditional impulse response of non-linear functions of the shock using the method of \cite{Goncalves2021}. The linear specification finds significant impacts of the pure monetary shock on all three nominal variables, while the spread shock impacts the two real variables. We find that, when compared with the sign effect specification, the linear specification underestimates the response of the long term rate to monetary policy, the unemployment rate to the information shock and the REER, CPI and industrial production to the spread shock. The size effect specification estimates impulse response that are close to the linear specification. We then perform a battery of significance test on both the coefficient of the non-linear transformation of the shock and both the conditional and unconditional impulse response functions. We find evidence suggesting the presence of sign non-linearities in the response to these monetary policy shocks. The evidence for size non-linearities is much weaker.


\pagebreak 

\newpage
\bibliography{main}

\begin{thebibliography}{28}
\providecommand{\natexlab}[1]{#1}
\providecommand{\url}[1]{\texttt{#1}}
\expandafter\ifx\csname urlstyle\endcsname\relax
  \providecommand{\doi}[1]{doi: #1}\else
  \providecommand{\doi}{doi: \begingroup \urlstyle{rm}\Url}\fi

\bibitem[Akaike(1973)]{Akaike1973}
Hirotsugu Akaike.
\newblock Information theory and an extension of the maximum likelihood principle.
\newblock \emph{2nd International Symposium on Information Theory}, 1973.

\bibitem[Altavilla et~al.(2019)Altavilla, Brugnolini, Gürkaynak, Motto, and Ragusa]{Altavilla2019}
Carlo Altavilla, Luca Brugnolini, Refet~S. Gürkaynak, Roberto Motto, and Giuseppe Ragusa.
\newblock {Measuring euro area monetary policy}.
\newblock \emph{Journal of Monetary Economics}, 108\penalty0 (C):\penalty0 162--179, 2019.

\bibitem[Arellano(1987)]{Arellano1987}
Manuel. Arellano.
\newblock Practitioners’ corner: Computing robust standard errors for within-groups estimators*.
\newblock \emph{Oxford Bulletin of Economics and Statistics}, 49\penalty0 (4):\penalty0 431--434, 1987.
\newblock \doi{https://doi.org/10.1111/j.1468-0084.1987.mp49004006.x}.
\newblock URL \url{https://onlinelibrary.wiley.com/doi/abs/10.1111/j.1468-0084.1987.mp49004006.x}.

\bibitem[Atkinson et~al.(2020)Atkinson, Richter, and Throckmorton]{atkinsonRichterThrockmorton2020}
Tyler Atkinson, Alexander~W Richter, and Nathaniel~A Throckmorton.
\newblock The zero lower bound and estimation accuracy.
\newblock \emph{Journal of Monetary Economics}, 115:\penalty0 249--264, 2020.

\bibitem[Cabilio and Masaro(1996)]{CabilioMasaro1996}
Paul Cabilio and Joe Masaro.
\newblock A simple test of symmetry about an unknown median.
\newblock \emph{The Canadian Journal of Statistics / La Revue Canadienne de Statistique}, 24\penalty0 (3):\penalty0 349--361, 1996.
\newblock ISSN 03195724.
\newblock URL \url{http://www.jstor.org/stable/3315744}.

\bibitem[Caggiano et~al.(2017)Caggiano, Castelnuovo, and Pellegrino]{caggianocastelnuovopellegrino2017}
Giovanni Caggiano, Efrem Castelnuovo, and Giovanni Pellegrino.
\newblock Estimating the real effects of uncertainty shocks at the zero lower bound.
\newblock \emph{European Economic Review}, 100:\penalty0 257--272, 2017.

\bibitem[Camara(R\&R, 2023)]{Camara2023US}
Santiago Camara.
\newblock Spillovers of us interest rates.
\newblock \emph{Revise \& Resubmit, Journal of International Economcis}, R\&R, 2023.

\bibitem[Caravello and Martinez-Bruera(Working Paper, 2024)]{CaravelloMartinezWP}
Tomas Caravello and Pedro Martinez-Bruera.
\newblock Disentangling sign and size non-linearities.
\newblock \emph{Working Paper}, Working Paper, 2024.

\bibitem[Dabrowski(2022)]{Dabrowski2022}
Marek Dabrowski.
\newblock Anti-fragmentation: an incomplete diagnosis and wrong solution.
\newblock \emph{European Parliament, ECON Commitee Study, Monetary Dialogue}, 2022.

\bibitem[Dees et~al.(2007)Dees, Mauro, Pesaran, and Smith]{PesaranGVAR2007}
Stephane Dees, Filippo~di Mauro, M.~Hashem Pesaran, and L.~Vanessa Smith.
\newblock Exploring the international linkages of the euro area: a global var analysis.
\newblock \emph{Journal of Applied Econometrics}, 22\penalty0 (1):\penalty0 1--38, 2007.
\newblock \doi{https://doi.org/10.1002/jae.932}.
\newblock URL \url{https://onlinelibrary.wiley.com/doi/abs/10.1002/jae.932}.

\bibitem[Dornbusch(1976)]{dornbusch1976expectations}
Rudiger Dornbusch.
\newblock Expectations and exchange rate dynamics.
\newblock \emph{Journal of political Economy}, 84\penalty0 (6):\penalty0 1161--1176, 1976.

\bibitem[Dufour and Renault(1998)]{dufour_reneault_1998}
Jean-Marie Dufour and Eric Renault.
\newblock Short run and long run causality in time series: Theory.
\newblock \emph{Econometrica}, 1998.

\bibitem[Fanelli and Marsi(2022)]{fanelli2022sovereign}
Luca Fanelli and Antonio Marsi.
\newblock Sovereign spreads and unconventional monetary policy in the euro area: A tale of three shocks.
\newblock \emph{European Economic Review}, 150:\penalty0 104281, 2022.

\bibitem[Fern{\'a}ndez-Villaverde et~al.(2015)Fern{\'a}ndez-Villaverde, Gordon, Guerr{\'o}n-Quintana, and Rubio-Ramirez]{fernandezGordonGuerronRubio2015}
Jes{\'u}s Fern{\'a}ndez-Villaverde, Grey Gordon, Pablo Guerr{\'o}n-Quintana, and Juan~F Rubio-Ramirez.
\newblock Nonlinear adventures at the zero lower bound.
\newblock \emph{Journal of Economic Dynamics and Control}, 57:\penalty0 182--204, 2015.

\bibitem[Fleming(1962)]{fleming1962domestic}
J~Marcus Fleming.
\newblock Domestic financial policies under fixed and under floating exchange rates.
\newblock \emph{Staff Papers}, 9\penalty0 (3):\penalty0 369--380, 1962.

\bibitem[Gon\c{c}alves et~al.(2021)Gon\c{c}alves, Herrera, Kilian, and Pesavento]{Goncalves2021}
Silvia Gon\c{c}alves, Ana~Maria Herrera, Lutz Kilian, and Elena Pesavento.
\newblock {Impulse response analysis for structural dynamic models with nonlinear regressors}.
\newblock \emph{Journal of Econometrics}, pages 107--130, 2021.

\bibitem[Hajek and Horvath(2018)]{hajekHorvath2018}
Jan Hajek and Roman Horvath.
\newblock International spillovers of (un) conventional monetary policy: The effect of the ecb and the us fed on non-euro eu countries.
\newblock \emph{Economic Systems}, 42\penalty0 (1):\penalty0 91--105, 2018.

\bibitem[Jarociński(2022)]{JAROCINSKI2022}
Marek Jarociński.
\newblock Central bank information effects and transatlantic spillovers.
\newblock \emph{Journal of International Economics}, 139:\penalty0 103683, 2022.
\newblock ISSN 0022-1996.
\newblock \doi{https://doi.org/10.1016/j.jinteco.2022.103683}.
\newblock URL \url{https://www.sciencedirect.com/science/article/pii/S0022199622001155}.

\bibitem[Jord{\`a}(2005)]{jorda2005estimation}
{\`O}scar Jord{\`a}.
\newblock Estimation and inference of impulse responses by local projections.
\newblock \emph{American economic review}, 95\penalty0 (1):\penalty0 161--182, 2005.

\bibitem[Miao et~al.(2006)Miao, Gel, and Gastwirth]{MiaoYuliaGastwirth2006}
Weiwen Miao, Yulia Gel, and Joseph Gastwirth.
\newblock A new test of symmetry about an unknown median.
\newblock \emph{Random Walk, Sequential Analysis and Related Topics-A Festschrift in Honor of Yuan-Shih Chow}, 12 2006.
\newblock \doi{10.1142/9789812772558_0013}.

\bibitem[Mira(1999)]{Mira1999}
Antonietta Mira.
\newblock Distribution-free test for symmetry based on bonferroni's measure.
\newblock \emph{Journal of Applied Statistics}, 26\penalty0 (8):\penalty0 959--972, 1999.
\newblock \doi{10.1080/02664769921963}.
\newblock URL \url{https://doi.org/10.1080/02664769921963}.

\bibitem[Miranda-Agrippino and Nenova(2022)]{Miranda-Agrippo-Nenova2022}
Silvia Miranda-Agrippino and Tsvetelina Nenova.
\newblock A tale of two global monetary policies.
\newblock \emph{Journal of International Economics}, 136:\penalty0 103606, 2022.

\bibitem[Mundell(1963)]{mundell1963capital}
Robert~A Mundell.
\newblock Capital mobility and stabilization policy under fixed and flexible exchange rates.
\newblock \emph{Canadian Journal of Economics and Political Science/Revue canadienne de economiques et science politique}, 29\penalty0 (4):\penalty0 475--485, 1963.

\bibitem[Schwarz(1978)]{Schwarz1978}
Gideon Schwarz.
\newblock {Estimating the Dimension of a Model}.
\newblock \emph{The Annals of Statistics}, 6\penalty0 (2):\penalty0 461 -- 464, 1978.
\newblock \doi{10.1214/aos/1176344136}.

\bibitem[Wald(1943)]{Wald1943}
Abraham Wald.
\newblock Tests of statistical hypotheses concerning several parameters when the number of observations is large.
\newblock \emph{Transactions of the American Mathematical Society}, 54\penalty0 (3):\penalty0 426--482, 1943.
\newblock ISSN 00029947.
\newblock URL \url{http://www.jstor.org/stable/1990256}.

\bibitem[Wu and Xia(2016)]{wuXia2016}
Jing~Cynthia Wu and Fan~Dora Xia.
\newblock Measuring the macroeconomic impact of monetary policy at the zero lower bound.
\newblock \emph{Journal of Money, Credit and Banking}, 48\penalty0 (2-3):\penalty0 253--291, 2016.

\bibitem[Wyplosz(2022)]{Wyplosz2022}
Charles Wyplosz.
\newblock The tpi: a useful step, just a step.
\newblock \emph{European Parliament, ECON Commitee Study, Monetary Dialogue}, 2022.

\bibitem[Yitzhaki(1996)]{Yitzhaki1996}
Shlomo Yitzhaki.
\newblock On using linear regressions in welfare economics.
\newblock \emph{Journal of Business \& Economic Statistics}, 14\penalty0 (4):\penalty0 478--486, 1996.
\newblock ISSN 07350015.
\newblock URL \url{http://www.jstor.org/stable/1392256}.

\end{thebibliography}
\pagebreak
\section*{Datasets Used}

\textbf{1) IMF dataset}, was used to source the index of industrial production, the consumer price index, the long term rate and the unemployment rate for all 5 countries as well as the aggregated CPI for the Euro-Zone. \url{https://www.imf.org/en/Data}\\[\baselineskip]
\textbf{2) OECD dataset}, was used to source the unemployment rate and an index of industrial production for the Euro-Zone. \url{https://data-explorer.oecd.org}\\[\baselineskip]
\textbf{3) Bruegel}'s real effective exchange rate database was used to source the real exchange rate for all 5 countries and the euro-area.
\url{https://www.bruegel.org/publications/datasets/real-effective-exchange-rates-for-178-countries-a-new-database}.\\[\baselineskip]
\textbf{4) The Euro Area Monetary Policy Event-Study Database}, was used to get the high-frequency movement in the overnight indexed swaps, the STOXX50 index and the Italian-OIS bond rate spreads, used to compute the monetary policy shocks. \url{https://www.ecb.europa.eu/pub/pdf/annex/Dataset_EA-MPD.xlsx}\\[\baselineskip]

\pagebreak
\section*{Appendix}
\subsection*{Appendix A: AIC selection Results}
In this section we provide the full model selection results with the information criterion of \cite{Akaike1973}. The columns represent the horizons, while the first three rows are the optimal lag selection q, p, r. The last row is the optimal lag selection. That is, $t^2 \equiv I_1=I_2=1$, $t \equiv I_1=1, I_2=0$ and $0\equiv I_1=I_2=0$.

\begin{table}[h!]
\centering
\scalebox{0.85}{
\captionof{table}{AIC Selection for Real Exchange Rate}
\begin{tabular}{r|lllllllllllllllllllllllll}
  \hline
 h=& 0 & 1 & 2 & 3 & 4 & 5 & 6 & 7 & 8 & 9 & 10 & 11 & 12 & 13 & 14 & 15 & 16 & 17 & 18 & 19 & 20 & 21 & 22 & 23 & 24 \\ 
  \hline
q & 2 & 4 & 3 & 3 & 3 & 2 & 2 & 2 & 2 & 2 & 2 & 2 & 2 & 2 & 2 & 2 & 2 & 2 & 2 & 2 & 2 & 2 & 2 & 2 & 3 \\ 
  p & 2 & 4 & 3 & 3 & 3 & 2 & 2 & 2 & 2 & 2 & 2 & 2 & 2 & 2 & 2 & 2 & 2 & 2 & 2 & 2 & 2 & 2 & 2 & 2 & 3 \\ 
  r & 2 & 4 & 3 & 3 & 3 & 2 & 2 & 2 & 2 & 2 & 2 & 2 & 2 & 2 & 2 & 2 & 2 & 2 & 2 & 2 & 2 & 2 & 2 & 2 & 2 \\ 
  T & 0 & $t^2$& $t^2$& $t^2$& $t^2$& $t^2$& $t^2$& $t^2$& $t^2$& $t^2$& $t^2$& $t^2$& $t^2$& $t^2$& $t^2$& $t^2$& $t^2$& $t^2$& $t^2$& $t^2$& $t^2$& $t^2$& $t^2$& $t^2$& $t^2$\\ 
   \hline
\end{tabular}}
\end{table}

\begin{table}[h!]
\centering
\scalebox{0.85}{
\captionof{table}{AIC Selection for the Unemployment Rate}
\begin{tabular}{r|lllllllllllllllllllllllll}
  \hline
 h=& 0 & 1 & 2 & 3 & 4 & 5 & 6 & 7 & 8 & 9 & 10 & 11 & 12 & 13 & 14 & 15 & 16 & 17 & 18 & 19 & 20 & 21 & 22 & 23 & 24 \\ 
  \hline
q & 2 & 2 & 2 & 2 & 2 & 2 & 2 & 2 & 3 & 3 & 2 & 2 & 2 & 2 & 2 & 2 & 2 & 2 & 2 & 2 & 2 & 2 & 2 & 2 & 2 \\ 
  p & 2 & 2 & 2 & 2 & 2 & 2 & 2 & 2 & 2 & 2 & 2 & 2 & 2 & 2 & 2 & 2 & 2 & 2 & 2 & 2 & 2 & 2 & 2 & 2 & 2 \\ 
  r & 2 & 2 & 2 & 2 & 2 & 2 & 2 & 2 & 3 & 3 & 2 & 2 & 2 & 2 & 2 & 2 & 2 & 2 & 2 & 2 & 2 & 2 & 2 & 2 & 2 \\ 
  T & $t^2$& $t^2$& $t^2$& $t^2$& $t^2$& $t^2$& $t^2$& $t^2$& $t$ & $t$ & $t$ & $t$ & $t$ & $t$ & $t$ & $t$ & $t$ & $t$ & $t^2$& $t^2$& $t^2$& $t^2$& $t^2$& $t^2$& $t^2$\\ 
   \hline
\end{tabular}}
\end{table}
\begin{table}[h!]
\centering
\scalebox{0.85}{
\captionof{table}{AIC Selection for the Consumer Price Index}
\begin{tabular}{r|lllllllllllllllllllllllll}
  \hline
 h=& 0 & 1 & 2 & 3 & 4 & 5 & 6 & 7 & 8 & 9 & 10 & 11 & 12 & 13 & 14 & 15 & 16 & 17 & 18 & 19 & 20 & 21 & 22 & 23 & 24 \\ 
  \hline
q & 2 & 2 & 2 & 2 & 2 & 2 & 2 & 2 & 2 & 2 & 2 & 2 & 2 & 2 & 2 & 2 & 2 & 2 & 2 & 2 & 2 & 3 & 2 & 2 & 2 \\ 
  p & 2 & 2 & 2 & 2 & 2 & 2 & 2 & 2 & 2 & 2 & 2 & 2 & 2 & 2 & 2 & 2 & 2 & 2 & 2 & 2 & 2 & 2 & 2 & 2 & 2 \\ 
  r & 2 & 2 & 2 & 2 & 2 & 2 & 2 & 2 & 2 & 2 & 2 & 2 & 2 & 2 & 2 & 2 & 2 & 2 & 2 & 2 & 2 & 3 & 2 & 2 & 2 \\ 
  T & $t^2$& $t^2$& $t^2$& $t^2$& $t^2$& $t^2$& $t^2$& $t^2$& $t^2$& $t^2$& $t^2$& $t^2$& $t^2$& $t^2$& $t^2$& $t^2$& $t^2$& $t^2$& $t^2$& $t^2$& $t^2$& $t^2$& $t^2$& $t^2$& $t^2$\\ 
   \hline
\end{tabular}}
\end{table}
\begin{table}[h!]
\centering
\scalebox{0.85}{
\captionof{table}{AIC Selection for the Industrial production}
\begin{tabular}{r|lllllllllllllllllllllllll}
  \hline
 h=& 0 & 1 & 2 & 3 & 4 & 5 & 6 & 7 & 8 & 9 & 10 & 11 & 12 & 13 & 14 & 15 & 16 & 17 & 18 & 19 & 20 & 21 & 22 & 23 & 24 \\ 
  \hline
q & 2 & 2 & 2 & 2 & 2 & 2 & 2 & 2 & 2 & 3 & 3 & 3 & 3 & 2 & 3 & 2 & 2 & 2 & 2 & 2 & 2 & 2 & 2 & 2 & 2 \\ 
  p & 2 & 2 & 2 & 2 & 2 & 2 & 2 & 2 & 2 & 2 & 2 & 2 & 2 & 2 & 2 & 2 & 2 & 2 & 2 & 2 & 2 & 2 & 2 & 2 & 2 \\ 
  r & 2 & 2 & 2 & 2 & 2 & 2 & 3 & 3 & 3 & 3 & 3 & 3 & 3 & 2 & 3 & 2 & 2 & 2 & 2 & 2 & 2 & 2 & 2 & 2 & 2 \\ 
  T & $t^2$& $t^2$& $t^2$& $t^2$& $t^2$& $t^2$& $t^2$& $t^2$& $t^2$& $t^2$& $t^2$& $t^2$& $t^2$& $t^2$& $t^2$& $t^2$& $t^2$& $t^2$& $t^2$& $t^2$& $t$ & $t$ & $t$ & $t$ & $t$ \\ 
   \hline
\end{tabular}}
\end{table}
\begin{table}[h!]
\centering
\scalebox{0.85}{
\captionof{table}{AIC Selection for the Long Term Interest Rate}
\begin{tabular}{r|lllllllllllllllllllllllll}
  \hline
 h=& 0 & 1 & 2 & 3 & 4 & 5 & 6 & 7 & 8 & 9 & 10 & 11 & 12 & 13 & 14 & 15 & 16 & 17 & 18 & 19 & 20 & 21 & 22 & 23 & 24 \\ 
  \hline
q & 2 & 2 & 2 & 7 & 7 & 7 & 7 & 7 & 7 & 7 & 7 & 7 & 7 & 7 & 7 & 7 & 7 & 7 & 7 & 6 & 5 & 4 & 4 & 4 & 3 \\ 
  p & 2 & 2 & 2 & 2 & 6 & 6 & 5 & 5 & 3 & 2 & 2 & 2 & 2 & 2 & 2 & 2 & 2 & 2 & 2 & 2 & 2 & 2 & 4 & 4 & 2 \\ 
  r & 2 & 2 & 2 & 3 & 7 & 7 & 7 & 7 & 4 & 2 & 2 & 7 & 7 & 7 & 5 & 4 & 3 & 3 & 2 & 3 & 3 & 2 & 2 & 2 & 2 \\ 
  T & $t^2$& $t^2$& $t^2$& $t^2$& $t^2$& $t^2$& $t^2$& $t^2$& $t^2$& $t^2$& $t^2$& $t^2$& $t^2$& $t^2$& $t^2$& $t^2$& $t^2$& $t^2$& $t^2$& $t^2$& $t^2$& $t^2$& $t^2$& $t^2$& $t^2$\\ 
   \hline
\end{tabular}}
\end{table}
\pagebreak
\section*{Appendix B: Comparaison with BIC}
In this section we provide the full model selection results with the information criterion of \cite{Schwarz1978}. The columns represent the horizons, while the first three rows are the optimal lag selection q, p, r. The last row is the optimal lag selection. That is, $t^2 \equiv I_1=I_2=1$, $t \equiv I_1=1, I_2=0$ and $0\equiv I_1=I_2=0$.
\begin{table}[h!]
\centering
\scalebox{0.85}{
\captionof{table}{BIC Selection for the Real Exchange Rate}
\begin{tabular}{r|lllllllllllllllllllllllll}
  \hline
 h=& 0 & 1 & 2 & 3 & 4 & 5 & 6 & 7 & 8 & 9 & 10 & 11 & 12 & 13 & 14 & 15 & 16 & 17 & 18 & 19 & 20 & 21 & 22 & 23 & 24 \\ 
  \hline
q & 2 & 2 & 2 & 2 & 2 & 2 & 2 & 2 & 2 & 2 & 2 & 2 & 2 & 2 & 2 & 2 & 2 & 2 & 2 & 2 & 2 & 2 & 2 & 2 & 2 \\ 
  p & 2 & 2 & 2 & 2 & 2 & 2 & 2 & 2 & 2 & 2 & 2 & 2 & 2 & 2 & 2 & 2 & 2 & 2 & 2 & 2 & 2 & 2 & 2 & 2 & 2 \\ 
  r & 2 & 2 & 2 & 2 & 2 & 2 & 2 & 2 & 2 & 2 & 2 & 2 & 2 & 2 & 2 & 2 & 2 & 2 & 2 & 2 & 2 & 2 & 2 & 2 & 2 \\ 
  T & 0 & 0 & 0 & $t^2$& $t^2$& $t^2$& $t^2$& $t^2$& $t^2$& $t^2$& $t^2$& $t^2$& $t^2$& $t^2$& $t^2$& $t^2$& $t^2$& $t^2$& $t^2$& $t^2$& $t^2$& $t^2$& $t^2$& $t^2$& $t^2$\\ 
   \hline
\end{tabular}}
\end{table}
\begin{table}[h!]
\centering
\scalebox{0.85}{
\captionof{table}{BIC Selection for the Unemployement Rate}
\begin{tabular}{r|lllllllllllllllllllllllll}
  \hline
 h= & 0 & 1 & 2 & 3 & 4 & 5 & 6 & 7 & 8 & 9 & 10 & 11 & 12 & 13 & 14 & 15 & 16 & 17 & 18 & 19 & 20 & 21 & 22 & 23 & 24 \\ 
  \hline
q & 2 & 2 & 2 & 2 & 2 & 2 & 2 & 2 & 2 & 2 & 2 & 2 & 2 & 2 & 2 & 2 & 2 & 2 & 2 & 2 & 2 & 2 & 2 & 2 & 2 \\ 
  p & 2 & 2 & 2 & 2 & 2 & 2 & 2 & 2 & 2 & 2 & 2 & 2 & 2 & 2 & 2 & 2 & 2 & 2 & 2 & 2 & 2 & 2 & 2 & 2 & 2 \\ 
  r & 2 & 2 & 2 & 2 & 2 & 2 & 2 & 2 & 2 & 2 & 2 & 2 & 2 & 2 & 2 & 2 & 2 & 2 & 2 & 2 & 2 & 2 & 2 & 2 & 2 \\ 
  T & $t$ & $t$ & $t^2$& $t$ & $t$ & $t$ & $t$ & $t$ & $t$ & $t$ & $t$ & $t$ & $t$ & $t$ & $t$ & $t$ & $t$ & $t$ & $t$ & $t$ & $t$ & $t$ & $t$ & $t$ & $t^2$\\ 
   \hline
\end{tabular}}
\end{table}
\begin{table}[h!]
\centering
\scalebox{0.85}{
\captionof{table}{BIC Selection for the Consumer Price Index}
\begin{tabular}{r|lllllllllllllllllllllllll}
  \hline
 h=& 0 & 1 & 2 & 3 & 4 & 5 & 6 & 7 & 8 & 9 & 10 & 11 & 12 & 13 & 14 & 15 & 16 & 17 & 18 & 19 & 20 & 21 & 22 & 23 & 24 \\ 
  \hline
q & 2 & 2 & 2 & 2 & 2 & 2 & 2 & 2 & 2 & 2 & 2 & 2 & 2 & 2 & 2 & 2 & 2 & 2 & 2 & 2 & 2 & 2 & 2 & 2 & 2 \\ 
  p & 2 & 2 & 2 & 2 & 2 & 2 & 2 & 2 & 2 & 2 & 2 & 2 & 2 & 2 & 2 & 2 & 2 & 2 & 2 & 2 & 2 & 2 & 2 & 2 & 2 \\ 
  r & 2 & 2 & 2 & 2 & 2 & 2 & 2 & 2 & 2 & 2 & 2 & 2 & 2 & 2 & 2 & 2 & 2 & 2 & 2 & 2 & 2 & 2 & 2 & 2 & 2 \\ 
  T & 0 & $t^2$& $t^2$& $t^2$& $t^2$& $t^2$& $t^2$& $t^2$& $t^2$& $t^2$& $t^2$& $t^2$& $t^2$& $t^2$& $t^2$& $t^2$& $t^2$& $t^2$& $t^2$& $t^2$& $t^2$& $t^2$& $t^2$& $t^2$& $t^2$\\ 
   \hline
\end{tabular}}
\end{table}

\begin{table}[h!]
\centering
\scalebox{0.85}{
\captionof{table}{BIC Selection for the Industrial Production}
\begin{tabular}{r|lllllllllllllllllllllllll}
  \hline
 h=& 0 & 1 & 2 & 3 & 4 & 5 & 6 & 7 & 8 & 9 & 10 & 11 & 12 & 13 & 14 & 15 & 16 & 17 & 18 & 19 & 20 & 21 & 22 & 23 & 24 \\ 
  \hline
q & 2 & 2 & 2 & 2 & 2 & 2 & 2 & 2 & 2 & 2 & 2 & 2 & 2 & 2 & 2 & 2 & 2 & 2 & 2 & 2 & 2 & 2 & 2 & 2 & 2 \\ 
  p & 2 & 2 & 2 & 2 & 2 & 2 & 2 & 2 & 2 & 2 & 2 & 2 & 2 & 2 & 2 & 2 & 2 & 2 & 2 & 2 & 2 & 2 & 2 & 2 & 2 \\ 
  r & 2 & 2 & 2 & 2 & 2 & 2 & 2 & 2 & 2 & 2 & 2 & 2 & 2 & 2 & 2 & 2 & 2 & 2 & 2 & 2 & 2 & 2 & 2 & 2 & 2 \\ 
  T & $t$ & $t^2$& $t^2$& $t^2$& $t^2$& $t^2$& $t^2$& $t^2$& $t^2$& $t^2$& $t^2$& $t^2$& $t^2$& $t^2$& $t^2$& $t^2$& $t$ & $t$ & $t$ & $t$ & $t$ & $t$ & $t$ & $t$ & $t$ \\ 
   \hline
\end{tabular}}
\end{table}
\begin{table}[h!]
\centering
\scalebox{0.85}{
\captionof{table}{BIC Selection for the Long Term Rate}
\begin{tabular}{r|lllllllllllllllllllllllll}
  \hline
 h=& 0 & 1 & 2 & 3 & 4 & 5 & 6 & 7 & 8 & 9 & 10 & 11 & 12 & 13 & 14 & 15 & 16 & 17 & 18 & 19 & 20 & 21 & 22 & 23 & 24 \\ 
  \hline
q & 2 & 2 & 2 & 2 & 2 & 2 & 2 & 2 & 2 & 2 & 2 & 2 & 2 & 2 & 2 & 2 & 2 & 2 & 2 & 2 & 2 & 2 & 2 & 2 & 2 \\ 
  p & 2 & 2 & 2 & 2 & 2 & 2 & 2 & 2 & 2 & 2 & 2 & 2 & 2 & 2 & 2 & 2 & 2 & 2 & 2 & 2 & 2 & 2 & 2 & 2 & 2 \\ 
  r & 2 & 2 & 2 & 2 & 2 & 2 & 2 & 2 & 2 & 2 & 2 & 2 & 2 & 2 & 2 & 2 & 2 & 2 & 2 & 2 & 2 & 2 & 2 & 2 & 2 \\ 
  T & 0 & $t^2$& $t^2$& $t^2$& $t^2$& $t^2$& $t^2$& $t^2$& $t^2$& $t^2$& $t^2$& $t^2$& $t^2$& $t^2$& $t^2$& $t^2$& $t^2$& $t^2$& $t^2$& $t^2$& $t^2$& $t^2$& $t^2$& $t^2$& $t^2$\\ 
   \hline
\end{tabular}}
\end{table}
\pagebreak

\subsection*{Appendix C: Symmetry of the Shocks }
In this section we plot the histogram of every shock and provide some summary statistics as well as the result of the symmetry tests of \cite{CabilioMasaro1996} (CM), \cite{MiaoYuliaGastwirth2006} (M1) and \cite{Mira1999} (M2). All these tests have an alternative hypothesis of asymmetry of the population distribution.
\begin{figure}[h!]
    \centering
    \includegraphics[scale=0.25]{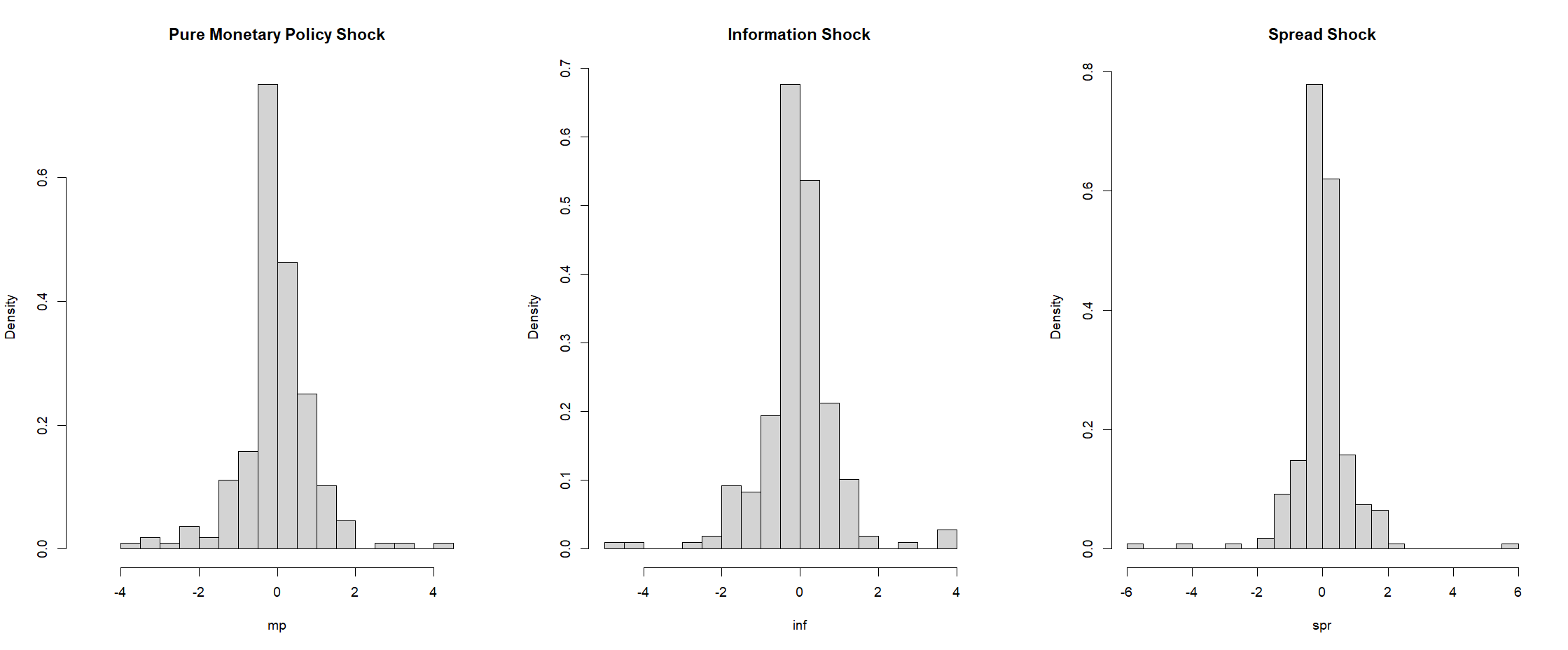}
    \caption{Histograms of the Three Shocks}
\end{figure}
\begin{table}[h!]
    \centering
    \begin{tabular}{c|c|c|c}
      &  Monetary Policy Shock   & Information Shock & Spread Shock  \\
      Mean   &  0.0077   &  -0.0349 & 0.0284 \\
      Standard Deviation & 0.9186 & 0.9751 & 0.8906 \\
      Skewness & -0.2338 & -0.3155 & -0.7736\\
      $80^{th}$ Quantile & 0.539 & 0.447 & 0.395 \\
     CM  Test Statistic & 0.1640 & -0.6955 &  0.6203 \\
      CM  p-value & 0.888  & 0.526 & 0.564 \\
           M1  Test Statistic & 0.2045 &  -0.9095 &  0.9005 \\
      M1  p-value & 0.896 & 0.478 & 0.554 \\
           M2  Test Statistic &  -0.1613 & -0.6517 & 0.5599\\
      M2  p-value &   0.88  & 0.526 &  0.64
    \end{tabular}
    \caption{Caption}
    \label{tab:my_label}
\end{table}\\
The monetary policy and information shock are approximately symmetric as their skewness is small. The spread shock is a bit less symmetric due to the presence of many high magnitude negative shocks. However, all three tests fail to reject that the sample comes from a symmetric distribution. All three shocks were standardized and then assigned to a monthly time series therefore, their means are close to 0 and their standard deviations are close to 1.

\pagebreak
\subsection*{Appendix D: IRFs and Inference for the pre-covid period} \label{sec:main_results} 
In this section we provide the estimated impulse response functions and the inference results for a sub-sample that ends in December 2019. That is:
\begin{table}[h!]
    \centering
    \small
    \begin{tabular}{c|c|c}
\textbf{Country} &\textbf{Entry Date}			&	\textbf{Exit Date} \\ \hline \hline
Czechia	&	2002M01	&	2019M12	\\
Hungary	&	2002M01	&	2019M12	\\
Poland	&	2002M01	&	2019M12	\\
Romania	&	2005M04	&	2017M09	\\
Sweden	&	2002M01	&	2017M05\\ \hline \hline
    \end{tabular}
    \caption{Composition of the Panel for the pre-pandemic period}
    \label{tab:country_list_Adv}
\end{table}\\
We start with the linear specification and move to both the sign and size non-linear specifications.
\subsection*{The Linear Model}
\begin{figure}[b!]
    \centering
    \includegraphics[scale=0.25]{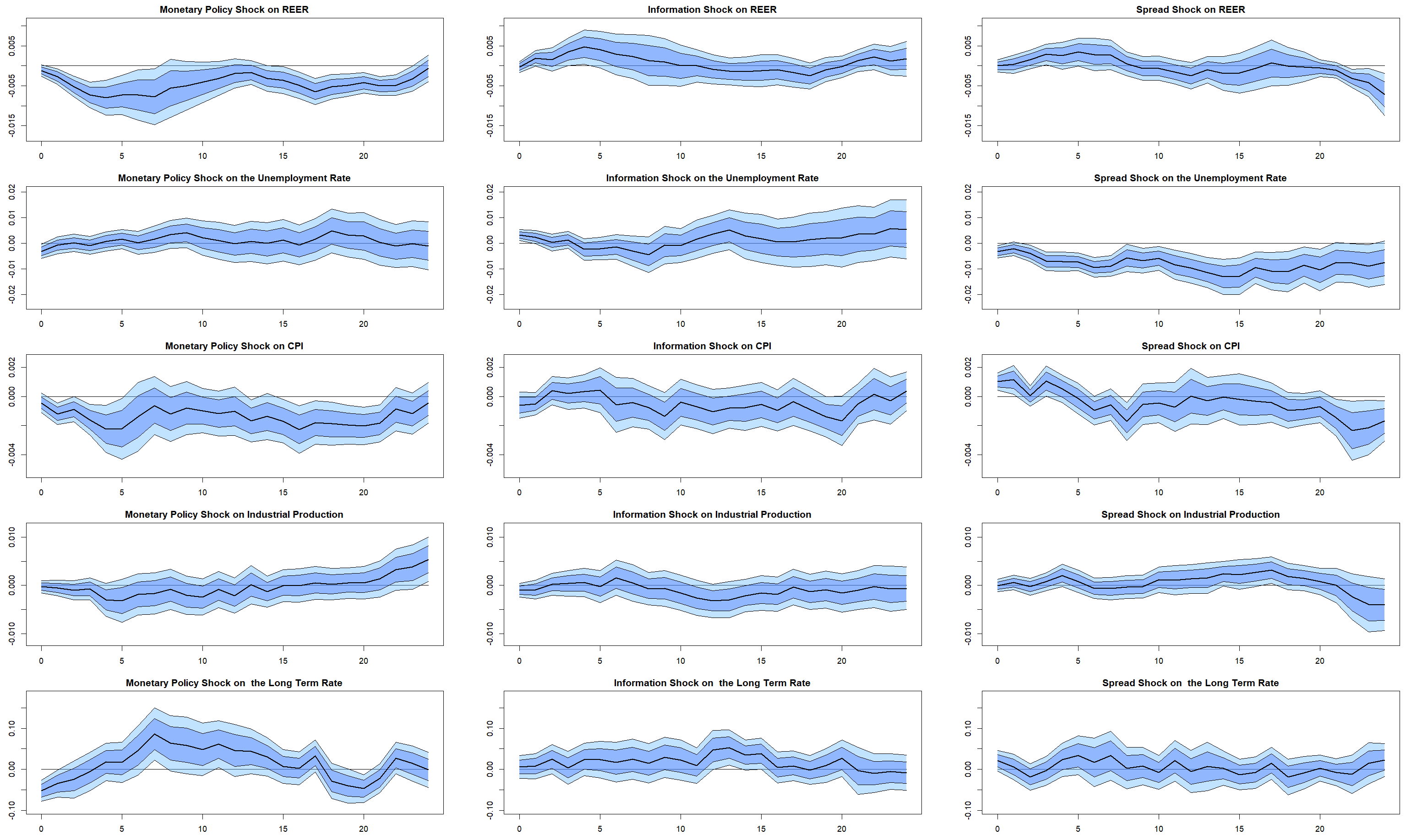}
    \caption{Results from the linear specification (6)}
\floatfoot{The horizontal axis represents the months since the shock (h). The rows are in order: the real exchange rate, the unemployment rate, the consumer price index, the index of industrial production and the long term interest rate. The columns are in order: the pure monetary policy shock, the information shock and the spread shock. The real exchange rate, consumer price index and industrial production index are in log points, while the unemployment rate and long term interest rate are in percentages. The dark blue lines represent the 1 standard deviation confidence interval while the light blue are $90\%$. The standard errors are computed using the method of \cite{Arellano1987} with the ``Jackknife" $HC_3$, HCCME.}.
\end{figure}
\noindent
\textbf{Figure 7}, provides the estimated impulse response functions of the linear specification. The pure monetary policy shock has significant, but temporary, effects on the real exchange rate and the CPI. The negative effects on industrial production are only significant at the $68\%$ significance level. The information shock has an quick but temporary impact on the real exchange rate and a delayed (and less significant) impact on CPI. Lastly, the spread shock has a significant negative impact on the unemployment rate and positive impact on industrial production. Its impact on CPI is positive at first (increase in CPI) but in the long run, this impact becomes negative.
\subsection*{The Non-Linear Model, Sign Non-Linearities}

\begin{figure}[!b]
    \centering
    \includegraphics[scale=0.25]{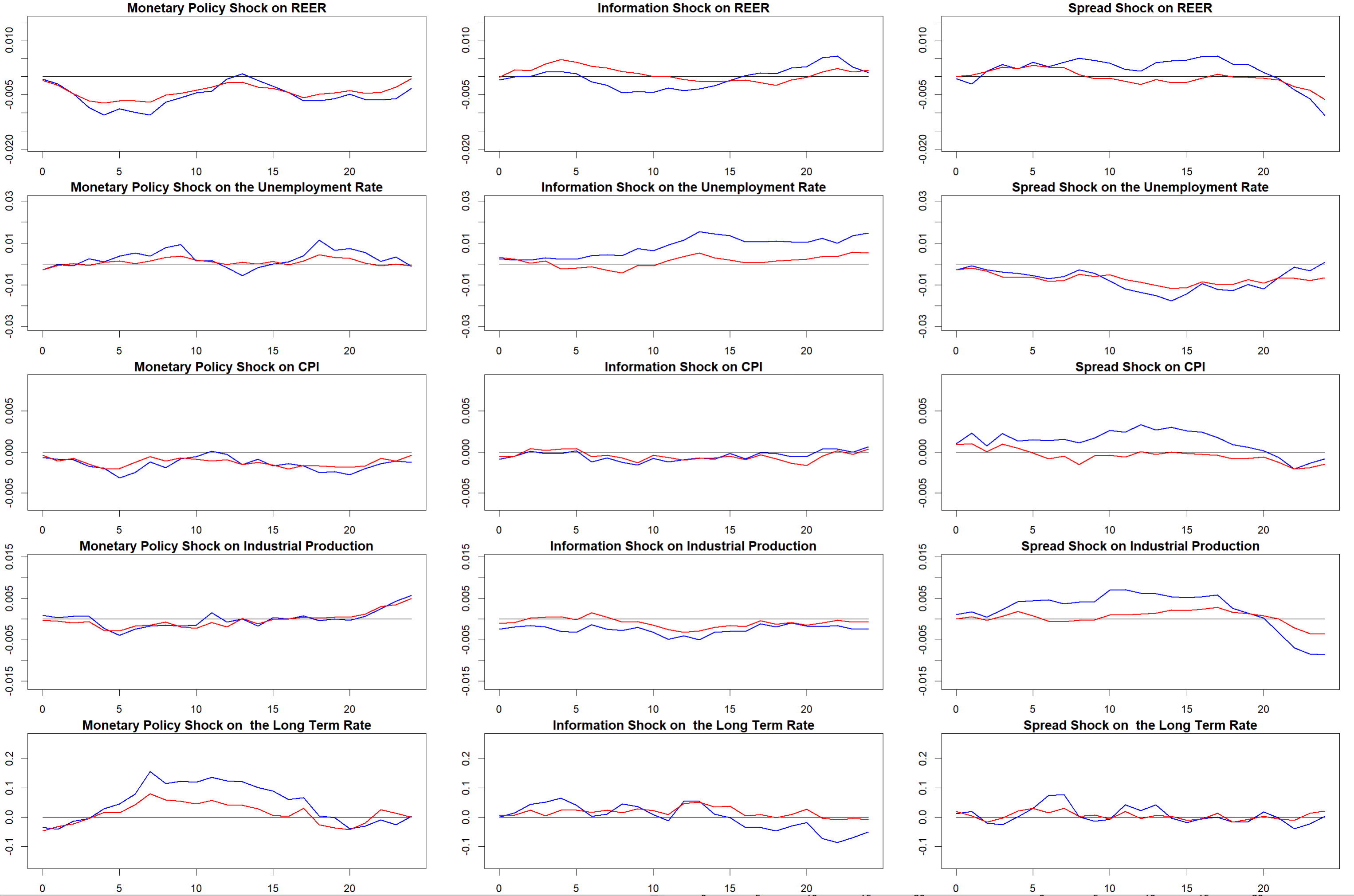}
    \caption{Impulse responses of the sign specification à la \cite{Goncalves2021}}
    \floatfoot{The red line is the original linear impulse response function, the blue line is the \cite{Goncalves2021} non-linear IRF. The horizontal axis represents the months since the shock (h). The rows are in order: the real exchange rate, the unemployment rate, the consumer price index, the index of industrial production and the long term interest rate. The columns are in order: the pure monetary policy shock, the information shock and the spread shock. The real exchange rate, consumer price index and industrial production index are in log points, while the unemployment rate and long term interest rate are in percentages.}
\end{figure}
\noindent
In \textbf{figure 8}, we plot in blue our estimate of the unconditional impulse response function to a 1 standard deviation disturbance in the sign-effect non-linear specification. In red is the original linear impulse response function to a 1 standard deviation shock. When comparing both we see that for most graphs the two IRFs nearly coincide, implying for them that on average both specification will estimate very similar IRFs. However, the monetary shock on the long term rate, is a lot stronger (temporarily) starting from horizon 5, when estimated with the sign-specification. For the information shock, the linear model estimates an effect on unemployment and industrial production that is close to zero, while the non-linear model estimates a much larger and positive IRF for unemployment and a negative IRF for industrial production.  This implies that the effect of the information shock on real variables contains a sign effect. Lastly, the non-linear model estimates a larger, but temporary effect of the spread shock on industrial production, the real exchange rate and the consumer price index.

\begin{figure}[!b]
    \centering
    \includegraphics[scale=0.25]{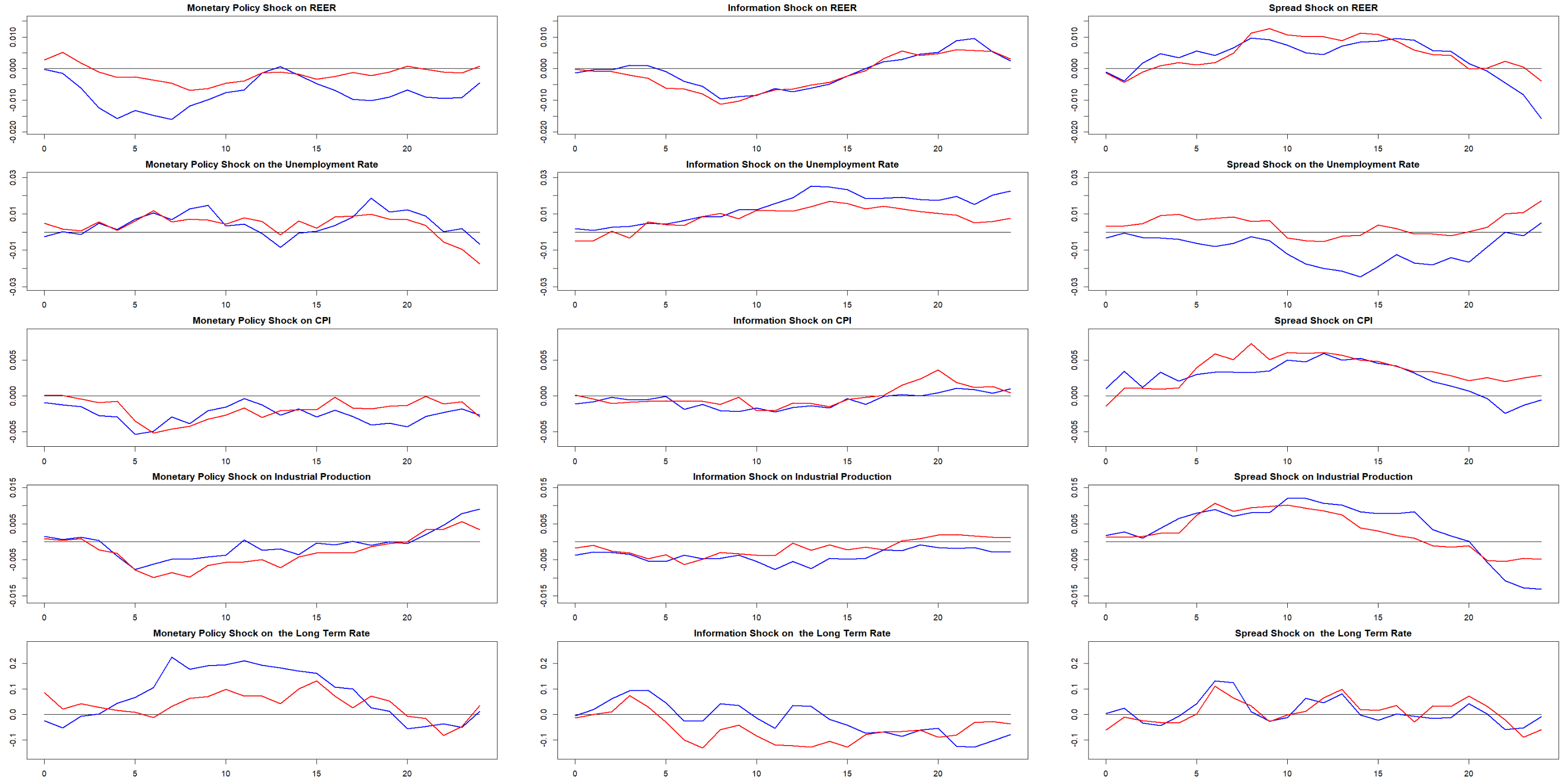}
    \caption{Conditional IRFs of the sign-effect specification }
    \floatfoot{The rows are in order: the real exchange rate, the unemployment rate, the consumer price index, the index of industrial production and the long term interest rate. The columns are in order: the pure monetary policy shock, the information shock and the spread shock. The real exchange rate, consumer price index and industrial production index are in log points, while the unemployment rate and long term interest rate are in percentages.}
\end{figure}
\noindent
\textbf{Figure 9}, compares two conditional impulse response functions estimated with the sign-effect specification. These are:
\begin{align*}
    \text{In blue:  IRF}(h,j|x_t \vargeq 0, \delta>0)= & \hat{\Gamma}_{j,h,0}+\hat{\psi}_{j,h,0} \\
       \text{In red:  IRF}(h,j|x_t \varleq 0, \delta<0)= & \hat{\Gamma}_{j,h,0}-\hat{\psi}_{j,h,0} 
\end{align*}
The blue IRFs are very similar to the unconditional \cite{Goncalves2021} IRFs as the standard deviation and the estimated $\hat{A}$ are both close to 1. We can see that for monetary policy shocks, a (conditional) positive shock has a large negative effect on REER and a large positive effect on the long term rate, while a negative shock has a more restrained effect on both. For industrial production, it is the opposite, negative monetary policy shocks have more negative impacts. Positive information shocks have stronger permanent effects on unemployment, while negative shocks have stronger negative impacts on the long term rate. For the spread shock the main difference is in its effect on the unemployment rate. A negative spread shock, has an effect on the unemployment rate close to 0. On the other hand, positive shocks have large negative (decrease) on the unemployment rate. All other conditional Impulse response functions are similar when comparing negative and positive shocks.

\subsection*{The Non-Linear Model, Size Non-Linearities}

\begin{figure}[!ht]
    \centering
    \includegraphics[scale=0.22]{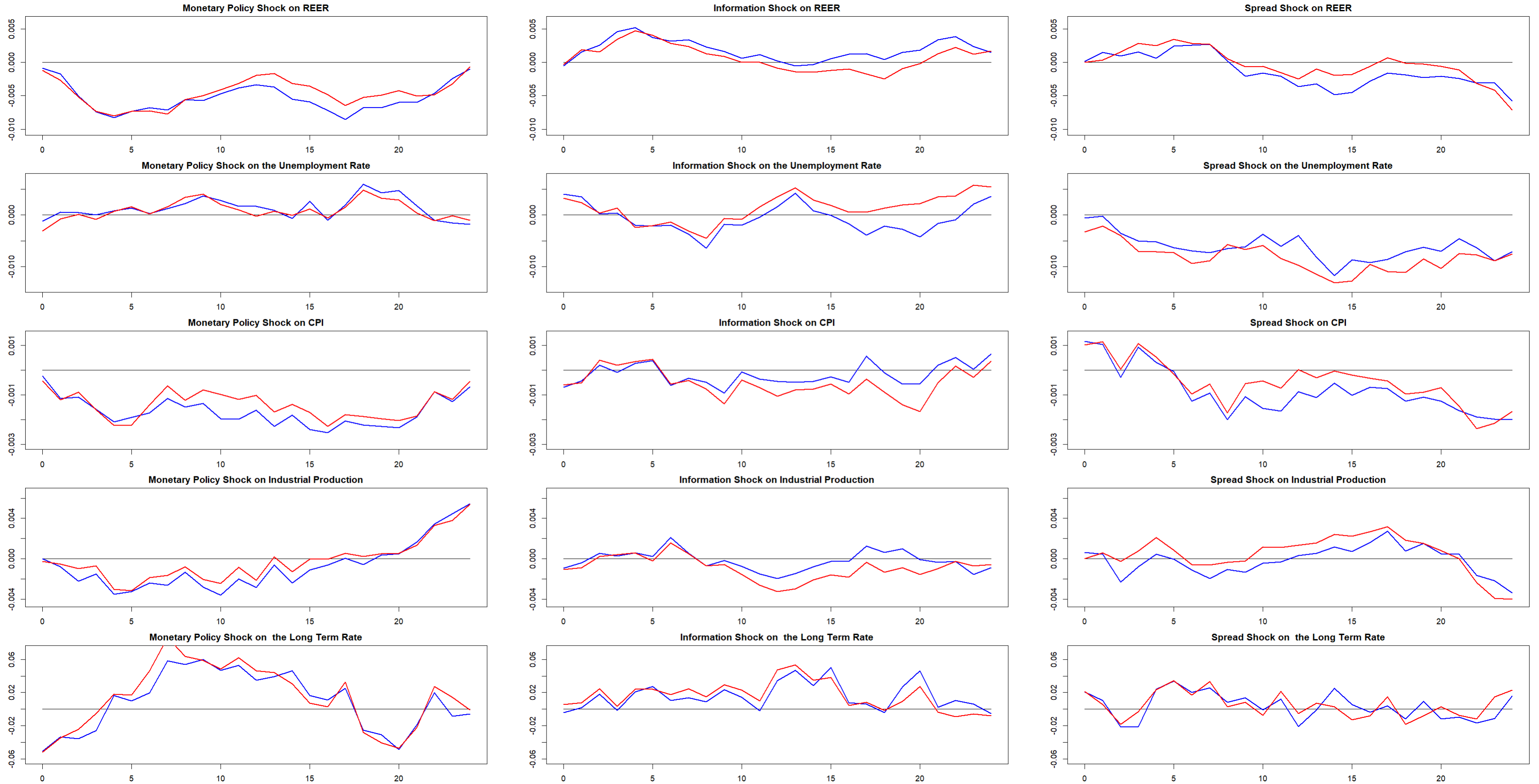}
    \caption{Impulse responses of the size specification à la \cite{Goncalves2021}}
    \floatfoot{The rows are in order: the real exchange rate, the unemployment rate, the consumer price index, the index of industrial production and the long term interest rate. The columns are in order: the pure monetary policy shock, the information shock and the spread shock. The real exchange rate, consumer price index and industrial production index are in log points, while the unemployment rate and long term interest rate are in percentages.}
\end{figure}
\noindent
\textbf{Figure 10} plots the \cite{Goncalves2021} unconditional impulse response functions for the size-effect specification in blue. We set the threshold $\bar{b}$ for each shock to be such that $P(|x_t|\varleq\bar{b})\approx 0.6$. The shock used in computing of $\hat{A}$ is still 1 standard deviation. In red is the original linear impulse response function. The monetary policy shock has a slightly larger temporary effect on the REER for horizon 13 to 22. The information shock has a smaller effect on the unemployment rate, CPI and Industrial production when it is estimated with the non-linear specification. For all other impulse response functions, the linear and non-linear specification coincide. This suggests that there are no important size effects for a 1 standard deviation shock.

\begin{figure}[!t]
    \centering
    \includegraphics[scale=0.25]{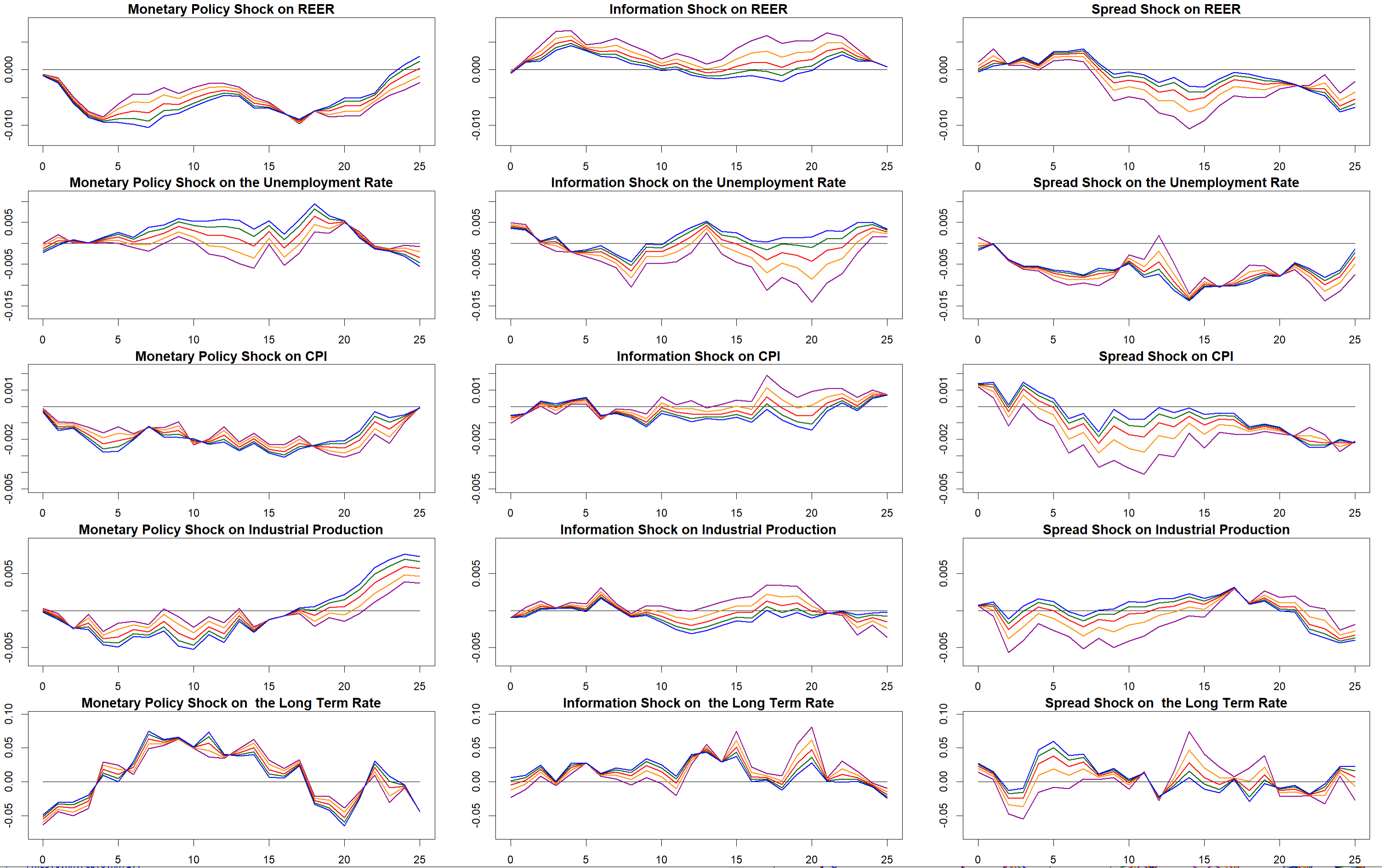}
    \caption{Impulse responses of the size specification à la \cite{Goncalves2021}}
    \floatfoot{ The magenta, orange, red, green and blue line are for shocks that are 0.5, 0.75, 1, 1.25 and 1.5 times the standard deviation respectively.
    The rows are in order: the real exchange rate, the unemployment rate, the consumer price index, the index of industrial production and the long term interest rate. The columns are in order: the pure monetary policy shock, the information shock and the spread shock. The real exchange rate, consumer price index and industrial production index are in log points, while the unemployment rate and long term interest rate are in percentages.}
\end{figure}
\noindent
\textbf{Figure 11} presents an alternative way of illustrating the size effect using the method of \cite{Goncalves2021}. Each line is an impulse response of the following form:
\begin{align*}
    IRF(h,a)=\frac{1}{a}\left(\hat{\psi}_{j,h,0}*a*\sigma+\sum_{t=1}^T \frac{1}{T}\left[f(x_t+a*\sigma)-f(x_t) \right]*a*\hat{\Gamma}_{j,h,0}  \right)
\end{align*}
Where $a\in \{0.5,0.75,1,1.25,1.5\}$.
We can see that for many impulse responses the shape depends greatly on the size of the shock. This suggest that even if we did not see any differences in the impulse response between the linear and non-linear specification for the average shock (1 standard deviation), it could appear for different size shocks. For example, monetary policy shocks on the REER have bigger marginal effects for larger shocks for the first year ($h<12$), but smaller effects for the second year.  The information shock's marginal effect on REER and Unemployment is larger for smaller shocks. Lastly, the spread shock's marginal effect on the REER, the CPI and Industrial Production is larger for smaller magnitude shocks. Therefore, if there are size effects, they are concave in the magnitude of the shock.
\clearpage

\subsection*{Inference Result}
\begin{table}[h!]
\begin{footnotesize}
\centering
\scalebox{0.8}{ \begin{tabular}{|r|r|rrrrrrrrrrrrr|}
  \hline
&  h= & 0 & 1 & 2 & 3 & 4 & 5 & 6 & 7 & 8 & 9 & 10 & 11 & 12 \\ 

\hline \multirow{ 3}{*}{REER} &Monetary &  & \cellcolor{yellow!50} &  & \cellcolor{green!50} & \cellcolor{green!50} & \cellcolor{green!50} & \cellcolor{green!50} & \cellcolor{green!50} & \cellcolor{green!50} & \cellcolor{green!50} & \cellcolor{green!50} & \cellcolor{green!50} &  \\ 
&  Information &  &  &  &  &  &  & \cellcolor{green!50} & \cellcolor{green!50} & \cellcolor{green!50} & \cellcolor{green!50} & \cellcolor{green!50} & \cellcolor{green!50} & \cellcolor{green!50} \\ 
 & Spread &  & \cellcolor{green!50} &  &  &  & \cellcolor{yellow!50} &  & \cellcolor{green!50} & \cellcolor{green!50} & \cellcolor{green!50} & \cellcolor{green!50} & \cellcolor{green!50} & \cellcolor{green!50} \\ 
\hline \multirow{ 3}{*}{Unemp.} &  Monetary &  &  &  & \cellcolor{green!50} &  & \cellcolor{green!50} & \cellcolor{green!50} & \cellcolor{green!50} & \cellcolor{green!50} & \cellcolor{green!50} &  &  &  \\ 
 & Information &  &  &  &  & \cellcolor{green!50} & \cellcolor{yellow!50} & \cellcolor{green!50} & \cellcolor{green!50} & \cellcolor{green!50} & \cellcolor{green!50} & \cellcolor{green!50} & \cellcolor{green!50} & \cellcolor{green!50} \\ 
 & Spread &  &  &  &  &  &  &  &  &  &  & \cellcolor{green!50} & \cellcolor{green!50} & \cellcolor{green!50} \\ 
 \hline \multirow{ 3}{*}{CPI} & Monetary &  &  &  & \cellcolor{green!50} & \cellcolor{green!50} & \cellcolor{green!50} & \cellcolor{green!50} & \cellcolor{green!50} & \cellcolor{green!50} & \cellcolor{green!50} & \cellcolor{green!50} &  & \cellcolor{green!50} \\ 
&  Information &  &  &  &  &  &  &  &  & \cellcolor{yellow!50} &  & \cellcolor{yellow!50} & \cellcolor{green!50} &  \\ 
&  Spread &  & \cellcolor{green!50} & \cellcolor{green!50} & \cellcolor{green!50} & \cellcolor{green!50} & \cellcolor{green!50} & \cellcolor{green!50} & \cellcolor{green!50} & \cellcolor{green!50} & \cellcolor{green!50} & \cellcolor{green!50} & \cellcolor{green!50} & \cellcolor{green!50} \\ 
\hline \multirow{ 3}{*}{Industry}  &  Monetary &  &  &  &  & \cellcolor{green!50} & \cellcolor{green!50} & \cellcolor{green!50} & \cellcolor{green!50} & \cellcolor{green!50} & \cellcolor{green!50} & \cellcolor{green!50} &  &  \\ 
&  Information & \cellcolor{green!50} &  & \cellcolor{green!50} & \cellcolor{green!50} & \cellcolor{green!50} & \cellcolor{green!50} & \cellcolor{green!50} & \cellcolor{green!50} & \cellcolor{yellow!50} &  & \cellcolor{green!50} & \cellcolor{green!50} &  \\ 
&  Spread & \cellcolor{yellow!50} & \cellcolor{yellow!50} &  & \cellcolor{green!50} & \cellcolor{green!50} & \cellcolor{green!50} & \cellcolor{green!50} & \cellcolor{green!50} & \cellcolor{green!50} & \cellcolor{green!50} & \cellcolor{green!50} & \cellcolor{green!50} & \cellcolor{green!50} \\ 
\hline \multirow{ 3}{*}{LT Rate}  & Monetary & \cellcolor{yellow!50} &  &  &  &  &  &  & \cellcolor{green!50} & \cellcolor{green!50} & \cellcolor{green!50} & \cellcolor{green!50} & \cellcolor{green!50} & \cellcolor{green!50} \\ 
 & Information &  &  &  & \cellcolor{green!50} & \cellcolor{green!50} &  & \cellcolor{green!50} & \cellcolor{green!50} &  &  & \cellcolor{yellow!50} & \cellcolor{green!50} &  \\ 
 & Spread & \cellcolor{yellow!50} &  &  &  &  &  & \cellcolor{green!50} & \cellcolor{green!50} &  &  &  &  & \cellcolor{green!50} \\ 
   \hline
 &  h= & 13 & 14 & 15 & 16 & 17 & 18 & 19 & 20 & 21 & 22 & 23 & 24 &  \\

\hline \multirow{ 3}{*}{REER} &Monetary &  &  & \cellcolor{yellow!50} & \cellcolor{green!50} & \cellcolor{green!50} & \cellcolor{green!50} & \cellcolor{green!50} &  & \cellcolor{green!50} & \cellcolor{green!50} & \cellcolor{green!50} &  &\\ 
&  Information & \cellcolor{green!50} & \cellcolor{green!50} &  &  &  & \cellcolor{green!50} & \cellcolor{green!50} & \cellcolor{green!50} & \cellcolor{green!50} & \cellcolor{green!50} & \cellcolor{green!50} &  &\\ 
&  Spread & \cellcolor{green!50} & \cellcolor{green!50} & \cellcolor{green!50} & \cellcolor{green!50} & \cellcolor{green!50} & \cellcolor{green!50} & \cellcolor{green!50} &  &  &  & \cellcolor{yellow!50} & \cellcolor{green!50} &\\ 
\hline \multirow{ 3}{*}{Unemp.} &  Monetary &  &  &  &  & \cellcolor{yellow!50} & \cellcolor{green!50} & \cellcolor{yellow!50} & \cellcolor{yellow!50} &  &  &  & \cellcolor{yellow!50} &\\ 
&  Information & \cellcolor{green!50} & \cellcolor{green!50} & \cellcolor{green!50} & \cellcolor{green!50} & \cellcolor{green!50} & \cellcolor{green!50} & \cellcolor{green!50} & \cellcolor{green!50} & \cellcolor{green!50} &  & \cellcolor{yellow!50} & \cellcolor{green!50} &\\ 
&  Spread & \cellcolor{green!50} & \cellcolor{green!50} &  &  & \cellcolor{yellow!50} &  &  &  &  &  &  & \cellcolor{green!50} &\\ 
\hline \multirow{ 3}{*}{CPI} &  Monetary & \cellcolor{green!50} & \cellcolor{yellow!50} & \cellcolor{green!50} &  & \cellcolor{green!50} & \cellcolor{green!50} & \cellcolor{green!50} & \cellcolor{green!50} &  & \cellcolor{yellow!50} &  & \cellcolor{green!50} & \\ 
&  Information &  &  &  &  &  &  &  & \cellcolor{yellow!50} & \cellcolor{yellow!50} &  &  &  &\\ 
 & Spread & \cellcolor{green!50} & \cellcolor{green!50} & \cellcolor{green!50} & \cellcolor{green!50} & \cellcolor{green!50} & \cellcolor{green!50} & \cellcolor{green!50} & \cellcolor{yellow!50} &  &  &  &  &\\ 
\hline \multirow{ 3}{*}{Industry}  &  Monetary & \cellcolor{yellow!50} & \cellcolor{yellow!50} &  &  &  &  &  &  &  &  & \cellcolor{green!50} & \cellcolor{green!50} &\\ 
&  Information & \cellcolor{green!50} &  &  &  &  &  &  &  &  &  &  &  &\\ 
 & Spread & \cellcolor{green!50} & \cellcolor{green!50} & \cellcolor{green!50} & \cellcolor{green!50} & \cellcolor{green!50} &  &  &  & \cellcolor{green!50} & \cellcolor{green!50} & \cellcolor{green!50} & \cellcolor{green!50} &\\ 
\hline \multirow{ 3}{*}{LT Rate} &  Monetary & \cellcolor{green!50} & \cellcolor{green!50} & \cellcolor{green!50} & \cellcolor{green!50} & \cellcolor{green!50} & \cellcolor{yellow!50} &  &  &  & \cellcolor{green!50} & \cellcolor{yellow!50} &  &\\ 
&  Information & \cellcolor{yellow!50} & \cellcolor{green!50} & \cellcolor{green!50} & \cellcolor{green!50} & \cellcolor{green!50} & \cellcolor{green!50} & \cellcolor{green!50} & \cellcolor{green!50} & \cellcolor{green!50} & \cellcolor{green!50} & \cellcolor{green!50} & \cellcolor{green!50} &\\ 
&  Spread & \cellcolor{green!50} &  &  &  &  &  &  & \cellcolor{green!50} &  & \cellcolor{yellow!50} & \cellcolor{green!50} &  &\\ 
   \hline
   \end{tabular}}
\end{footnotesize}
    \caption{Significance of the Non-Linear Component of the Sign Effect Specification}
\floatfoot{We report the significance level of the estimated coefficient $\hat{\Gamma}_{j,h,0}$. The non-linear transformation is the absolute value of the shocks. White cells refer to p-values greater than 0.1, yellow cells are p-values between 0.1 and 0.05 and green cells are p-values smaller than 0.05.}
\end{table}
\noindent
\textbf{Table 20} reports the results of Wald tests (\cite{Wald1943}) of the significance of the coefficient on the non-linear transformation of the shocks. As in \cite{CaravelloMartinezWP}, the results of these tests allow us to establish the presence of non-linearities at horizon h. We find multi-horizon significant sigh-effects of every type of shock on every outcome variable with the exception of the information shock on CPI. We describe the structure of the test bellow as we will present results on different version of it. We test the hypotheses, $$H_{0}^{1,h}:\hat{\Gamma}^{monetary}_{j,h,0}=0  \quad H_0^{2,h}: \hat{\Gamma}^{information}_{j,h,0}=0 \quad H_0^{3,h}:\hat{\Gamma}^{spread}_{j,h,0} =0$$
To do so we conduct the Wald test:
\begin{align*}
\hat{\beta}_{j,h}= & \langle\hat{\psi}^{monetary}_{j,h,0},\hat{\psi}^{information}_{j,h,0},\hat{\psi}^{spread}_{j,h,0},\hat{\Gamma}^{monetary}_{j,h,0},\hat{\Gamma}^{information}_{j,h,0},\hat{\Gamma}^{spread}_{j,h,0}, ... \rangle\\
  R_1 = & \langle 0,0,0,1,0,0,...\rangle, \quad R_2 =  \langle 0,0,0,0,1,0,...\rangle, \quad R_3 =  \langle 0,0,0,0,0,1,...\rangle\\
W_{i,j,h}= & (R_i\hat{\beta_{j,h}}^\prime)^\prime (R_i \hat{\Omega} R_i^\prime)^{-1}(R\hat{\beta_{j,h}}^\prime)
\end{align*}
Then for the Wald statistic $W_{i,j,h}\overset{a}{\sim}\chi(1)$, for hypotheses $i=\{1,2,3\}$ and every horizon $h$. The matrix $\hat{\Omega}$ is the covariance matrix of $\beta$ estimated using the method of \cite{Arellano1987} with $HC_3$.

\begin{table}[h!]
\begin{footnotesize}
\centering
\scalebox{0.8}{ \begin{tabular}{|r|r|rrrrrrrrrrrrr|}
  \hline
&  h= & 0 & 1 & 2 & 3 & 4 & 5 & 6 & 7 & 8 & 9 & 10 & 11 & 12 \\ 

\hline \multirow{ 3}{*}{REER} &Monetary &  &  &  &  &  & \cellcolor{green!50} & \cellcolor{green!50} & \cellcolor{green!50} & \cellcolor{green!50} &  \cellcolor{yellow!50} & \cellcolor{green!50} &  \cellcolor{yellow!50} &  \\ 
 & Information &  &  & \cellcolor{green!50} & \cellcolor{green!50} & \cellcolor{green!50} &  &  &  \cellcolor{yellow!50} &  &  &  &  &  \cellcolor{yellow!50} \\ 
&  Spread & \cellcolor{green!50} & \cellcolor{green!50} &  &  &  &  &  &  &  & \cellcolor{green!50} &  \cellcolor{yellow!50} & \cellcolor{green!50} & \cellcolor{green!50} \\ 
\hline \multirow{ 3}{*}{Unemp.} &  Monetary &  \cellcolor{yellow!50} &  &  &  &  &  &  & \cellcolor{green!50} & \cellcolor{green!50} &  \cellcolor{yellow!50} &  \cellcolor{yellow!50} & \cellcolor{green!50} & \cellcolor{green!50} \\ 
&  Information &  &  &  &  &  &  &  &  & \cellcolor{green!50} &  &  &  \cellcolor{yellow!50} &  \\ 
&  Spread &  &  &  &  &  &  &  &  &  &  &  &  & \cellcolor{green!50} \\ 
\hline \multirow{ 3}{*}{CPI} &  Monetary &  &  &  &  \cellcolor{yellow!50} & \cellcolor{green!50} & \cellcolor{green!50} &  &  &  &  &  &  &  \\ 
&  Information &  &  &  &  &  &  &  &  &  &  &  &  &  \cellcolor{yellow!50} \\ 
 & Spread &  &  \cellcolor{yellow!50} & \cellcolor{green!50} & \cellcolor{green!50} & \cellcolor{green!50} & \cellcolor{green!50} & \cellcolor{green!50} & \cellcolor{green!50} & \cellcolor{green!50} & \cellcolor{green!50} & \cellcolor{green!50} & \cellcolor{green!50} & \cellcolor{green!50} \\ 
\hline \multirow{ 3}{*}{Industry}  &  Monetary &  &  &  & \cellcolor{green!50} &  & \cellcolor{green!50} &  &  & \cellcolor{green!50} & \cellcolor{green!50} &  \cellcolor{yellow!50} &  &  \\ 
 & Information &  &  &  &  &  &  &  &  &  &  &  &  &  \\ 
 & Spread &  &  \cellcolor{yellow!50} & \cellcolor{green!50} & \cellcolor{green!50} & \cellcolor{green!50} & \cellcolor{green!50} & \cellcolor{green!50} & \cellcolor{green!50} & \cellcolor{green!50} & \cellcolor{green!50} & \cellcolor{green!50} &  \cellcolor{yellow!50} &  \\ 
\hline \multirow{ 3}{*}{LT Rate} &  Monetary &  &  &  &  &  &  &  &  &  &  &  &  &  \\ 
&  Information & \cellcolor{green!50} &  &  &  &  &  &  &  &  &  &  &  &  \\ 
&  Spread &  &  &  &  &  \cellcolor{yellow!50} &  \cellcolor{yellow!50} &  &  &  &  &  &  &  \\ 
   \hline
&  h= & 13 & 14 & 15 & 16 & 17 & 18 & 19 & 20 & 21 & 22 & 23 & 24 &  \\ 

\hline \multirow{ 3}{*}{REER} &Monetary &  &  &  &  &  &  &  & \cellcolor{green!50} & \cellcolor{green!50} &  & \cellcolor{green!50} & \cellcolor{green!50} & \cellcolor{green!50} \\ 
&  Information &  &  \cellcolor{yellow!50} & \cellcolor{green!50} & \cellcolor{green!50} & \cellcolor{green!50} & \cellcolor{green!50} & \cellcolor{green!50} & \cellcolor{green!50} & \cellcolor{green!50} &  \cellcolor{yellow!50} &  &  &  \\ 
 & Spread & \cellcolor{green!50} & \cellcolor{green!50} & \cellcolor{green!50} & \cellcolor{green!50} &  \cellcolor{yellow!50} &  \cellcolor{yellow!50} &  \cellcolor{yellow!50} &  &  &  &  \cellcolor{yellow!50} &  &  \cellcolor{yellow!50} \\ 
 \hline \multirow{ 3}{*}{Unemp.} & Monetary & \cellcolor{green!50} & \cellcolor{green!50} &  \cellcolor{yellow!50} & \cellcolor{green!50} & \cellcolor{green!50} &  \cellcolor{yellow!50} &  &  &  &  &  &  &  \\ 
 & Information &  &  &  &  & \cellcolor{green!50} &  \cellcolor{yellow!50} &  \cellcolor{yellow!50} & \cellcolor{green!50} & \cellcolor{green!50} &  \cellcolor{yellow!50} &  &  &  \\ 
 & Spread &  &  &  &  &  &  &  &  &  &  &  &  &  \\ 
\hline \multirow{ 3}{*}{CPI}  & Monetary &  &  &  &  &  &  &  &  &  \cellcolor{yellow!50} & \cellcolor{green!50} & \cellcolor{green!50} &  &  \\ 
 & Information &  &  &  &  & \cellcolor{green!50} & \cellcolor{green!50} & \cellcolor{green!50} & \cellcolor{green!50} &  \cellcolor{yellow!50} &  &  &  &  \\ 
 & Spread & \cellcolor{green!50} &  &  \cellcolor{yellow!50} &  &  &  &  &  &  &  &  &  &  \\ 
\hline \multirow{ 3}{*}{Industry}   & Monetary &  &  &  &  &  &  &  & \cellcolor{green!50} & \cellcolor{green!50} & \cellcolor{green!50} & \cellcolor{green!50} & \cellcolor{green!50} &  \cellcolor{yellow!50} \\ 
&  Information &  &  &  &  &  &  \cellcolor{yellow!50} &  &  &  &  &  &  &  \\ 
 & Spread &  &  &  &  &  &  &  &  &  &  &  &  &  \\ 
\hline \multirow{ 3}{*}{LT Rate} &  Monetary &  &  &  &  &  &  &  &  &  &  & \cellcolor{green!50} &  &  \\ 
&  Information &  &  &  \cellcolor{yellow!50} &  &  &  & \cellcolor{green!50} & \cellcolor{green!50} &  &  &  &  &  \\ 
&  Spread &  & \cellcolor{green!50} &  \cellcolor{yellow!50} &  &  & \cellcolor{green!50} &  \cellcolor{yellow!50} &  &  &  &  &  &  \cellcolor{yellow!50} \\ 
   \hline
   \end{tabular}}
\end{footnotesize}
    \caption{Significance of the Non-Linear Component of the Size Effect Specification}
\floatfoot{We report the significance level of the estimated coefficient $\hat{\Gamma}_{j,h,0}$. The non-linear transformation is $f(x_t)=\mathbf{I}\{x_{t} \varleq -\bar{b}\}(x_{t}+\bar{b})+\mathbf{I}\{x_{t} \vargeq \bar{b}\}(x_{t}-\bar{b})$ with $\bar{b}$ chosen such that $P(|x_t|<\bar{b})=0.6$. White cells refer to p-values greater than 0.1, yellow cells are p-values between 0.1 and 0.05 and green cells are p-values smaller than 0.05.}
\end{table}
\noindent
\textbf{Table 21} reports the result of the same Wald test as in table 2, but on the size effect specification. That is: 
\begin{align*}
     R_1 = & \langle 0,0,0,1,0,0,...\rangle, \quad R_2 =  \langle 0,0,0,0,1,0,...\rangle, \quad R_3 =  \langle 0,0,0,0,0,1,...\rangle
\end{align*}
The results are more muted then with the sign effect. For example, there does not seem to be any significant (and persistent over many horizon) size effect of any shock on the long term rate. The information shock has no size effect on industrial production. The persistent and significant non-linearities are that of the spread shock on CPI and Industry and all three shocks on the REER.

\begin{table}[h!]
\begin{footnotesize}
\centering
\scalebox{0.8}{ \begin{tabular}{|r|r|rrrrrrrrrrrrrr|}
  \hline
&h= & 0 & 1 & 2 & 3 & 4 & 5 & 6 & 7 & 8 & 9 & 10 & 11 & 12 & \\ 
  
\hline \multirow{ 3}{*}{REER} &Monetary &&& \cellcolor{green!50} & \cellcolor{green!50} & \cellcolor{green!50} & \cellcolor{green!50} & \cellcolor{green!50} & \cellcolor{green!50} & \cellcolor{green!50} & \cellcolor{yellow!50} &&& & \\ 
&  Information &&&&&&&&&&&&& & \\ 
&  Spread &&&&&& \cellcolor{yellow!50} &&& \cellcolor{yellow!50} &&&& & \\ 
\hline  \multirow{ 3}{*}{Unemp.} & Monetary &&&&&&&&& \cellcolor{yellow!50} & \cellcolor{green!50} &&& & \\ 
&  Information &&&&&&&&&&&& \cellcolor{yellow!50} & \cellcolor{green!50} & \\ 
&  Spread &&&&& \cellcolor{yellow!50} & \cellcolor{green!50} & \cellcolor{green!50} &&&& \cellcolor{yellow!50} & \cellcolor{green!50} & \cellcolor{green!50} & \\ 
\hline  \multirow{ 3}{*}{CPI} &  Monetary &&&& \cellcolor{green!50} & \cellcolor{yellow!50} & \cellcolor{green!50} & \cellcolor{yellow!50} &&&&&& & \\ 
&  Information &&&&&&&&&&&&&& \\ 
&  Spread & \cellcolor{green!50} & \cellcolor{green!50} && \cellcolor{green!50} & \cellcolor{yellow!50} & \cellcolor{yellow!50} &&&&& \cellcolor{green!50} & \cellcolor{green!50} & \cellcolor{green!50} & \\ 
\hline \multirow{ 3}{*}{Industry} &  Monetary &&&&&&&&&&&&& &  \\ 
&  Information & \cellcolor{yellow!50} &&&&&&&&&&& \cellcolor{yellow!50} & &  \\ 
&  Spread &&&& \cellcolor{yellow!50} & \cellcolor{green!50} & \cellcolor{green!50} & \cellcolor{green!50} & \cellcolor{yellow!50} & \cellcolor{green!50} & \cellcolor{green!50} & \cellcolor{green!50} & \cellcolor{green!50} & \cellcolor{green!50} &  \\ 
\hline  \multirow{ 3}{*}{LT Rate} &  Monetary & \cellcolor{yellow!50} &&&&&&& \cellcolor{green!50} & \cellcolor{green!50} & \cellcolor{green!50} & \cellcolor{green!50} & \cellcolor{green!50} & \cellcolor{green!50} &  \\ 
&  Information &&&&& \cellcolor{yellow!50} &&&&&&&& &  \\ 
&  Spread &&&&&&& \cellcolor{green!50} & \cellcolor{green!50} &&&&& & \\ 
   \hline

& h= & 13 & 14 & 15 & 16 & 17 & 18 & 19 & 20 & 21 & 22 & 23 & 24 & 25 & \\ 
  \hline
\multirow{ 3}{*}{REER} &Monetary &&&&& \cellcolor{green!50} & \cellcolor{green!50} & \cellcolor{green!50} & \cellcolor{green!50} & \cellcolor{green!50} & \cellcolor{green!50} & \cellcolor{green!50} && &  \\ 
&  Information &&&&&&&&& \cellcolor{green!50} & \cellcolor{yellow!50} &&& &  \\ 
&  Spread &&&&&&&&&& \cellcolor{yellow!50} & \cellcolor{yellow!50} & \cellcolor{green!50} & \cellcolor{green!50} & \\ 
\hline  \multirow{ 3}{*}{Unemp.} &  Monetary &&&&&&&&&&&&& & \\ 
&  Information & \cellcolor{green!50} & \cellcolor{green!50} & \cellcolor{green!50} & \cellcolor{yellow!50} &&&&&&&&& &\\ 
&  Spread & \cellcolor{green!50} & \cellcolor{green!50} & \cellcolor{yellow!50} &&&&&&&&&& & \\ 
\hline  \multirow{ 3}{*}{CPI} &  Monetary &&&&&& \cellcolor{green!50} & \cellcolor{yellow!50} & \cellcolor{green!50} & \cellcolor{yellow!50} &&&& & \\ 
&  Information &&&&&&&&&&&&& & \\ 
&  Spread & \cellcolor{green!50} & \cellcolor{green!50} & \cellcolor{green!50} & \cellcolor{green!50} & \cellcolor{green!50} &&&&&&&& &\\ 
\hline \multirow{ 3}{*}{Industry} &  Monetary &&&&&&&&&&&&& \cellcolor{yellow!50} &\\ 
&  Information & \cellcolor{yellow!50} &&&&&&&&&&&& &\\ 
&  Spread & \cellcolor{green!50} & \cellcolor{green!50} & \cellcolor{green!50} & \cellcolor{green!50} & \cellcolor{green!50} &&&&& \cellcolor{green!50} & \cellcolor{yellow!50} & \cellcolor{green!50} & \cellcolor{green!50} &\\ 
\hline  \multirow{ 3}{*}{LT Rate} &  Monetary & \cellcolor{green!50} & \cellcolor{green!50} & \cellcolor{green!50} && \cellcolor{yellow!50} &&&&&&&& & \\ 
&  Information &&&&&& \cellcolor{yellow!50} &&& \cellcolor{green!50} & \cellcolor{green!50} & \cellcolor{green!50} && & \\ 
& Spread &&&&&&&&&&&&& &\\ 
   \hline
   \end{tabular}}
\end{footnotesize}
    \caption{Sign-Effect Impulse Response Function \`a la \cite{Goncalves2021}}
    \floatfoot{We report the significance level of the \cite{Goncalves2021} IRF with the sign effect specification.  White cells refer to p-values greater than 0.1, yellow cells are p-values between 0.1 and 0.05 and green cells are p-values smaller than 0.05.}
\end{table}

\noindent
\textbf{Table 22}, reports our attempt at testing the significance of the \cite{Goncalves2021} with a Wald test. This corresponds to the the following three hypotheses at each horizon, $$H_0^{i,h}:\hat{\psi}_{j,h,0}^i\delta^{i}+\hat{A}^i\hat{\Gamma}_{j,h,0}^i=0, \quad i \in \{monetary,information,spread\}$$
To do so, we construct the following matrix of linear restrictions:
\begin{align*}
    R_1= & \langle \delta^{monetary},0,0,\hat{A}^{monetary}(\delta^{monetary}),0,0,...\rangle \\
    R_2= & \langle 0,\delta^{information},0,0,\hat{A}^{information}(\delta^{information}),0,...\rangle \\ 
     R_3= & \langle 0,0,\delta^{spread},0,0,\hat{A}^{spread}(\delta^{spread}),...\rangle  
     \end{align*}
Where: $\delta^i$ is the sample standard deviation of shock $i$ and $\hat{A}^i= \frac{1}{T}\sum_{t=1}^T \left[f(x_t^i+\delta^i)-f(x_t^i)\right] $, and recall that for the sign effect specification $f(x)=|x|$. The Wald Statistic is then constructed in the same way as before. This procedure is not exact as it does not take into account the estimation uncertainty in $\hat{A}^i$. Although we previously established the presence (or not) of non-linearities, the goal of this exercise is to see if these non-linearities do not ``cancel out" the linear effect such that the weighted sum of both is not significant. The \cite{Goncalves2021} IRF with the sign effect specification is significant for monetary shocks to the REER and the long term rate and the spread shock to CPI, unemployment and industrial production. The sign and magnitude of the impulse response is reported in \textbf{figure 8}.

\begin{table}[h!]
\begin{footnotesize}
\centering
\scalebox{0.8}{ \begin{tabular}{|r|r|rrrrrrrrrrrrrr|}
  \hline
& h=& 0 & 1 & 2 & 3 & 4 & 5 & 6 & 7 & 8 & 9 & 10 & 11 & 12 &\\ 
  
\hline \multirow{ 3}{*}{REER}&Monetary &  &  & \cellcolor{green!50} & \cellcolor{green!50} & \cellcolor{green!50} & \cellcolor{green!50} & \cellcolor{green!50} & \cellcolor{green!50} & \cellcolor{yellow!50} & \cellcolor{green!50} & \cellcolor{yellow!50} &  &  & \\ 
&  Information &  &  &  & \cellcolor{green!50} & \cellcolor{green!50} &  &  &  &  &  &  &  &  & \\ 
&  Spread &  &  &  &  &  &  &  &  &  &  &  &  &  & \\ 
\hline \multirow{ 3}{*}{Unemp.} &  Monetary &  &  &  &  &  &  &  &  &  &  &  &  &  & \\ 
&  Information & \cellcolor{green!50} & \cellcolor{yellow!50} &  &  &  &  &  &  &  &  &  &  &  & \\ 
&  Spread &  &  &  &  &  & \cellcolor{yellow!50} & \cellcolor{yellow!50} & \cellcolor{yellow!50} &  &  &  &  &  & \\ 
\hline \multirow{ 3}{*}{CPI} &  Monetary &  & \cellcolor{green!50} & \cellcolor{yellow!50} & \cellcolor{green!50} & \cellcolor{green!50} & \cellcolor{yellow!50} &  &  &  &  & \cellcolor{green!50} & \cellcolor{green!50} &  & \\ 
&  Information &  &  &  &  &  &  &  &  &  &  &  &  &  & \\ 
&  Spread & \cellcolor{green!50} & \cellcolor{yellow!50} &  &  &  &  &  &  & \cellcolor{green!50} &  &  & \cellcolor{yellow!50} &  & \\ 
\hline \multirow{ 3}{*}{Industry}  &  Monetary &  &  & \cellcolor{yellow!50} &  & \cellcolor{yellow!50} &  &  &  &  &  & \cellcolor{yellow!50} &  &  & \\ 
&  Information &  &  &  &  &  &  &  &  &  &  &  &  &  & \\ 
&  Spread &  &  & \cellcolor{yellow!50} &  &  &  &  &  &  &  &  &  &  & \\ 
\hline \multirow{ 3}{*}{LT Rate} &  Monetary & \cellcolor{green!50} &  &  &  &  &  &  &  &  &  &  &  &  & \\ 
&  Information &  &  &  &  &  &  &  &  &  &  &  &  &  & \\ 
&  Spread &  &  &  &  &  &  &  &  &  &  &  &  &  & \\ 
    \hline
&h= & 13 & 14 & 15 & 16 & 17 & 18 & 19 & 20 & 21 & 22 & 23 & 24 & 25 & \\ 

\hline \multirow{ 3}{*}{REER} & Monetary & \cellcolor{yellow!50} & \cellcolor{green!50} & \cellcolor{green!50} & \cellcolor{green!50} & \cellcolor{green!50} & \cellcolor{green!50} & \cellcolor{green!50} & \cellcolor{green!50} & \cellcolor{green!50} & \cellcolor{green!50} &  &  &  & \\ 
&  Information &  &  &  &  &  &  &  &  & \cellcolor{yellow!50} &  &  &  &  & \\ 
&  Spread &  & \cellcolor{green!50} & \cellcolor{green!50} &  &  &  &  &  &  &  &  & \cellcolor{green!50} & \cellcolor{yellow!50} & \\ 
\hline \multirow{ 3}{*}{Unemp.} &  Monetary &  &  &  &  &  &  &  &  &  &  &  &  &  & \\ 
&  Information &  &  &  &  &  &  &  &  &  &  &  &  &  & \\ 
&  Spread & \cellcolor{yellow!50} & \cellcolor{green!50} & \cellcolor{yellow!50} & \cellcolor{yellow!50} &  &  &  &  &  &  &  &  &  & \\ 
\hline \multirow{ 3}{*}{CPI} &  Monetary & \cellcolor{green!50} & \cellcolor{yellow!50} & \cellcolor{green!50} & \cellcolor{green!50} & \cellcolor{green!50} & \cellcolor{green!50} & \cellcolor{green!50} & \cellcolor{green!50} & \cellcolor{green!50} &  &  &  &  & \\ 
&  Information &  &  &  &  &  &  &  &  &  &  &  &  &  & \\ 
&  Spread &  &  &  &  &  &  &  &  &  &  & \cellcolor{yellow!50} & \cellcolor{yellow!50} & \cellcolor{green!50} & \\ 
\hline \multirow{ 3}{*}{Industry} &  Monetary &  &  &  &  &  &  &  &  &  &  &  & \cellcolor{yellow!50} & \cellcolor{yellow!50} & \\ 
&  Information &  &  &  &  &  &  &  &  &  &  &  &  &  & \\ 
&  Spread &  &  &  &  &  &  &  &  &  &  &  &  &  & \\ 
\hline \multirow{ 3}{*}{LT Rate} &  Monetary &  & \cellcolor{yellow!50} &  &  &  &  &  & \cellcolor{green!50} &  &  &  &  &  & \\ 
&  Information &  &  & \cellcolor{yellow!50} &  &  &  &  & \cellcolor{yellow!50} &  &  &  &  &  & \\ 
&  Spread &  &  &  &  &  &  &  &  &  &  &  &  &  & \\ 
   \hline
   \end{tabular}}
\end{footnotesize}
    \caption{Size-Effect Impulse Response Function \`a la \cite{Goncalves2021}}
\floatfoot{We report the significance level of the \cite{Goncalves2021} IRF with the size effect specification.  White cells refer to p-values greater than 0.1, yellow cells are p-values between 0.1 and 0.05 and green cells are p-values smaller than 0.05.}
\end{table}
\pagebreak
\noindent
\textbf{Table 23} reports results of the same test as in table 4, albeit on the size-effect specification, with $\bar{b}$ such that $P(|x_t|<\bar{b})=0.6$. That is we test the hypothesis:
$$H_0^{i,h}:\hat{\psi}_{j,h,0}^i\delta^{i}+\hat{A}^i\hat{\Gamma}_{j,h,0}^i=0, \quad i \in \{monetary,information,spread\}$$
The wald statistic is constructed in the same way as in table 4. As when we tested the significance of $\hat{\Gamma}_{j,h,0}^i$ alone, we find more muted results. The impulse response function is only significant (for multiple horizons) for monetary shocks to REER and CPI.

\begin{table}[h!]
\begin{footnotesize}
\centering
\scalebox{0.8}{ \begin{tabular}{|r|r|rrrrrrrrrrrrrr|}
  \hline
& h=& 0 & 1 & 2 & 3 & 4 & 5 & 6 & 7 & 8 & 9 & 10 & 11 & 12 &\\ 

\hline \multirow{ 3}{*}{REER} &Monetary &  &  & \cellcolor{green!50} & \cellcolor{green!50} & \cellcolor{green!50} & \cellcolor{green!50} & \cellcolor{green!50} & \cellcolor{green!50} & \cellcolor{green!50} & \cellcolor{green!50} &  &  &  & \\ 
&  Information &  &  &  &  &  &  &  &  & \cellcolor{yellow!50} &  & \cellcolor{yellow!50} &  & \cellcolor{yellow!50} & \\ 
 & Spread &  & \cellcolor{yellow!50} &  &  &  &  &  &  & \cellcolor{green!50} & \cellcolor{yellow!50} &  &  &  & \\ 
\hline \multirow{ 3}{*}{Unemp.} &  Monetary &  &  &  &  &  &  &  &  & \cellcolor{yellow!50} & \cellcolor{green!50} &  &  &  & \\ 
&  Information &  &  &  &  &  &  &  &  &  &  &  & \cellcolor{green!50} & \cellcolor{green!50} & \\ 
&  Spread &  &  &  &  &  &  &  &  &  &  & \cellcolor{yellow!50} & \cellcolor{green!50} & \cellcolor{green!50} & \\ 
\hline \multirow{ 3}{*}{CPI} &  Monetary &  &  &  & \cellcolor{green!50} & \cellcolor{yellow!50} & \cellcolor{green!50} & \cellcolor{green!50} &  & \cellcolor{yellow!50} &  &  &  &  & \\ 
&  Information &  &  &  &  &  &  &  &  &  &  &  &  &  & \\ 
&  Spread & \cellcolor{yellow!50} & \cellcolor{green!50} &  & \cellcolor{green!50} & \cellcolor{yellow!50} & \cellcolor{green!50} & \cellcolor{green!50} & \cellcolor{yellow!50} & \cellcolor{green!50} & \cellcolor{green!50} & \cellcolor{green!50} & \cellcolor{green!50} & \cellcolor{green!50} & \\ 
\hline \multirow{ 3}{*}{Industry}  &  Monetary &  &  &  &  &  & \cellcolor{yellow!50} &  &  &  &  &  &  &  & \\ 
&  Information & \cellcolor{yellow!50} &  &  &  & \cellcolor{yellow!50} & \cellcolor{yellow!50} &  &  &  &  &  & \cellcolor{yellow!50} &  & \\ 
&  Spread &  &  &  & \cellcolor{yellow!50} & \cellcolor{green!50} & \cellcolor{green!50} & \cellcolor{green!50} & \cellcolor{green!50} & \cellcolor{green!50} & \cellcolor{green!50} & \cellcolor{green!50} & \cellcolor{green!50} & \cellcolor{green!50} & \\ 
\hline \multirow{ 3}{*}{LT Rate} &  Monetary &  &  &  &  &  &  &  & \cellcolor{green!50} & \cellcolor{green!50} & \cellcolor{green!50} & \cellcolor{green!50} & \cellcolor{green!50} & \cellcolor{green!50} & \\ 
&  Information &  &  &  & \cellcolor{yellow!50} & \cellcolor{yellow!50} &  &  &  &  &  &  &  &  & \\ 
&  Spread &  &  &  &  &  &  & \cellcolor{green!50} & \cellcolor{green!50} &  &  &  &  &  & \\ 
   \hline

&h=  & 13 & 14 & 15 & 16 & 17 & 18 & 19 & 20 & 21 & 22 & 23 & 24 & & \\ 

\hline \multirow{ 3}{*}{REER} &Monetary &  &  &  &  & \cellcolor{green!50} & \cellcolor{green!50} & \cellcolor{green!50} &  & \cellcolor{green!50} & \cellcolor{green!50} & \cellcolor{green!50} &  & & \\ 
&  Information &  &  &  &  &  &  &  &  & \cellcolor{green!50} & \cellcolor{green!50} &  &  & & \\ 
 & Spread & \cellcolor{yellow!50} & \cellcolor{yellow!50} &  & \cellcolor{yellow!50} &  &  &  &  &  &  & \cellcolor{yellow!50} & \cellcolor{green!50} & & \\ 
\hline \multirow{ 3}{*}{Unemp.}  & Monetary &  &  &  &  &  &  &  &  &  &  &  &  & & \\ 
&  Information & \cellcolor{green!50} & \cellcolor{green!50} & \cellcolor{green!50} & \cellcolor{yellow!50} & \cellcolor{yellow!50} & \cellcolor{yellow!50} &  &  &  &  &  &  & & \\ 
 & Spread & \cellcolor{green!50} & \cellcolor{green!50} &  &  &  &  &  &  &  &  &  &  & & \\ 
\hline \multirow{ 3}{*}{CPI} &  Monetary &  &  &  &  &  & \cellcolor{green!50} & \cellcolor{yellow!50} & \cellcolor{green!50} &  &  &  &  & & \\ 
&  Information &  &  &  &  &  &  &  &  &  &  &  &  & & \\ 
&  Spread & \cellcolor{green!50} & \cellcolor{green!50} & \cellcolor{green!50} & \cellcolor{green!50} & \cellcolor{green!50} &  &  &  &  &  &  &  & & \\ 
\hline \multirow{ 3}{*}{Industry}  &  Monetary &  &  &  &  &  &  &  &  &  &  &  &  & & \\ 
&  Information & \cellcolor{yellow!50} &  &  &  &  &  &  &  &  &  &  &  & & \\ 
 & Spread & \cellcolor{green!50} & \cellcolor{green!50} & \cellcolor{green!50} & \cellcolor{green!50} & \cellcolor{green!50} &  &  &  &  & \cellcolor{green!50} & \cellcolor{green!50} & \cellcolor{green!50} & & \\ 
\hline \multirow{ 3}{*}{LT Rate} &  Monetary & \cellcolor{green!50} & \cellcolor{green!50} & \cellcolor{green!50} & \cellcolor{yellow!50} & \cellcolor{yellow!50} &  &  &  &  &  &  &  & & \\ 
&  Information &  &  &  &  &  & \cellcolor{green!50} &  &  & \cellcolor{green!50} & \cellcolor{green!50} & \cellcolor{green!50} &  & & \\ 
&  Spread &  &  &  &  &  &  &  &  &  &  &  &  & & \\ 
   \hline
   \end{tabular}}
\end{footnotesize}
    \caption{Conditional IRFs Positive Sign-Effect}
    \floatfoot{We report the significance level of the conditional IRFs for a positive shock with the sign effect specification.  White cells refer to p-values greater than 0.1, yellow cells are p-values between 0.1 and 0.05 and green cells are p-values smaller than 0.05.}
\end{table}
\noindent
\textbf{Table 24}, provides the results for significance tests on the sign-effect, positive-shock, conditional impulse response functions. That is we test the hypotheses:
$$H_0^{i,h}:\hat{\psi}_{j,h,0}^i+\hat{\Gamma}_{j,h,0}^i=0, \quad i \in \{monetary,information,spread\}$$
To do so we construct the restriction matrix:
$$R_1=\langle 1,0,0,1,0,0,...\rangle \quad R_2=\langle 0,1,0,0,1,0,...\rangle \quad R_3=\langle 0,0,1,0,0,1,...\rangle $$
We then construct the Wald statistic in the same way. We find that conditional on $x_t\vargeq0$ and $\delta>0$, there are significant effects of the monetary policy shock to REER, CPI and the Long term Rate, significant effects of the information shock to unemployment at horizon 11 to 18 and significant effects of the spread shocks to CPI, Industry and Unemployment.
\pagebreak
\begin{table}[t]
\begin{footnotesize}
\centering
\scalebox{0.8}{ \begin{tabular}{|r|r|rrrrrrrrrrrrrr|}
  \hline
& h= & 0 & 1 & 2 & 3 & 4 & 5 & 6 & 7 & 8 & 9 & 10 & 11 & 12 &\\ 
 
\hline \multirow{ 3}{*}{REER} &Monetary & \cellcolor{yellow!50} & \cellcolor{green!50} &  &  &  &  &  &  &  &  &  &  &  & \\ 
&  Information &  &  &  &  &  &  & \cellcolor{yellow!50} & \cellcolor{green!50} & \cellcolor{green!50} & \cellcolor{green!50} & \cellcolor{yellow!50} & \cellcolor{yellow!50} & \cellcolor{yellow!50} & \\ 
&  Spread &  & \cellcolor{green!50} &  &  &  &  &  &  & \cellcolor{green!50} & \cellcolor{green!50} & \cellcolor{green!50} & \cellcolor{green!50} & \cellcolor{green!50} & \\ 
\hline \multirow{ 3}{*}{Unemp.} &  Monetary &  &  &  &  &  &  & \cellcolor{yellow!50} &  &  &  &  &  &  & \\ 
&  Information & \cellcolor{yellow!50} &  &  &  &  &  &  & \cellcolor{yellow!50} & \cellcolor{green!50} &  & \cellcolor{green!50} & \cellcolor{green!50} & \cellcolor{yellow!50} & \\ 
&  Spread &  &  &  & \cellcolor{green!50} & \cellcolor{green!50} &  &  &  &  &  &  &  &  & \\ 
\hline \multirow{ 3}{*}{CPI} &  Monetary &  &  &  &  &  & \cellcolor{green!50} & \cellcolor{green!50} & \cellcolor{green!50} & \cellcolor{green!50} & \cellcolor{yellow!50} &  &  &  & \\ 
&  Information &  &  &  &  &  &  &  &  &  &  &  &  &  & \\ 
&  Spread &  &  &  &  &  & \cellcolor{green!50} & \cellcolor{green!50} & \cellcolor{green!50} & \cellcolor{green!50} & \cellcolor{green!50} & \cellcolor{green!50} & \cellcolor{green!50} & \cellcolor{green!50} & \\ 
\hline \multirow{ 3}{*}{Industry}  &  Monetary &  &  &  &  &  & \cellcolor{green!50} & \cellcolor{green!50} & \cellcolor{yellow!50} & \cellcolor{green!50} &  &  &  &  & \\ 
&  Information &  &  &  &  &  &  & \cellcolor{green!50} &  &  &  &  &  &  & \\ 
&  Spread &  &  &  &  &  & \cellcolor{green!50} & \cellcolor{green!50} & \cellcolor{green!50} & \cellcolor{green!50} & \cellcolor{green!50} & \cellcolor{green!50} & \cellcolor{green!50} & \cellcolor{green!50} & \\ 
\hline \multirow{ 3}{*}{LT Rate} &  Monetary & \cellcolor{green!50} &  &  &  &  &  &  &  &  &  &  &  &  & \\ 
&  Information &  &  &  & \cellcolor{yellow!50} &  &  & \cellcolor{green!50} & \cellcolor{green!50} &  &  & \cellcolor{yellow!50} & \cellcolor{green!50} & \cellcolor{green!50} & \\ 
&  Spread & \cellcolor{green!50} &  &  &  &  &  & \cellcolor{yellow!50} &  &  &  &  &  &  & \\ 
   \hline

&h= & 13 & 14 & 15 & 16 & 17 & 18 & 19 & 20 & 21 & 22 & 23 & 24 & & \\ 

\hline \multirow{ 3}{*}{REER} &Monetary &  &  &  &  &  &  &  &  &  &  &  &  & & \\ 
&  Information &  &  &  &  &  &  &  & \cellcolor{yellow!50} & \cellcolor{green!50} & \cellcolor{green!50} &  &  & & \\ 
&  Spread & \cellcolor{green!50} & \cellcolor{green!50} & \cellcolor{green!50} & \cellcolor{green!50} &  &  &  &  &  &  &  &  & & \\ 
\hline \multirow{ 3}{*}{Unemp.} &  Monetary &  &  &  &  &  &  &  &  &  &  &  &  & & \\ 
&  Information & \cellcolor{yellow!50} & \cellcolor{green!50} & \cellcolor{yellow!50} &  &  &  &  &  &  &  &  &  & & \\ 
&  Spread &  &  &  &  &  &  &  &  &  &  &  & \cellcolor{yellow!50} & & \\ 
\hline \multirow{ 3}{*}{CPI} &  Monetary &  &  &  &  &  &  &  &  &  &  &  &  & & \\ 
&  Information &  &  &  &  &  &  & \cellcolor{yellow!50} & \cellcolor{green!50} &  &  &  &  & & \\ 
&  Spread & \cellcolor{green!50} & \cellcolor{green!50} & \cellcolor{green!50} & \cellcolor{green!50} & \cellcolor{yellow!50} & \cellcolor{green!50} & \cellcolor{yellow!50} &  &  &  &  &  & & \\ 
\hline \multirow{ 3}{*}{Industry}  &  Monetary &  &  &  &  &  &  &  &  &  &  &  &  & & \\ 
&  Information &  &  &  &  &  &  &  &  &  &  &  &  & & \\ 
&  Spread & \cellcolor{green!50} &  &  &  &  &  &  &  &  &  &  &  & & \\ 
\hline \multirow{ 3}{*}{LT Rate} &  Monetary &  & \cellcolor{yellow!50} & \cellcolor{green!50} &  &  &  &  &  &  & \cellcolor{yellow!50} &  &  & & \\ 
&  Information & \cellcolor{green!50} & \cellcolor{green!50} & \cellcolor{green!50} & \cellcolor{yellow!50} & \cellcolor{yellow!50} & \cellcolor{yellow!50} & \cellcolor{yellow!50} & \cellcolor{green!50} &  &  &  &  & & \\ 
& Spread & \cellcolor{yellow!50} &  &  &  &  &  &  &  &  &  & \cellcolor{yellow!50} &  & & \\ 
   \hline
   \end{tabular}}
\end{footnotesize}
    \caption{Conditional IRFs Negative Sign-Effect}
    \floatfoot{We report the significance level of the conditional IRFs for a negative shock with the sign effect specification. Yellow cells are p-values between 0.1 and 0.05 and green cells are p-values smaller than 0.05.}
\end{table}

\noindent
\textbf{Table 25}, provides the results for significance tests on the sign-effect, negative-shock, conditional impulse response functions. That is we test the hypotheses:
$$H_0^{i,h}:\hat{\Gamma}_{j,h,0}^i-\hat{\psi}_{j,h,0}^i=0, \quad i \in \{monetary,information,spread\}$$
Which imply the restriction matrices:
$$R_1=\langle -1,0,0,1,0,0,...\rangle \quad R_2=\langle 0,-1,0,0,1,0,...\rangle \quad R_3=\langle 0,0,-1,0,0,1,...\rangle $$
From the wald test we find that if $\delta<0$ and $x_t \varleq 0$, then there are significant effects of a monetary policy shock to the CPI and Industry, significant effects of an information shock to the REER, unemployement rate and the long term rate and significant effects of the spread shock to the REER, the CPI and industrial production.
\end{document}